\documentclass[preprint2]{exoplanet} 

\pdfoutput=1   

\usepackage{times}

\def\m{\rm \,m}

\def\K{\rm \,K}

\def\sec{\rm \, sec}
\def\km{\rm \, km}

\def\del{{\partial}}

\begin{document}

\title{\textbf{\LARGE Atmospheric Circulation of Exoplanets}}

\author {\textbf{\large Adam P. Showman}}
\affil{\small\em University of Arizona}

\author {\textbf{\large James Y-K. Cho}}
\affil{\small\em Queen Mary, University of London}

\author {\textbf{\large Kristen Menou}}
\affil{\small\em Columbia University}

\begin{abstract}
\begin{list}{ } {\rightmargin 1in}
\baselineskip = 11pt
\parindent=1pc
{\small We survey the basic principles of atmospheric dynamics relevant 
to explaining existing and future observations of exoplanets, 
both gas giant and terrestrial. Given the paucity of data on 
exoplanet atmospheres, 
our approach is to emphasize fundamental principles and insights gained
from Solar-System studies that are likely to be generalizable to
exoplanets.  We begin by presenting the hierarchy of  
basic equations used
in atmospheric dynamics, including the Navier-Stokes, primitive,
shallow-water, and two-dimensional nondivergent
models.  We  then survey key concepts in atmospheric dynamics, 
including the importance of 
planetary rotation, the concept of balance, and simple scaling arguments 
to show how turbulent interactions generally produce large-scale east-west 
banding on rotating planets.  We next turn to issues specific to
giant planets, including their expected interior and atmospheric thermal 
structures, the implications for their wind patterns, and mechanisms 
to pump their east-west jets.  Hot Jupiter atmospheric dynamics are given
particular attention, as these close-in planets have been the subject of 
most of the concrete developments in the study of exoplanetary atmospheres.  
We then turn to the basic elements of circulation on terrestrial
planets as inferred from Solar-System studies, including Hadley cells, 
jet streams, processes that govern the large-scale horizontal 
temperature contrasts, and climate, and we discuss how these insights may
apply to terrestrial exoplanets.   Although exoplanets
surely possess a greater diversity of circulation regimes
than seen on the planets in our Solar System,
our guiding philosophy is that the multi-decade study of 
Solar-System planets reviewed here provides a foundation upon which 
our understanding of more exotic exoplanetary meteorology must build.
 \\~\\~\\~}
 
\end{list}
\end{abstract}

\section{INTRODUCTION}
\label{Intro}

The study of atmospheric circulation and climate began
hundreds of years ago with attempts to understand the processes 
that determine the distribution of surface winds on the Earth 
\citep[e.g.,][]{hadley-1735}.  As theories of Earth's general circulation
became more sophisticated \citep[e.g.,][]{lorenz-1967}, the 
characterization of Mars, Venus, Jupiter, and other Solar-System
planets by spacecraft starting in the 1960s demonstrated
that the climate and circulation of other atmospheres 
differ, sometimes radically, from that of Earth. 
Exoplanets, occupying a far greater range of physical and 
orbital characteristics than planets in our Solar System,
likewise plausibly span an even greater diversity of 
circulation and climate regimes.  This diversity provides
a motivation for extending the theory of atmospheric 
circulation beyond our terrestrial experience.  Despite
continuing questions, our understanding of the circulation 
of the modern Earth atmosphere is now well developed 
\citep[see, e.g.,][]{held-2000, schneider-2006, vallis-2006}, 
but attempts to unravel the atmospheric dynamics of Venus, Jupiter, 
and other Solar-System planets remain ongoing, and the
study of atmospheric circulation of exoplanets is in its
infancy.

For exoplanets, driving questions fall into
several overlapping categories.  First, we wish to understand 
and explain new observations constraining atmospheric
structure, such as light curves, photometry,
and spectra obtained with the {\it Spitzer, } {\it Hubble}, or 
{\it James Webb Space Telescopes (JWST)}, thus helping to
characterize specific exoplanets as remote worlds.   Second, we wish
to extend the theory of atmospheric circulation to the
wide range of planetary parameters encompassed by exoplanets.
Existing theory was primarily developed for conditions
relevant to Earth, and our understanding of how atmospheric 
circulation depends on atmospheric mass,
composition, stellar flux, planetary rotation rate, 
orbital eccentricity, and other parameters remains rudimentary. 
Significant progress is possible with theoretical,
numerical, and laboratory investigations that span a 
wider range of planetary parameters.  Third, we wish
to understand the conditions under which planets are habitable,
and answering this question requires addressing the
intertwined issues of atmospheric circulation and climate.

\begin{figure*}
 \epsscale{1.0}
\plotone{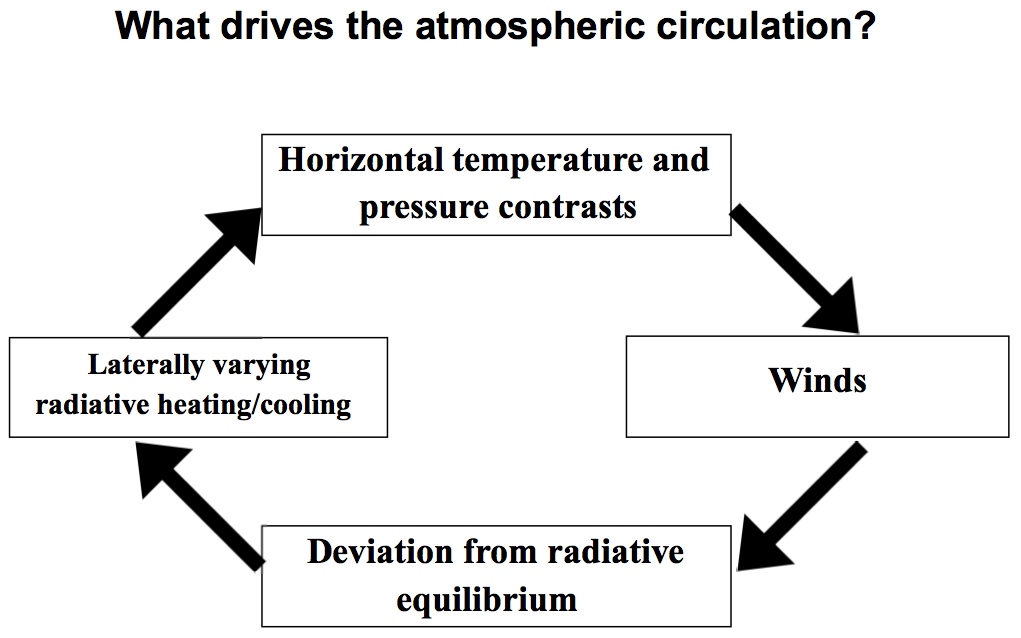}
 \caption{\small Atmospheric circulation results from a coupled interaction
between radiation and hydrodynamics:  horizontal temperature and pressure
contrasts generate winds, which drive the atmosphere away 
from local radiative equilibrium.  This in turn
allows the spatially variable thermodynamic (radiative)
heating and cooling that maintains the horizontal temperature and
pressure contrasts.}
\label{flow-chart}
 \end{figure*}

\begin{figure*}
 \epsscale{1.0}
\plotone{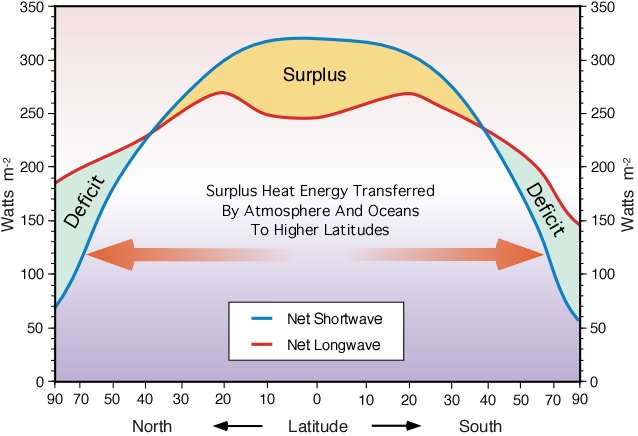}
 \caption{\small Earth's energy balance.  The Earth absorbs 
more sunlight at the equator than the poles (blue curve, denoted
``shortwave'').  The Earth also radiates more infrared energy
to space at the equator than the poles (red curve, denoted ``longwave'').
However, because the atmospheric/oceanic circulation act to mute
the latitudinal temperature contrasts (relative to radiative equilibrium),
the longwave radiation exhibits less latitudinal variation than the
shortwave absorption.  Thus, the circulation leads to net heating at the 
equator and net cooling at the poles, which in turn drives the circulation.  
Data are an annual-average for 1987 obtained from the NASA Earth Radiation 
Budget Experiment (ERBE) project.  Copyright M. Pidwirny, 
www.physicalgeography.net, used with permission.}
\label{earth-radiation-balance}
 \end{figure*}

What drives atmospheric circulation?  
Horizontal temperature contrasts imply the existence of 
horizontal pressure contrasts, which drive winds.  
The winds in turn push the atmosphere
away from radiative equilibrium by transporting heat from
hot regions to cold regions (e.g., from the equator to the 
poles on Earth).  This deviation from radiative equilibrium
allows net radiative heating and cooling to occur, thus
helping to maintain the horizontal temperature and pressure
contrasts that drive the winds (see Fig.~\ref{flow-chart}).  
Spatial contrasts in thermodynamic heating/cooling thus
fundamentally drive the circulation, yet it is the existence 
of the circulation that allows these heating/cooling patterns
to exist. (In the absence of a circulation, the atmosphere
would relax into a radiative-equilibrium state with a
net heating rate of zero.)  The atmospheric circulation is 
thus a coupled radiation-hydrodynamics problem.  On the
Earth, for example (see Fig.~\ref{earth-radiation-balance}),
the equator and poles are not in radiative equilibrium.  The
equator is subject to net heating, the poles to net cooling,
and it is the mean latitudinal heat transport that is both
responsible for and driven by these net imbalances.

The mean climate (e.g., the global-mean surface
temperature of a planet) depends foremost on the
absorbed stellar flux and the atmosphere's need to 
reradiate that energy to space.  Yet even the global-mean
climate is strongly affected by the atmospheric mass, 
composition, and circulation.  On a terrestrial planet,
for example, the circulation helps to control the distribution
of clouds and surface ice, which in turn determine the
planetary albedo and the mean surface temperature.
In some cases, a planetary climate can have multiple
equilibria (e.g., a warm, ice-free state or a cold,
ice-covered ``snowball Earth'' state), and in such
cases the circulation plays an important role in
determining the relative stability of these equilibria.

Understanding the atmosphere/climate system is
challenging because of its nonlinearity, which involves
multiple positive and negative feedbacks between
radiation, clouds, dynamics, surface processes, 
planetary interior, and life (if any).  The inherent 
nonlinearity of fluid motion further implies that even 
atmospheric-circulation models neglecting the radiative, 
cloud, and surface/interior components can exhibit 
a large variety of behaviors.

From the perspective of studying the atmospheric circulation, 
transiting exoplanets are particularly intriguing because they
allow constraints on key planetary attributes that 
are a prerequisite to characterizing an atmosphere's
circulation regime.  When
combined with Doppler velocity data, transit observations permit a
direct measurement of the exoplanet's radius, mass and thus surface
gravity\footnote{Combining Doppler velocity and transit
measurements lifts the mass-inclination degeneracy.}.  With the
additional expectation that close-in exoplanets are
tidally locked if on a circular orbit, or
pseudo-synchronized\footnote{Pseudo-synchronization refers to a state
of tidal synchronization achieved only at periastron passage (=closest
approach), as expected from the strong dependence of tides with
orbital separation.} if on an eccentric orbit, the planetary rotation
rate is thus indirectly known as well.  Knowledge of
the radius, surface gravity, rotation rate and external irradiation
conditions for several exoplanets, together with the availability of
direct observational constraints on their emission, absorption and
reflection properties, opens the way for the development of
comparative atmospheric science beyond the reach of our own Solar
System.

The need to interpret these astronomical data reliably, by accounting
for the effects of atmospheric circulation and understanding its
consequences for the resulting planetary emission, absorption and
reflection properties, is the central theme of this chapter. Tidally
locked close-in exoplanets, for example, are subject to an
unusual situation of permanent day/night radiative forcing, which
does not exist in our Solar System\footnote{Venus may provide a
partial analogy, which has not been fully exploited yet.}.  To address
the new regimes of forcings and responses of these exoplanetary
atmospheres, a discussion of fundamental principles of atmospheric
fluid dynamics and how they are implemented in multi-dimensional,
coupled radiation-hydrodynamics numerical models of the GCM (General
Circulation Model) type is required.

Contemplating the wide diversity of exoplanets raises a number of
fundamental questions.  What determines the mean wind speeds, direction, 
and 3D flow geometry in atmospheres?  What controls the equator-to-pole
and day-night temperature differences?  What controls the 
frequencies and spatial scales of temporal variability?  
What role does the circulation play in controlling the mean climate 
(e.g., global-mean surface temperature, composition) of an atmosphere?  
How do these answers depend on parameters such as the planetary 
rotation rate, gravity, atmospheric mass and composition, and 
stellar flux?  And, finally, what are the implications for
observations and habitability of exoplanets?

At present, only partial answers to these questions exist
\citep[see reviews by][]{showman-etal-2008b, cho-2008}.  
With upcoming observations of exoplanets, constraints from Solar-System
atmospheres, and careful theoretical work, significant progress is
possible over the next decade.
While a rich variety of atmospheric flow behaviors is 
realized in the Solar System alone---and an even wider
diversity is possible on exoplanets---the fundamental physical
principles obeyed by all planetary atmospheres are nonetheless
universal. With this unifying notion in mind, this chapter provides a
basic description of atmospheric circulation principles developed on
the basis of extensive Solar-System studies and discusses the
prospects for using these principles to better understand physical
conditions in the atmospheres of remote worlds.

The plan of this chapter is as follows.
In Section~\ref{equations}, we introduce several of the equation sets 
that are used to investigate atmospheric circulation at
varying levels of complexity. This is followed (Section~\ref{basic-concepts})
by a tutorial on basic ideas in atmospheric dynamics, including 
atmospheric energetics, timescale arguments, force
balances relevant to the large-scale circulation,
the important role of rotation in generating east-west 
banding, and the role of waves and eddies in shaping
the circulation.  In Section~\ref{giants}, we survey the 
atmospheric dynamics of giant planets, beginning with
generic arguments to constrain the thermal and dynamical
structure and proceeding to specific models for 
understanding the circulation of our ``local'' giant
planets (Jupiter, Saturn, Uranus, Neptune) as well
as hot Jupiters and hot Neptunes.\footnote{The terms hot
Jupiter and hot Neptune refer to giant exoplanets with masses
comparable to those of Jupiter and Neptune, respectively, with
orbital semi-major axes less than $\sim0.1\,$AU, leading to high
temperatures.}  In Section~\ref{terrestrial}, we turn
to the climate and circulation of terrestrial exoplanets.
Observational constraints in this area do not yet exist,
and so our goal is simply to summarize basic concepts that
we expect to become relevant as this field expands
over the next decade.  This includes a description
of climate feedbacks (Section~\ref{climate}), global circulation
regimes (Section~\ref{regimes}), Hadley-cell dynamics (Section~\ref{hadley}),
the dynamics of the so-called midlatitude ``baroclinic'' zones
where baroclinic instabilities dominate 
(Section~\ref{baroclinic-zone}), the slowly rotating regime relevant to
Venus and Titan (Section~\ref{slowly-rotating}), and finally a survey
of how the circulation responds to the unusual forcing
associated with synchronous rotation, extreme obliquities,
or extreme orbital eccentricities (Section~\ref{unusual-forcing}).  
The latter topics, while perhaps the most relevant, are the
least understood theoretically.  In Section~\ref{highlights} we
summarize recent highlights, both observational and
theoretical, and in Section~\ref{future} we finish with a survey
of future prospects.

\bigskip
\section{EQUATIONS GOVERNING ATMOSPHERIC CIRCULATION}
\bigskip
\label{equations}

A wide range of dynamical models has been developed to explore 
atmospheric circulation and climate.   Such models are
used to explain observations, understand mechanisms 
that govern the circulation/climate system, determine
the sensitivity of a planet's circulation/climate to changes 
in parameters, test hypotheses about how the system works,
and make predictions.  

Developing a good understanding of atmospheric 
circulation requires the use of a hierarchy of atmospheric fluid dynamics 
models.  Complex models that properly represent the full range of 
physical processes may be required for detailed predictions or comparisons
with observations, but their very complexity can obscure the
specific physical mechanisms causing a given phenomenon.  In contrast,
simpler models contain less physics, but they are easier to diagnose 
and can often lead to a better understanding of cause-and-effect in an 
idealized setting.  Whether a given model contains sufficient physics
to explain a given phenomenon is a question that can only be
answered by exploring a hierarchy of models with a range of complexity.
Exploring a hierarchy of models is therefore invaluable because 
it allows one to determine the minimal set of physical ingredients
that are needed to generate a specific atmospheric behavior---insight
that typically cannot be obtained from one type of model alone.

The equations governing atmospheric behavior derive from 
conservation of momentum, mass, and energy for a fluid, which we
here assume to be an electrically neutral continuum.  For three-dimensional
models, where momentum is a three-dimensional vector, this implies
five governing equations, which are generally represented as five coupled 
partial differential equations for the three-dimensional velocity,
density, and internal energy per mass (with other thermodynamic
state variables determined from density and internal energy by the
equation of state).  The {\it Navier-Stokes equations}, described in
\S~\ref{navier-stokes}, constitute the canonical example and
provide a complete representation of a continuum, electrically
neutral, viscous fluid in three dimensions.

So-called {\it reduced} models simplify the dynamics in one or 
more ways, for example by reducing specified equations to their 
leading-order balances.  For example, because most atmospheres
have large aspect ratios (with characteristic horizontal length scales
for the global circulation typically 10--100 times the characteristic
vertical scales), the vertical momentum balance is typically close
to a local hydrostatic balance, with the local weight of fluid parcels
balancing the local vertical pressure gradient [see, e.g.,
\citet[][pp.~41-42]{holton-2004} or \citet[][pp.~80-84]{vallis-2006}
for a derivation].  The {\it primitive equations}, described in
\S~\ref{tpe}, formalize this fact by replacing the full vertical
momentum equation with local vertical hydrostatic balance.  Although the
system is still governed by five equations, this alteration simplifies
the dynamics by removing vertically propagating sound waves, which
are unimportant for most meteorological phenomena.  It also leads
to mathematical simplification, making it easier to obtain
analytic and numerical solutions. This is the equation set that
forms the basis for most cutting-edge global-scale climate models used
for studying atmospheres of Solar-System planets, 
although some global-scale high-resolution models
now include non-hydrostatic effects.

A further common reduction is to simplify the dynamics to a one-layer
model representing (for example) the vertically averaged flow.
The most important example is the {\it shallow-water model,}
described in \S~\ref{shallow-water}, which govern the behavior of
a thin layer of constant-density fluid of variable thickness.
This implies three coupled equations
for horizontal momentum and mass conservation (governing the evolution 
of the two horizontal velocity components and the layer thickness) 
as a function of longitude, latitude, and time.  Although 
highly idealized, the shallow-water model has proven
surprisingly successful at capturing a wide range of atmospheric
phenomena and has become a time-honored process model in
atmospheric dynamics \citep[see, for example,][Chapter 3]{pedlosky-1987}.

A further reduction results from assuming the fluid-layer thickness
is constant in the shallow-water model.  Given that density
is also constant, the mass conservation equation then becomes a statement 
that horizontal convergence/divergence is zero.  This constraint,
which has the effect of removing gravity (buoyancy) waves from
the system, allows the horizontal velocity components to be represented
using a streamfunction, leading finally to a single governing partial
differential equation for the streamfunction as a function of longitude,
latitude, and time.  \S~\ref{nondivergent} describes this 
{\it two-dimensional, non-divergent model}.  The 
impressive reduction from five coupled equations in five dependent variables
(as for the Navier-Stokes or primitive equations)
to one equation in one variable leads to great mathematical simplification,
enabling analytic solutions in cases when they are otherwise difficult
to obtain.  Moreover, the exlusion of buoyancy effects, gravity
waves, and vertical structure leads to a conceptual simplification, 
allowing the exploration of (for example) vortex and jet formation 
in the most idealized possible setting.  

A comparison of results from the full range of models described 
here provides a path toward identifying the relative roles
of acoustic waves, vertical structure, buoyancy effects, and gravity waves
in affecting any given meteorological phenomenon of interest.
We now present the equations associated with each of these models.

\subsection{Navier-Stokes Equations}
\label{navier-stokes}
Let ${\bf u} = {\bf u}({\bf x},t)$ be velocity at position ${\bf x}$
and time $t$, where ${\bf x},{\bf u}\in\mathbb{R}^3$.  If the
frictional force per unit area of the fluid is linearly proportional
to shear in the fluid, then it is a Newtonian fluid
\citep[e.g.,][]{batchelor-1967}.  Such fluids are described by the
Navier-Stokes equations:
\begin{mathletters}\label{nse}
\begin{eqnarray}
  \frac{D{\bf u}}{D t}\ =\ -\frac{1}{\rho}\nabla p + {\bf f}_b + 
\frac{1}{\rho}\,\nabla\!\cdot\!\left\{2\mu\left[\mathsf{e} -       
\frac{1}{3}(\nabla\!\cdot\!{\bf u})\,\mathbb{I}\right]\right\},
\end{eqnarray}
\mbox{where}
\begin{eqnarray}
\frac{D}{D t}\ =\ \frac{\del}{\del t} + {\bf u}\!\cdot\!\nabla
\end{eqnarray}
\noindent
is the material derivative (i.e., the deriative following the motion
of a fluid element).  Here, $\rho$ is density, $p$ is pressure,
${\bf f}_b$ represents various body forces per mass (e.g., gravity and
Coriolis), $\mu$~is molecular dynamic viscosity, and $\mathsf{e} =
\frac{1}{2}[(\nabla{\bf u}) + (\nabla{\bf u})^{\mathsf{T}}]$ and
$\mathbb{I}$ are the strain-rate and unit tensors, respectively.  In
Eq.~(\ref{nse}a) the quantity inside the braces is the viscous stress
tensor.  Here, as in Eq.~(\ref{energy}) below, the average normal
viscous stress (bulk viscosity) has been assumed to be zero.
\end{mathletters}

Eq.~(\ref{nse}a) is closed with the following equations for mass (per
unit volume), internal energy (per unit mass), and state:
\begin{eqnarray}\label{mass}
  \frac{D\rho}{Dt}\ =\  -\rho\,\nabla\!\cdot\!{\bf u},
\end{eqnarray}
\begin{eqnarray}\label{energy} \nonumber
  \frac{D\epsilon}{D t}\ =\ 
  -\frac{p}{\rho}\left(\nabla\cdot{\bf u}\right) + 
  \frac{2\mu}{\rho}\left[\mathsf{e}\! :\! (\nabla{\bf u})^{\mathsf{T}} 
    - \frac{1}{3}(\nabla\cdot{\bf u})^2\,\right]\\    
  + \frac{1}{\rho}\nabla\cdot\left(K_{\mbox{\tiny T}}\nabla T\right) 
  + {\cal Q},
\end{eqnarray}
\begin{eqnarray}\label{eos}
  p\ =\ p(\rho,T),
\end{eqnarray}
where $\epsilon = \epsilon(T,s)$ is specific internal energy, $s$ is
specific entropy, $K_{\mbox{\tiny T}}$ is heat conduction coefficient,
$T$ is temperature, and ${\cal Q}$ is thermodynamic heating rate per
mass.  In Eq.~(\ref{energy}) ``\,:\,'' is scalar-product (i.e.,
component-wise multiplication) operator for two tensors.
Eqs.~(\ref{nse}--\ref{eos}) constitute 6 equations for 6 independent
unknowns, $\{{\bf u},p,\rho,T\}$.  Note that for a homogeneous
thermodynamic system, which involves a single phase, only two state
variables can vary independently; hence, there are only two
thermodynamic degrees of freedom for such a system.

The neutral atmosphere is well described by Eqs.~(\ref{nse}--\ref{eos}) when 
the characteristic length scale $L$ is much larger than the mean free path of 
the constituents that make up the atmosphere.  Hence, the equations are valid 
up to heights where ionization is not significant and the continuum hypothesis 
does not break down.  Under normal conditions, 
the atmosphere behaves like an ideal gas.  The parameters 
$\mu$, $K_{\mbox{\tiny T}}$, and other physical properties of the fluid depend on 
$T$, as well as $\rho$.  When appreciable temperature differences exists in the 
flow field, these properties must be regarded as a function of position.  For 
large-scale atmosphere applications, however, the terms involving $\mu$ in 
Eqs.~(\ref{nse}a) and (\ref{energy}) are small and can be neglected in most 
cases. The typical boundary conditions are ${\bf u}\cdot{\bf n} = 0$ at 
the lower boundary, where ${\bf n}$ is the normal to the boundary, and 
$\rho, p\rightarrow 0$ as $z\rightarrow\infty$.  For local, limited area 
models, periodic boundary conditions are often used.

\subsection{The Primitive Equations}\label{tpe}

On the large scale (to be more precisely quantified below), the motion
of an atmosphere is governed by the primitive equations.  They read
\citep[e.g.,][] {salby-1996}:
\begin{mathletters}\label{pe}
\begin{eqnarray}
  \frac{D{\bf v }}{D t}
  \! &\! =\! &\! -\nabla\! _p\,\Phi - f{\bf k}\!\times\!{\bf v} +   
  {\cal F} - {\cal D}\\
  \frac{\del\Phi}{\del p}\! &\!  =\! &\! -\frac{1}{\rho}\\
  \frac{\del\varpi}{\del p} &\! =\! &\! -\nabla\! _p\cdot{\bf v}\\ 
  \frac{D\theta}{D t} &\! =\! &\! \frac{\theta}{c_p T}\ 
  \dot{q}_{\rm net}\ ,
\end{eqnarray}
\mbox{where}
\begin{equation}
  \frac{D}{D t} \ = \ \frac{\del}{\del t} + 
  {\bf v}\!\cdot\!\nabla\! _p + \varpi\!\frac{\del}{\del p}\ .
\end{equation}
\noindent
Note here that $p$, rather than the geometric height $z$, is used as
the vertical coordinate.  This coordinate, which simplifies the
gradient term in Eq.~(\ref{pe}a), is common in atmospheric studies; it
renders $z = z({\bf x},p,t)$ a dependent variable, where now ${\bf
  x}\!\in\!\mathbb{R}^2$.  In Eq.~(\ref{pe}) ${\bf v}({\bf x},t)\! =\!
(u,v)$ is the (eastward\footnote{Cardinal directions are defined here consistent
with everyday usage; east is defined to be the prograde direction,
that is, the direction in which the planet rotates.  North is the direction
along which ${\bf \Omega}\cdot{\bf k}$ becomes more positive.}, northward) 
velocity in a frame rotating with
$\mathbf{\Omega}$, where $\mathbf{\Omega}$ is the planetary rotation vector as represented 
in inertial space; $\Phi = gz$ is the geopotential, where $g$ is the
gravitational acceleration \citep[assumed to be constant and to include 
the centrifugal acceleration contribution; see][pp. 13-14] {holton-2004}
and $z$ is the height above a fiducial geopotential surface; 
${\bf k}$ is the local upward unit vector;
$f = 2 \Omega \sin \phi$ is the Coriolis parameter,
the locally vertical component of the planetary vorticity vector 
$2{\bf\Omega}$; $\nabla\! _p$ is the horizontal gradient on a $p$-surface;
$\varpi = Dp/Dt$ is the vertical velocity; ${\cal F}$ and ${\cal D}$
represent the momentum sources and sinks, respectively; $\theta =
T(p_{\mbox{\tiny ref}}/p)^{\kappa}$ is the potential
temperature\footnote{The potential temperature $\theta$ is related to
  the entropy $s$ by $ds = c_p d\ln\theta$.  When $c_p$ is constant,
  this yields $\theta=T(p_{\rm ref}/p)^{\kappa}$.}, where
$p_{\mbox{\tiny ref}}$ is a reference pressure and ${\kappa} = R/c_p$
with $R$ the specific gas constant and $c_p$ the specific heat at
constant pressure; and $\dot{q}_{\rm net}$ is the {\it net}
diabatic heating rate (heating minus cooling). 
Note that $\dot{q}_{\rm net}$ can include not only
radiative heating/cooling but latent heating and, at low pressures
where the thermal conductivity becomes large, conductive heating.
The Newtonian cooling scheme, which relaxes temperature toward
a prescribed radiative-equilibrium temperature over a specified
radiative time constant, is one simple parameterization of 
$\dot{q}_{\rm net}$.
\end{mathletters} 

The fundamental presumption in the use of Eqs.~(\ref{pe}) is that
small scale processes are parameterizable within the framework of
large-scale dynamics.  Here by ``large'' scales, it is meant typically
$L \gtrsim a/10$, where $a$ is the planetary radius.  By ``small'' 
scales, it is meant those scales that are not resolvable numerically 
by global models---typically $\lesssim a/10$. Regions
of the atmosphere where small scale processes are important are often
highly concentrated (e.g., fronts and convective updrafts). Their
characteristic scales are $\ll\! a/10$.  Therefore, it is possible
that the Eq.~(\ref{pe}) set---as with all the other equation sets
discussed in this chapter---leaves out some processes important for
large-scale dynamics.

To arrive at Eq.~(\ref{pe}), one begins with
Eqs.~(\ref{nse}--\ref{eos}) in spherical geometry
\citep[e.g.,][]{batchelor-1967}. Two
approximations are then made.  These are the ``shallow
atmosphere'' and the ``traditional'' approximations (e.g.,
\citet{salby-1996}).  The first assumes $z/a\ll 1$.  The second is
formally valid in the limit of strong stratification, when the Prandtl
ratio $(N^2/\Omega^2) \gg 1$. Here, $N = N({\bf x},z,t)$ is the
Brunt-V\"ais\"al\"a (buoyancy) frequency, the oscillation frequency
for an air parcel that is displaced vertically under adiabatic
conditions:
\begin{eqnarray}\label{bvf}
  N = \left[g\,\frac{\del(\ln\theta)}{\del z}\right]^{1/2}.
\end{eqnarray}
These approximations allow the Coriolis terms involving vertical
velocity to be dropped from Eq.~(\ref{nse}a) and vertical
accelerations to be assumed small.  The latter is explicitly embodied
in Eq.~(\ref{pe}b), the hydrostatic balance, which we discuss further
below.

Hydrostatic balance renders the primitive equations valid only
when $N^2/\omega^2 \gg 1$, where $2\pi/\omega$ is the timescale of the
motion under consideration.  This condition, which is distinct from
the Prandtl ratio condition, restricts the vertical length scale of
motions to be small compared to the horizontal length
scale. Therefore, the hydrostatic balance approximation breaks down in
weakly stratified regions (here we refer to the dynamically evolving
hydrostatic balance associated with circulation-induced perturbations
in pressure and density; the mean background density and pressure---i.e.,
those that would exist in absence of dynamics---will remain hydrostatically
balanced even when the circulation-induced perturbations are not).
The hydrostatic assumption filters
vertically propagating sound waves from the equations.  

According to Eq.~(\ref{pe}d), when $\dot{q}_{\rm net} = 0$, individual
values of $\theta$ are retained by fluid elements as they move with
the flow.  In this case, Eq.~(\ref{pe}) also admit a dynamically
important conserved quantity, the potential vorticity:
\begin{mathletters}\label{epvc}
\begin{equation}
  q_{\mbox{\tiny PE}} = 
  \left[\frac{(\zeta + f)\,{\bf k}}{\rho}\right]\!\cdot\!\nabla\theta,
\end{equation}
where $\zeta = {\bf k}\cdot \nabla\!\times\!{\bf v}$ is the relative
vorticity. This quantity provides the crucial connection between the
primitive equations and the physically simpler models that follow.
For example, undulations of potential vorticity are often 
a direct manifestation of Rossby
waves, which are represented in all the models presented in this 
section.  The
conservation of the potential vorticity $q_{\mbox{\tiny PE}}$ following
the flow,
\begin{equation}
  \frac{Dq_{\mbox{\tiny PE}}}{Dt} = 0\, ,
\end{equation}
and the redistribution of $q_{\mbox{\tiny PE}}$ implied by it, 
is one of the most important properties in atmospheric dynamics.
\end{mathletters}

\subsection{Shallow-Water Model}
\label{shallow-water}

For many applications, Eq.~(\ref{pe}) is too complex and broad in
scope.  In the absence of observational information to properly
constrain the model parameters, reduction of the equations is
beneficial.  A commonly used approach is to collapse the 3D
primitive equations
to a two-dimensional (2D), one-layer model.  Such reduction 
allows investigation of horizontal
vortex and jet interactions in an idealized setting.  

Among the most widely used one-layer models is the shallow-water model.
Consider a thin layer of homogeneous (i.e., constant-density) fluid,
bounded above by a free surface and below by an impermeable boundary,
so that its thickness is $h({\bf x},t)$.  The dynamics of such a layer
is governed by the following equations \citep[e.g.,][chapter 3]
{pedlosky-1987}:
\begin{mathletters}\label{swe}
\begin{eqnarray}
\frac{D{\bf v }}{D t}\!  & = &\! -g\nabla h - f {\bf k \times v} \\
\frac{Dh}{D t}\!  & = &\! -h\,\nabla\!\cdot\!{\bf v},
\end{eqnarray}
\mbox{where}
\begin{equation}
\frac{D}{D t} \ = \ \frac{\del}{\del t} + {\bf v}\!\cdot\!\nabla. 
\end{equation}
Forcing and dissipation are not included in Eq.~(\ref{swe}), but they 
can be added in the usual way.
In the absence of forcing and dissipation, the equations preserve 
the potential vorticity,
\end{mathletters}
\begin{equation}\label{swpv}
  q_{\mbox{\tiny SW}} = \frac{\zeta + f}{h},
\end{equation} 
following the flow.

If Eq.~(\ref{swe}) is derived as the vertical mean of the flow of an
isentropic atmosphere with a free upper boundary, $h$ must be replaced
by $h^\kappa$ in the geopotential gradient term.  If they are derived
as a vertical mean of the flow of an isentropic atmosphere between
rigid upper and lower boundaries, $h$ must be replaced by 
$h^{\kappa/(1-\kappa)}$ in the geopotential gradient term.  Note that while
$\nabla\cdot{\bf v}\ne 0$ in Eq.~(\ref{swe}b), $\nabla\cdot{\bf u} =
0$, since the layer is homogeneous (i.e., density is
constant).  Hence, the sound speed $c_s\rightarrow\infty$, and the
sound waves are filtered out from the system.  However, the system
does retain gravity waves, which propagate at speed $c_g = \sqrt{gh}$.

The shallow-water equations are widely used as a process model
in geophysical fluid dynamics.  They are much simpler
than the primitive equations, yet they still describe a wealth of
phenomena---including vortices, jet streams, Rossby waves, gravity 
waves, and the interactions between them.  Excluded are any processes
that depend on the details of the vertical structure---including
vertically propagating waves, baroclinic instabilities (see \S\ref{3d-eddies}),
and depth-dependent flow.  However, because both rotational
and buoyancy processes are included (the latter via the variable
layer thickness), the shallow-water model---as well as all the models discussed
so far---includes a fundamental length scale
called the {\it Rossby radius of deformation} (often simply called
the deformation radius), which is a natural length scale for
a variety of phenomena that depend on both rotation and stratification.
In the shallow-water system, this length scale is
\begin{equation}
L_D={\sqrt{g h}\over f}.
\label{sw-deformation-radius}
\end{equation}

\subsection{Two-Dimensional, Nondivergent Model}
\label{nondivergent}

This is the simplest useful one-layer model for large-scale dynamics.
For large-scale weather systems characterized by $U/c_g \ll 1$, we can apply
a rigid upper boundary to the shallow-water model, since $c_g$
represents external gravity wave speed in the model.  Then, $H$ is
large and Eq.~(\ref{swe}b) implies $\nabla\!\cdot\!{\bf v}\ll
1$. Taking $\nabla\cdot{\bf v} = 0$ then gives the 2D nondivergent
equation:
\begin{eqnarray}\label{2de}
  \frac{D{\bf v }}{D t} = -g\nabla h - f{\bf k}\times{\bf v},
\end{eqnarray}
where $D/Dt$ is same as in (\ref{swe}c).  The $\nabla\cdot{\bf v} = 0$
restriction on the velocity implies that we can define a streamfunction,
$\psi({\bf x},t)$, such that 
\begin{equation}\label{streamfunction}
  {\bf v} = 
  \left(-\frac{\del\psi}{\del y},\frac{\del\psi}{\del x}\right).
\end{equation}

 Using this definition,
Eq.~(\ref{2de}) can finally be written as a single governing equation
for the evolution of the streamfunction:
\begin{eqnarray}\label{2dve}
  \frac{D}{Dt}(\nabla^2\psi + f)\ =\  0\, .
\end{eqnarray}
From Eq.~(\ref{2dve}), we see that $q_{\mbox{\tiny 2D}} = \nabla^2\psi
+ f$ is the materially conserved potential vorticity for the 2D
nondivergent model.  This also results simply by letting
$h\rightarrow$ constant in Eq.~(\ref{swpv}).  

In addition to
the nonlinear vorticity advection,
this equation---along with all the other equation sets described
in this section---represents the dynamical effects of latitudinally
varying Coriolis parameter.  
This is the so-called ``beta effect,'' where
$\beta\equiv df/dy$ is the northward gradient of the Coriolis
parameter.

Eq.~(\ref{2dve}) describes Rossby waves,
non-linear advection, and phenomena---such as
the formation of zonal jet streams---that require
the interaction of all these aspects (see \S\ref{rhines}).
However, it lacks a finite deformation radius ($L_D\to \infty$)
and does not possess gravity wave solutions.  Therefore, any phenomena
that depend on finite deformation radius, gravity waves, or buoyancy cannot
be captured.  (These assumptions render the equation valid only for
$U/c_g\ll 1$ and $L/L_D\ll 1$.)  As a result, the 2D non-divergent model 
cannot serve as an accurate predictive tool for most applications; however,
its very simplicity renders it a valuable process model for investigating
jet formation in the simplest possible setting.  For a review
of examples, see for example \citet{vasavada-showman-2005} or
\citet{vallis-2006}.

\subsection{Conserved Quantities}

Potential vorticity conservation has been emphasized throughout
because of its central importance in atmospheric dynamics.  There are
other useful conserved quantities.  For example, the full
Navier-Stokes equation gives
\begin{equation}
  \frac{D}{Dt}\,({\bf r}\!\times\!{\bf u})\ =\ 
  {\bf r} \times\left( -\frac{1}{\rho}\nabla p -         
2\mathbf{\Omega}\times{\bf u} - \nabla\Phi^\ast + {\bf F} \right),
\end{equation}
where $\Phi^\ast$ is effective geopotential and ${\bf F}$ represents
any additional forces on the fluid.  From this, we obtain the
conservation law for specific, axial angular momentum ${\cal M}$:
\begin{mathletters}\label{cam}
\begin{equation}
  \frac{D{\cal M}}{Dt}\ =\
  -\frac{1}{\rho}\frac{\del p}{\del\lambda} + {\cal         F}_\lambda\cos\phi,
\end{equation}
where
\begin{equation}
{\cal M} = (\Omega r\cos\phi + u)\, r\cos\phi\, .
\end{equation}
Eq.~(\ref{cam}) relates the material change of ${\cal M}$ to the axial
components of torques present.  For a thin atmosphere, $r$ can be
replaced with $a$.  
\end{mathletters}

Eq.~(\ref{nse}) also gives the material conservation law for the
specific total energy $E$:
\begin{equation}
  \frac{DE}{Dt}\ =\ -\frac{1}{\rho}\nabla\!\cdot\!(p{\bf u})\, +\, 
  (\dot{Q}_{\rm net} + {\bf u}\!\cdot\!{\bf F}),
\end{equation}
where $E$ is the total energy including kinetic, potential, and
internal contributions: 
$E = \onehalf{\bf u}^2 + \Phi + c_v T$ with $c_v$ the specific
heat at constant volume and $T$ the temperature.  In flux form, the
conservation law is:
\begin{equation}
  \frac{\del}{\del t}\,(\rho E) + 
  \nabla\!\cdot\![(\rho E + p){\bf u}]\ =\ 
  \rho\, (\dot{Q}_{\rm net} + {\bf u}\!\cdot\!{\bf F}). 
\end{equation}
As already noted for ${\cal M}$, the total energy reduces in the
appropriate way for the various simpler physical situations discussed
in previous subsections.  For example, $\Phi$ and $c_v T$ terms do not
exist for the 2D nondivergent case.  An important issue in the study
of atmospheric energetics is the extent to which $\Phi$ and $c_v T$ are
available to be converted to $\onehalf{\bf u}^2$.
\\

\section{BASIC CONCEPTS}
\label{basic-concepts}

The equation sets summarized in \S\ref{equations} describe nonlinear,
potentially turbulent flows with many degrees of freedom.  Unfortunately, 
due to the nonlinearity and complexity, analytic solutions rarely exist, and
one must resort to solving the equations numerically on a computer.
To represent the atmospheric circulation of a particular planet,
the chosen equation set is 
solved numerically with a specified spatial resolution and timestep,
subject to appropriate parameter values (e.g., composition, gravity,
planetary rotation rate), boundary conditions, and forcing/damping
(e.g., prescriptions for heating/cooling and friction). 

Such models vary greatly in complexity and numerical method.  General
Circulation Models (GCMs) in the Solar-System studies literature, for example,
typically solve the 3D primitive equations with sophisticated
representations of radiative transfer, cloud formation, surface/atmosphere
interactions, surface ice formation, and (if relevant) oceanic processes.
These models are useful for exploring the interaction of dynamics
with surface processes, radiation, and climate and are needed for
quantitative comparisons with observational records. 

However, because of their complexity, numerical simulations with full
GCMs are computationally expensive, limiting such simulations to only 
moderate spatial resolution and making it difficult to broadly survey 
the relevant parameter space.  Even more problematic, because of
the inherent complexity of nonlinear fluid dynamics and its possible 
interactions with radiation and surface processes, it is rarely
obvious {\it why} a given GCM simulation produces the output it does.
By itself, the output of a sophisticated 3D model often provides 
little more fundamental understanding than the observations of 
the actual atmosphere themselves.  
To understand how a given atmospheric circulation
would vary under different planetary parameters, for example, an
understanding of the {\it mechanisms} shaping the circulation is required.
Although careful diagnostics of GCM results can provide important insights
into the mechanisms that are at play, a deep mechanistic understanding 
does not always flow naturally from such simulations.

Rather, obtaining a robust understanding requires
a diversity of model types, ranging from simple to complex, in which various
processes are turned on and off and the results carefully diagnosed.
This is called a modeling hierarchy and its use forms the
backbone of forward progress in the field of atmospheric dynamics
of Earth and other Solar-System planets \citep[see, e.g.,][]{held-2005}.
For example, despite the existence of numerous full GCMs for modern
Earth climate, significant advances in our understanding of the 
{\it mechanisms} shaping the atmospheric circulation rely heavily 
on the usage of linear models, simplified one-layer non-linear models
(such as the 2D non-divergent or shallow-water models), and 3D models that 
do not include the sophisticated treatments of radiation and sub-gridscale
convective processes included in full GCMs.\footnote{To illustrate, a summary
of the results of such a hierarchy for understanding Jupiter's jet
streams can be found in \citet{vasavada-showman-2005}.}  
Even more fundamentally, obtaining
understanding requires the development of basic theory that can 
(at least qualitatively) explain the results of these various models as 
well as observations of actual atmospheres.  One of the major goals in  
performing simplified models is to aid in the construction of
such a theory \citep[see, e.g.,][]{schneider-2006}.

Exoplanet GCMs will surely be useful 
in the coming years. But, as with Solar-System planets, 
we expect that a fundamental understanding
will require use of a modeling hierarchy as well as basic theory.
In this section we survey
key concepts in atmospheric dynamics 
that provide insight into the expected atmospheric
circulation regimes. Emphasis is placed on presenting a conceptual 
understanding and as such we describe
not only GCM results but basic theory and
the results of highly simplified models as well.  Here
we focus on basic aspects relevant to both
gaseous and terrestrial planets.
Detailed presentations of issues specific to
giant and terrestrial exoplanets are deferred to \S\ref{giants} and 
\S\ref{terrestrial}.

\subsection{Energetics of atmospheric circulation}


Atmospheric circulations involve an energy cycle.  Absorption of starlight and 
emission of infrared energy to space creates potential energy, which
is converted to kinetic energy and then lost via friction.  Each
step in the process involves nonlinearities, and generally the
atmosphere self-adjusts so that, in a time mean sense, the conversion
rates balance.  

What matters for driving the circulation is not the {\it total} potential
energy but rather the {\it fraction} of the potential energy
that can be extracted by adiabatic atmospheric motions.  For example,
a stably stratified, horizontally uniform atmosphere can contain
vast potential energy, but none can be extracted---any adiabatic 
motions can only {\it increase} the potential energy of such a state.  
Uniformly heating the top layers of such an atmosphere would further 
increase its potential energy but would still preclude an atmospheric 
circulation. 

For convecting atmospheres, creating extractable potential energy requires 
heating the fluid at lower altitudes than it is cooled.  This creates buoyant 
air parcels (positively buoyant at the bottom, negatively buoyant at the top); 
vertical motion of these buoyant parcels releases potential energy and drives 
convection.\footnote{To emphasize the importance of the distinction, consider
a hot, isolated giant planet. The cooling caused by its radiation to space 
{\it decreases} its {\it total} potential energy, yet (because the cooling 
occurs near the top) this {\it increases} the fraction of the 
remaining potential energy that can be extracted by motions.  This is what 
can allow convection to occur on such objects.}
But, most atmospheres are stably stratified, and in this case extractable
energy---called {\it available potential energy}---only exists when density
varies horizontally on isobars \citep[][chapter 14]{peixoto-oort-1992}.  
In this case, the denser regions can slide
laterally and downward underneath the less-dense regions, decreasing the potential energy and
creating kinetic energy (winds).  Continual generation of
available potential energy (required to balance its continual conversion
to kinetic energy and loss via friction) requires heating the regions of the 
atmosphere that are already hot (e.g., the tropics on Earth) and cooling the 
regions that are already cold (e.g., the poles). For Earth, available potential 
energy is generated at a global-mean rate of $\sim2\rm\,W\,m^{-2}$, which is 
$\sim1$\% of the global-mean absorbed and radiated flux of $240\rm\,W\,m^{-2}$ 
\citep[][pp.~382-385]{peixoto-oort-1992}.

The rate of frictional dissipation can affect the mean state, but rigorously
representing such friction in models is difficult. For Solar-System planets,
kinetic-energy loss occurs via turbulence, waves, and friction 
against the surface (if any).  Ohmic dissipation may be important in the
deep interiors of gas giants \citep{kirk-stevenson-1987, liu-etal-2008},
as well as in the upper atmosphere where ionization becomes important.
These processes sometimes have length scales
much smaller (by up to several orders of magnitude) than
can easily be resolved in global, 3D numerical models.  In Earth GCMs,
such frictional dissipation mechanisms are therefore often {\it parameterized} 
by adding to the equations quasi-empirical damping terms (e.g., a vertical
diffusion to represent turbulent kinetic-energy losses by small-scale shear 
instabilities and breaking waves). A difficulty is that such prescriptions,
while physically motivated, are often non-rigorous and the extent to which
they can be extrapolated to other planetary environments is unclear.
Perhaps for this reason, models of hot Jupiters published to date 
do not include such parameterizations of frictional processes (although 
they all include small-scale viscosity for numerical reasons).\footnote{Note 
that a statistically steady (or quasi-steady) state can still occur in 
such a case; this requires the atmosphere to self-adjust so that the rates of
generation of available potential energy and its conversion to kinetic energy
become small.} Nevertheless, \citet{goodman-2009} has highlighted the 
possible importance that such processes could play in the hot-Jupiter 
context, and future models of hot Jupiters will surely explore the 
possible effect that friction may have on the mean states.

Solar-System planets offer interesting lessons on the role of friction.
Despite absorbing a greater solar flux than any other thick atmosphere in our
Solar System, Earth's winds are relatively slow, with a mean wind speed of
$\sim$~$20\rm\,m\,sec^{-1}$.  In contrast, Neptune absorbs a solar flux only
0.1\% as large, but has wind speeds reaching $400\rm\,m\,sec^{-1}$.
Presumably, Neptune can achieve such fast winds despite its weak radiative
forcing because its frictional damping is extremely weak.  Qualitatively,
this makes sense because Neptune lacks a surface, which is a primary source
of frictional drag on Earth.   More puzzling is the fact that Neptune
has significantly stronger winds than Jupiter 
(Table~\ref{planetary-parameters}) despite absorbing only 4\% the 
solar flux absorbed by Jupiter.  Possible explanations are that
Jupiter experiences greater frictional damping than Neptune
or that it has equilibrated to a state that has relatively slow wind speeds
despite weak damping. This is not well understood and
argues for humility in efforts to model the circulations of exoplanets.

\subsection{Timescale arguments for the coupled radiation-dynamics
problem}
\label{timescales}
The atmospheric circulation represents a coupled
radiation-hydrodynamics problem.  The circulation advects the temperature
field and thereby influences the radiation field; in turn, the
radiation field (along with atmospheric opacities and surface conditions)
determines the atmospheric heating and cooling rates that drive the
circulation. Rigorously attacking this problem requires coupled
treatment of both radiation and dynamics.  However, crude insight into
the thermal response of an atmosphere can be obtained with simple 
timescale arguments.  Suppose $\tau_{\rm advect}$ is an advection time
(e.g., the characteristic time for air to advect across a hemisphere) 
and $\tau_{\rm rad}$ is the radiative time (i.e., the characteristic 
time for radiation to induce large fractional entropy changes).  When
$\tau_{\rm rad} \ll \tau_{\rm advect}$, we expect temperature to deviate
only slightly from the (spatially varying) radiative equilibrium
temperature structure.  Because the radiative-equilibrium temperature
typically varies greatly from dayside to nightside (or from equator
to pole), this implies that such a planet would exhibit large fractional
temperature contrasts.   On the other hand, 
when $\tau_{\rm rad} \gg \tau_{\rm advect}$,
dynamical transport dominates and air will tend to homogenize its
entropy, implying that lateral temperature contrasts should be modest.

In estimating the advection time, one must distinguish north-south from
east-west advection; east-west advection (relative to the pattern of
stellar insolation) will often be dominated by the planetary rotation.  
For synchronously rotating planets, a characteristic horizontal 
advection time is
\begin{equation}
\tau_{\rm advect}\sim {a\over U},
\label{tau_advect}
\end{equation}
where $U$ is a characteristic horizontal wind speed.  A similarly
crude estimate of the radiative time can be obtained by considering a 
layer of pressure thickness $\Delta p$
that is slightly out of radiative equilibrium and radiates to space 
as a blackbody.  If the radiative equilibrium temperature is $T_{\rm rad}$
and the actual temperature is $T_{\rm rad}+\Delta T$, with $\Delta T\ll
T_{\rm rad}$, then the net flux
radiated to space is $4\sigma T_{\rm rad}^3 \Delta T$ and the radiative
timescale is \citep[][pp.~65-66]{showman-guillot-2002, james-1994}
\begin{equation}
\tau_{\rm rad} \sim {\Delta p\over g} {c_p\over 4\sigma T^3}.
\label{tau-rad}
\end{equation}
In deep, optically thick atmospheres where the radiative transport 
is diffusive, a more appropriate estimate might be a diffusion time, 
crudely given by $\tau_{\rm rad}\sim H^2/D$, where $H$ is the vertical height 
of a thermal perturbation and $D$ is the radiative diffusivity.

\citet{showman-etal-2008b} estimated advective and radiative time
constants for Solar-System planets and found that, as expected,
planets with $\tau_{\rm rad} \gg \tau_{\rm advect}$ generally have
small horizontal temperature contrasts and vice versa.  

For hot Jupiters, most
models suggest peak wind speeds of $\sim$1--$3\rm\,km\,sec^{-1}$ 
(\S\ref{hot-jupiters}), implying advection times of $\sim$$10^5\rm\,sec$
based on the peak speed.  
Eq.~(\ref{tau-rad}) would then suggest that $\tau_{\rm rad}\ll 
\tau_{\rm advect}$ at $p\ll 1\,$bar whereas $\tau_{\rm rad}\gg 
\tau_{\rm advect}$ at $p\gg 1\,$bar.  Thus, one might crudely expect 
large day-night temperature differences at low pressure and small
day-night temperature difference at high pressure, with the transition
occurring at $\sim0.1$--$1\,$bar.  These estimates are generally
consistent with the observational inference of \citet{barman-2008} and 
3D numerical simulations \citep[e.g.,][]{showman-etal-2009,
dobbs-dixon-lin-2008} of hot Jupiters---though some uncertainties
still exist with modeling and interpretation.

For synchronously rotating terrestrial planets in the habitable zones
of M dwarfs, a mean wind speed of $20\rm\,m\,sec^{-1}$
(typical for terrestrial planets in our Solar System; 
see Table~\ref{planetary-parameters}) would imply an advection time
of $\sim$$3$ Earth days.  For a temperature of $300\,$K, Eq.~(\ref{tau-rad})
would then imply that 
$\tau_{\rm rad}$ is much smaller (greater) than $\tau_{\rm advect}$
when the surface pressure is much less (greater) than $\sim$$0.2\,$bars.
This argument suggests that synchronously rotating terrestrial exoplanets with 
a surface pressure much less than
$\sim$0.2 bars should develop large day-night temperature differences,
whereas if the surface pressure greatly exceeds 
$\sim$0.2 bars, day-night temperature differences would be modest.
As with hot Jupiters,
these estimates are consistent with 3D GCM simulations \citep{joshi-etal-1997},
which suggest that this transition should occur at $\sim$$0.1\,$bars.
These estimates may have relevance for whether CO$_2$ atmospheres
would collapse due to nightside condensation and hence whether
such planets are habitable.

\subsection{Basic force balances: importance of rotation}
\label{basic-force-balances}
Planets rotate, and this typically constitutes a dominant
factor in shaping the circulation.    The importance of rotation
can be estimated by performing a {\it scale analysis} on the
equation of motion.  Suppose the circulation has a mean speed
$U$ and that we are interested in flows with characteristic
length scale $L$ (this might approach a planetary radius for 
global-scale flows).  To order-of-magnitude, the strength
of the acceleration term is $U^2/L$ (namely, $U$ divided by
a time $L/U$ to advect fluid across a distance $L$), while the
magnitude of the Coriolis term is $f U$.   The ratio of the 
acceleration term to the Coriolis term can therefore 
be represented by the {\it Rossby number},

\begin{equation}
Ro\equiv {U\over f L}.
\label{rossby}
\end{equation}
Whenever $Ro\ll 1$, the
acceleration terms $D{\bf v}/Dt$ are weak compared to the Coriolis 
force per unit mass in the horizontal momentum equation.   Because friction
is generally weak, the only other term that can balance the horizontal
Coriolis force is the pressure-gradient force, which is just
$-\nabla_p \Phi$ in pressure coordinates.  The resulting balance,
called {\it geostrophic balance}, is given by
\begin{equation}
f u = -\left({\partial \Phi\over\partial y}\right)_p \qquad\qquad
f v = \left({\partial\Phi\over\partial x}\right)_p
\label{geostrophy}
\end{equation}
where $x$ and $y$ are eastward and northward distance, respectively and
the derivatives are evaluated at constant pressure.
In our Solar System, geostrophic balance holds at large scales in the mid- and 
high-latitude atmospheres of Earth, Mars, Jupiter, Saturn, Uranus, and Neptune.  
Rossby numbers range from 0.01--0.1 for these rapidly rotating planets, but 
exceed unity in the stratosphere of Titan\footnote{Titan's Rossby number is 
smaller near the surface, where winds are weak.} and reach $\sim$10 for Venus, 
implying in the latter case that the Coriolis force plays a less important 
role in the force balance (Table~\ref{planetary-parameters}).
Note that, even on rapidly rotating planets, horizontal geostrophy breaks 
down at the equator, where the horizontal Coriolis forces go to zero.

Determining Rossby numbers for exoplanets requires estimates of wind speeds, which
are unknown.  Some models of hot Jupiter atmospheres
have generally suggested peak winds of several $\rm km\,sec^{-1}$ near the
photosphere,\footnote{Defined here as the approximate pressure 
at which infrared photons can escape directly to space.}
with mean values perhaps a factor of several smaller.    
To illustrate the possibilities, Table~\ref{planetary-parameters} presents
$Ro$ values for several hot Jupiters assuming a range of wind speeds of 
100--$4000\rm\,m\,sec^{-1}$.  Generally, if mean wind speeds are fast (several 
$\rm\,km \,sec^{-1}$), Rossby numbers approach or exceed unity.  
If mean wind speeds
are hundreds of $\rm m\,sec^{-1}$ or less, Rossby numbers should be much
less than one, implying that geostrophy approximately holds.  One might thus
plausibly expect a situation where the Coriolis force plays an important
but not overwhelming role (i.e. $Ro\sim1$) near photosphere levels, 
with the flow transitioning to geostrophy in the interior if winds are 
weaker there.

\begin{deluxetable}{lcrllllllcrl}
\tabletypesize{\footnotesize}
\tablecaption{Planetary parameters\label{planetary-parameters}}
\tablewidth{0pt}
\tablehead  
{Planet &$a^{*}$          &Rotation period$^{\sharp}$ &$\Omega$            &gravity$^{\aleph}$ &$F_*^{\Box}$ &$T_{\rm e}^{\spadesuit}$        &$H_p^{\dagger}$  &$U^{\ddagger}$                 &$Ro^{\P}$ &$L_D/a^{\clubsuit}$ &$L_{\rm \beta}/a^{\Diamond}$\\   
        &($10^3$ km)  &(Earth days)    &($\rm rad\,sec^{-1}$)   &($\rm m\,sec^{-2}$)  &($\rm W\,m^{-2}$) &(K) &(km)  &($\rm m\,sec^{-1}$) &     &    & }
\startdata
Venus   &6.05    &243    &$3\times10^{-7}$      &8.9     &2610   &232   &5   &$\sim20$   &10    &70     &7 \\
Earth   &6.37    &1      &$7.27\times10^{-5}$   &9.82     &1370   &255   &7                     &$\sim20$   &0.1   &0.3    &0.5 \\
Mars    &3.396   &1.025  &$7.1\times10^{-5}$    &3.7      &590   &210   &11                    &$\sim20$   &0.1   &0.6    &0.6 \\
Titan   &2.575   &16     &$4.5\times10^{-6}$    &1.4      &15   &85     &18                    &$\sim20$   &2     &10     &3  \\
Jupiter &71.4    &0.4    &$1.7\times10^{-4}$    &23.1     &50   &124    &20                    &$\sim40$   &0.02  &0.03   &0.1 \\
Saturn  &60.27   &0.44   &$1.65\times10^{-4}$   &8.96     &15   &95     &39                    &$\sim150$  &0.06  &0.03   &0.3 \\
Uranus  &25.56   &0.72   &$9.7\times10^{-5}$    &8.7      &3.7   &59    &25                    &$\sim100$  &0.1   &0.1    &0.4 \\
Neptune &24.76   &0.67   &$1.09\times10^{-4}$   &11.1    &1.5   &59   &20                    &$\sim200$  &0.1   &0.1    &0.6 \\ 
\\
WASP-12b &128    &1.09   &$6.7\times10^{-5}$    &11.5     &$8.8\times10^6$ &2500 &800 &-      &0.01--0.3   &0.1   &0.2--1.5 \\
HD 189733b &81  &2.2    &$3.3\times10^{-5}$     &22.7     &$4.7\times10^5$ &1200 &200 &-      &0.03--1   &0.3    &0.4--3    \\
HD 149026b &47  &2.9    &$2.5\times10^{-5}$     &21.9     &$1.8\times10^6$ &1680 &280 &-      &0.06--2    &0.8    &0.6--4  \\
HD 209458b &94  &3.5    &$2.1\times10^{-5}$     &10.2     &$1.0\times10^6$ &1450 &520 &-      &0.04--1    &0.4    &0.5--3   \\
TrES-2    &87   &2.4    &$2.9\times10^{-5}$     &21       &$1.1\times10^6$ &1475 &260 &-      &0.03--1    &0.3    &0.4--3  \\
TrES-4    &120  &3.5    &$2.0\times10^{-5}$     &7.8      &$2.5\times10^6$ &1825 &870 &-      &0.03--1    &0.4    &0.4--3  \\
HAT-P-7b  &97   &2.2    &$3.3\times10^{-5}$     &25       &$4.7\times10^6$ &2130 &320 &-      &0.02--1    &0.3    &0.4--3\\
GJ 436b   &31   &2.6    &$2.8\times10^{-5}$     &9.8      &$4.3\times10^4$ &660  &250 &-      &0.1--3     &0.7    &0.8--5\\
HAT-P-2b   &68  &5.6    &$1.3\times10^{-5}$     &248      &$9.5\times10^5$ &1400 &21  &-      &0.1--3     &1      &0.8--5\\
Corot-Exo-4b &85 &9.2   &$7.9\times10^{-6}$     &13.2     &$3.0\times10^5$ &1080  &300 &-     &0.1--4     &1      &0.9--5\\

\enddata
\tablecomments{$^*$Equatorial planetary radius. $^{\sharp}$Assumes synchronous rotation for exoplanets.  $^{\aleph}$ Equatorial gravity at the surface.
$^{\Box}$Mean incident stellar flux.  $^{\spadesuit}$Global-average blackbody emission temperature, which for exoplanets
is calculated from Eq.~(\ref{teff}) assuming zero albedo.
$^{\dagger}$Pressure scale height, evaluated at temperature $T_{\rm e}$.  $^{\ddagger}$Rough estimates of characteristic 
horizontal wind speed.  Estimates for Venus and Titan are in the high-altitude 
superrotating jet; both planets have weaker winds (few $\rm m\,sec^{-1}$) 
in the bottom scale height.  In all cases, peak winds exceed the 
listed values by factors of two or more.
$^{\P}$Rossby number, evaluated in mid-latitudes using wind values 
listed in Table and $L\sim2000\,$km for Earth, Mars, and Titan, 
$6000\,$km for Venus, and $10^4\,$km for Jupiter, Saturn, Uranus, 
and Neptune. For exoplanets, we present a range of possible values evaluated with $L=a$ and winds from 100 to $4000\rm\,m\,sec^{-1}$.  
$^{\clubsuit}$Ratio of Rossby deformation radius 
to planetary radius, evaluted in mid-latitudes with $H$ equal to 
the pressure scale height and $N$ appropriate for a vertically isothermal 
temperature profile.  $^{\Diamond}$Ratio of Rhines length 
(Eq.~\ref{rhines-scale}) to planetary radius, calculated using the 
equatorial value of $\beta$ and the wind speeds listed in the Table.}
\end{deluxetable}

Geostrophy implies that, rather than flowing from pressure highs
to lows as often occurs in a non-rotating fluid, the primary horizontal 
wind flows {\it perpendicular} to the horizontal pressure gradient.  
Thus, the primary flow does not erase 
the horizontal pressure contrasts; rather, the Coriolis forces
associated with that flow actually help preserve large-scale
pressure gradients in a rotating atmosphere.  Geostrophy explains
why the isobar contours included in most mid-latitude
weather maps (say of the U.S. or Europe) are so useful: the 
isobars describe not only the pressure field but the large-scale 
wind field, which flows along the isobar contours.

In many cases, 
rotation inhibits the ability of the circulation to equalize 
horizontal temperature differences.  
On rapidly rotating planets
like the Earth, which is heated by sunlight primarily at low latitudes,
the mean horizontal temperature gradients in the 
troposphere\footnote{The {\it troposphere} is the bottommost, 
optically thick layer 
of an atmosphere, where temperature decreases with altitude and convection
may play an important role; the {\it tropopause} defines the top of
the troposphere, and the {\it stratosphere} refers to the stably stratified,
optically thin region overlying the troposphere.  In some cases, a
stratosphere's temperature may increase with altitude due to absorption
of sunlight by gases or aerosols; in other cases, however, the 
stratosphere's temperature can be nearly constant with altitude.} 
generally point 
from the poles toward the equator.  Integration of the hydrostatic equation 
(Eq.~\ref{pe}b) implies that the mean pressure gradients also point north-south.
Thus, on a rapidly rotating planet where geostrophy holds and the 
primary temperature contrast is between equator and pole, the mean midlatitude 
winds will be {\it east-west} rather than {\it north-south}---thus limiting
the ability of the circulation to homogenize its temperature differences
in the north-south direction.  We might thus expect that, everything else 
being equal, a more rapidly rotating planet will harbor a greater 
equator-to-pole temperature difference.

In rapidly rotating atmospheres, a tight link exists between
horizontal temperature contrasts and the vertical gradients
of the horizontal wind.  This can be shown by taking the 
derivative with pressure of Eq.~(\ref{geostrophy}) and invoking the hydrostatic
balance equation (\ref{pe}b) and the ideal-gas law.  We obtain
the {\it thermal-wind equation} for a shallow atmosphere \citep[][pp. 70-75]{holton-2004}:
\begin{equation}
f {\partial u\over\partial \ln p}= {\partial(R T)\over\partial y}
\qquad\qquad
f {\partial v\over\partial \ln p}=-{\partial(R T)\over\partial x}
\label{thermal-wind}
\end{equation}
where $R$ is the specific gas constant (i.e., the universal gas
constant divided by the molar mass).
The equation states that, in a geostrophically balanced atmosphere,
north-south temperature gradients must be associated with
a vertical gradient in the zonal (east-west) wind, whereas east-west temperature
gradients must be associated with a vertical gradient in the meridional
(north-south) wind.  Given the primarily equatorward pointing midlatitude 
horizontal temperature gradient in the tropospheres of Earth and Mars, 
for example, and given the weak winds at the surface of a terrestrial 
planet (a result of surface friction), this equation 
correctly demonstrates that the mean mid-latitude winds in the upper
troposphere must flow to the east---as observed for the mid-latitude 
tropospheric jet streams on Earth and Mars.

Geostrophy relates the 3D structure of the winds and temperatures
at a given time but says nothing about the flow's time evolution.  In 
a rapidly rotating atmosphere, both the temperatures and winds often evolve
together, maintaining approximate geostrophic balance as they do so.
This time evolution depends on the {\it ageostrophic} component of
the circulation, which tends to be of order $Ro$ smaller than the
geostrophic component.  The fact that a time evolving flow maintains
approximate geostrophic balance implies that, in a rapidly rotating
atmosphere, adjustment mechanisms exist that tend to re-establish geostrophic
balance when departures from it occur.

What is the mechanism for establishing and maintaining geostrophy?  If
a rapidly rotating atmosphere deviates from geostrophic balance, it implies that 
the horizontal pressure-gradient and Coriolis forces only imperfectly cancel,
leaving an unbalanced residual force.  This force generates a component
of ageostrophic motion between pressure highs and lows.  The Coriolis force
on this ageostrophic motion, and the alteration of the pressure contrasts
caused by the ageostrophic wind, act to re-establish geostrophy.

To give a concrete example, imagine an atmosphere with zero winds and a 
localized circular region of high surface pressure surrounded on all sides 
by lower surface pressure.  This state has an unbalanced pressure-gradient 
force, which would induce a horizontal acceleration of fluid radially away 
from the high-pressure region.  The horizontal Coriolis force on this outward 
motion causes a lateral deflection (to the right
in the northern hemisphere and left in the southern hemisphere\footnote{Northern
and southern hemispheres are here defined as those where ${\bf \Omega}\cdot
{\bf k}>0$ and $<0$, respectively, where ${\bf \Omega}$ and ${\bf k}$ are the
planetary rotation vector and local vertical (upward) unit vector.}), leading
to a vortex surrounding the high-pressure region.  The Coriolis force
on this vortical motion points radially toward the high-pressure center,
resisting its lateral expansion.  This process continues until the 
inward-pointing Coriolis force balances the outward pressure-gradient
force---hence establishing geostrophy and inhibiting further expansion.  
Although this is an extreme example,
radiative heating/cooling, friction, and other forcings gradually push
the atmosphere away from geostrophy, and the process described above
re-establishes it.   

This process, called {\it geostrophic
adjustment}, tends to occur with a natural length scale comparable to the 
Rossby radius of deformation, given in a 3D system by
\begin{equation}
L_D = {N H\over f}
\label{deformation-radius}
\end{equation}
where $N$ is Brunt-V\"ais\"al\"a frequency (Eq.~\ref{bvf}), 
$H$ is the vertical scale of
the flow, and $f$ is the Coriolis parameter. Thus, geostrophic adjustment
naturally generates large-scale atmospheric flow structures with horizontal
sizes comparable to the deformation radius (see 
\citet{holton-2004} or \citet{vallis-2006} for a detailed treatment).

Equation~(\ref{thermal-wind}), as written, applies to atmospheres that
are vertically thin compared to their horizontal dimensions (i.e.,
shallow atmospheres).
However, a non-shallow analog of Eq.~(\ref{thermal-wind}) can
be obtained by considering the 3D vorticity equation.  In a 
geostrophically balanced fluid where the friction force is weak,
the vorticity balance is given by \citep[see][p. 43]{pedlosky-1987}
\begin{equation}
2({\bf \Omega}\cdot \nabla){\bf u} - 2 {\bf \Omega}(\nabla\cdot {\bf u})
= - {\nabla\rho\times\nabla p\over \rho^2},
\label{3d-vorticity}
\end{equation}
where ${\bf \Omega}$ is the planetary rotation vector and ${\bf u}$
is the 3D wind velocity.  The term on the right, called the baroclinic
term, is nonzero when density varies on constant-pressure surfaces.
In the interior of a giant planet, however, convection tends to
homogenize the entropy, in which case fractional density variations
on isobars are extremely small.   Such a fluid, where surfaces
of constant $p$ and $\rho$ align, is called a barotropic
fluid.  In this case the right side of Eq.~(\ref{3d-vorticity}) can
be neglected, leading to the compressible-fluid generalization of
the {\it Taylor-Proudman theorem:}
\begin{equation}
2({\bf \Omega}\cdot \nabla){\bf u} - 2 {\bf \Omega}(\nabla\cdot {\bf u})
=0
\label{taylor-proudman}
\end{equation}
Consider a Cartesian coordinate system ($x_*$, $y_*$, $z_*$) with the 
$z_*$ axis parallel to ${\bf \Omega}$ and the $x_*$ and $y_*$ axes
lying in the equatorial plane.  Eq.~(\ref{taylor-proudman}) can 
then be expressed in component form as
\begin{equation}
{\partial u_*\over \partial z_*} = {\partial v_*\over\partial z_*}
=0
\label{taylor-proudman2}
\end{equation}
\begin{equation}
{\partial u_*\over \partial x_*}+ {\partial v_*\over\partial y_*}=0
\label{taylor-proudman3}
\end{equation}
where $u_*$ and $v_*$ are the wind components along the $x_*$
and $y_*$ axes, respectively.  These equations state that the wind 
components parallel to the equatorial plane are independent of direction
along the axis perpendicular to the equatorial plane, and moreover
that the divergence of the winds in the equatorial plane must
be zero.  The flow thus exhibits a structure with winds constant
on columns, called Taylor columns, that are parallel to the rotation axis.
In the context of a giant planet, motion of Taylor columns toward
or away from the rotation axis is disallowed because the planetary
geometry would force the columns to stretch or contract, causing
a nonzero divergence in the equatorial plane---violating 
Eq.~(\ref{taylor-proudman3}).  Rather, the columns are free to move
only along latitude circles.  The flow then takes the form of concentric 
cylinders, centered about the rotation axis, which can move in the 
east-west direction.  This columnar flow provides one model for the
structure of the winds in the deep interiors of Jupiter, Saturn, Uranus,
Neptune, and giant exoplanets (see \S\ref{giants}).

\subsection{Other force balances: The case of slowly rotating planets}

The pressure-gradient force can be balanced by forces other than the
Coriolis force.  An important example is {\it cyclostrophic
balance}, which is a balance between horizontal centrifugal and
pressure-gradient force, expressed in its simplest form as:
\begin{equation}
{u_t^2\over r} = {1\over\rho}{\partial p\over \partial r},
\label{cyclostrophic}
\end{equation}
where we are considering circular flow around a central point.
Here, $u_t$ is the tangential speed of the circular flow 
and $r$ is the radius of curvature of the flow.  The left-hand side 
of the equation is the centrifugal force per unit mass  and the right-hand 
side is the radial pressure-gradient force per unit mass.  Cyclostrophic 
balance is the force balance that occurs within dust devils and tornadoes, 
for example.  

In the context of a global-scale planetary circulation, if the primary 
flow is east-west, the
centrifugal force manifests as the curvature term $u^2\tan\phi/a$, where
$a$ is planetary radius and $\phi$ is latitude \citep[see][pp.~31-38 for
a discussion of curvature terms]{holton-2004}.  Cyclostrophic balance can 
then be written
\begin{equation}
{u^2\tan\phi\over a}= -\left({\partial\Phi\over\partial y}\right)_p
\label{cyclostrophic2}
\end{equation}
This is the force balance relevant on Venus, for example, where a strong
zonal jet with peak speeds reaching $\sim$$100\rm\,m\,sec^{-1}$ dominates
the stratospheric circulation \citep{gierasch-etal-1997}.
Eq.~(\ref{cyclostrophic2}) can be differentiated with respect to 
pressure to yield a cyclostrophic version of the thermal-wind equation:
\begin{equation}
{\partial(u^2)\over\partial \ln p}=-{a\over\tan\phi}
{\partial(RT)\over\partial y},
\label{cyclostrophic3}
\end{equation}
where hydrostatic balance and the ideal-gas law have been invoked.

Thus, on a slowly rotating, cyclostrophically balanced planet like Venus, 
Eq.~(\ref{cyclostrophic3}) would indicate that variation of the zonal winds 
with height requires latitudinal temperature gradients 
as occur with geostrophic balance.  Note, however, that for zonal winds
whose strength increases with height, the latitudinal temperature gradients for
cyclostrophic balance point
equatorward {\it regardless} of whether the zonal winds are east or west.  This
differs from geostrophy, where latitudinal temperature gradients would point equatorward
for a zonal wind that becomes more eastward with height but poleward for a zonal 
wind that becomes more westward with height.

\subsection{What controls vertical velocities?}

In atmospheres, mean vertical velocities associated with the large-scale
circulation tend to be much smaller than horizontal velocities.  This results from
several factors:

\medskip

\noindent
{\it Large aspect ratio:} Atmospheres generally
have horizontal dimensions greatly exceeding their vertical
dimensions, which leads to a strong geometric constraint
on vertical motions.  To illustrate, suppose the continuity
equation can be approximated in 3D by the incompressibility condition 
$\nabla\cdot {\bf u}= 
\partial u/\partial x + \partial v/\partial y + 
\partial w/\partial z = 0$.  To order of magnitude, the
horizontal terms can be approximated as $U/L$ and the 
vertical term can be approximated as $W/H$, where $L$ and $H$
are the horizontal and vertical scales and $U$ and $W$ are the
characteristic magnitudes of the horizontal and vertical wind.  
This then suggests a vertical velocity of approximately
\begin{equation}
W \sim {H\over L}U
\label{vert-geometric}
\end{equation}
On a typical terrestrial planet, $U\sim10\m\sec^{-1}$, $L\sim1000\km$,
and $H\sim10\km$, which would suggest $W\sim0.1\m\sec^{-1}$, two orders
of magnitude smaller than horizontal velocities.  
However, this estimate of $W$ is an upper limit,
since partial cancellation can occur between $\partial u/\partial x$
and $\partial v/\partial y$, and indeed for most atmospheres 
Eq.~(\ref{vert-geometric}) greatly overestimates their mean vertical
velocities.  On Earth, for example, mean midlatitude vertical
velocities on large scales ($\sim$$10^3\rm\,km$) 
are actually $\sim$$10^{-2}\m\sec^{-1}$ in the troposphere and
$\sim$$10^{-3}\m\sec^{-1}$ in the stratosphere, much smaller
than suggested by Eq.~(\ref{vert-geometric}).

\medskip

\noindent
{\it Suppression of vertical motion by rotation:}
The geostrophic velocity flows perpendicular to the horizontal pressure gradient, 
which implies that a large fraction of this flow cannot cause horizontal
divergence or convergence (as necessary to allow vertical motions).  
However, the {\it ageostrophic} component 
of the flow, which is $O(Ro)$ smaller than the geostrophic component, can
cause horizontal convergence/divergence.  Thus, to order-of-magnitude
one might expect $W\sim Ro\, U/L$.  Considering the geostrophic flow
and taking the curl of Eq.~(\ref{geostrophy}), we can show that the 
horizontal divergence of the horizontal geostrophic wind is
\begin{equation}
\nabla_p\cdot {\bf v_G} = {\beta\over f}v = {v\over a\tan\phi}
\label{vert-rotation}
\end{equation}
where in the rightmost expression $\beta/f$ has been evaluated assuming
spherical geometry.  To order of magnitude, $v\sim U$.  Eq.~(\ref{vert-rotation}) 
then implies that to order-of-magnitude the horizontal flow divergence is not
$U/L$ but rather $U/a$.  On many planets, the dominant flow structures
are smaller than the planetary radius (e.g., $L/a \sim 0.1$--0.2 for Earth,
Jupiter, and Saturn).  This implies that the horizontal divergence
of the geostrophic flow is $\sim$$L/a$ smaller than the estimate in
Eq.~(\ref{vert-geometric}).   Taken together, these constraints 
suggest that rapid rotation can suppress vertical velocities by close to
an order of magnitude.  For Earth parameters, the estimates imply 
$W\sim0.01\m\sec^{-1}$, similar to the mean vertical velocities in Earth's 
troposphere.

\medskip
\noindent
{\it Suppression of vertical motion by stable stratification:}
Most atmospheres are stably stratified, implying that entropy
(potential temperature) increases with height.  Adiabatic
expansion/contraction in ascending (descending) air would cause
temperature at a given height to decrease (increase) over time.
In the absence of radiation, such steady flow patterns are
unsustainable because they induce density variations that
resist the motion---ascending air becomes denser and descending
air becomes less dense than the surroundings.  Thus, in stably 
stratified atmospheres, the radiative heating/cooling rate exerts
major control over the rate of vertical ascent/decent.  Steady
vertical motion can only occur as fast as radiation can remove
the temperature variations caused by the adiabatic ascent/descent
\citep{showman-etal-2008a}, when conduction is negligible.  
The idea can be quantified by
rewriting the thermodynamic energy equation (Eq.~\ref{pe}d) in terms
of the Brunt-V\"ais\"al\"a frequency:
\begin{equation}
{\partial T\over \partial t} + {\bf v}\cdot\nabla_p T
- \omega {H_p^2 N^2\over R p} = {\dot q_{\rm net}\over c_p}
\label{thermo}
\end{equation}
where $H_p$ is the pressure scale height.  In situations where the
flow is approximately steady and
vertical thermal advection is more important than (or comparable
to) horizontal thermal advection, an estimate of the magnitude of the 
vertical velocity in an overturning circulation can be 
obtained by equating the right side to the last term on the left side.
Converting to velocity expressed in height units, we obtain
\citep{showman-etal-2008a, showman-guillot-2002}\footnote{Oscillatory 
vertical motions associated with (for example) fast
waves can occur adiabatically and are not governed by this constraint.}
\begin{equation}
W\sim {\dot q_{\rm net}\over c_p}{R\over H_p N^2}.
\label{vert-thermo}
\end{equation}
For Earth's midlatitude troposphere, $H_p\sim10\km$, $N\sim0.01\sec^{-1}$,
and $\dot q_{\rm net}/c_p\sim3\times 10^{-5}\K\sec^{-1}$, implying 
$W\sim10^{-2}\m\sec^{-1}$.  In the stratosphere, however,
the heating rate is lower, and the stable stratification is greater,
leading to values of $W\sim10^{-3}\m\sec^{-1}$.

\medskip

For a canonical hot Jupiter with a 3-day rotation period, horizontal wind 
speeds of $\rm km\,sec^{-1}$ imply Rossby numbers of $\sim$1, and 
for scale heights of $\sim$$300\rm\,km$ (Table~\ref{planetary-parameters})
and horizontal length scales of a planetary radius, one thus
estimates $Ro \, U H/L \sim 10\rm\,m\,sec^{-1}$.  Likewise,
for $R=3700\rm\,J\,kg^{-1}\,K^{-1}$, $H_p\sim 300\rm\,km$, 
$N\approx 0.005\sec^{-1}$ (appropriate for an isothermal layer on
a hot Jupiter with surface gravity of $20\rm\,m\,sec^{-1}$), and 
$\dot q_{\rm net}/c_p\sim10^{-2}\rm K\,sec^{-1}$ (perhaps appropriate
for photospheres of a typical hot Jupiter),  Eq.~(\ref{vert-thermo})
suggests mean vertical speeds of $W\sim 10\rm\,m\,sec^{-1}$ 
\citep{showman-guillot-2002, showman-etal-2008a}.  Thus, despite the
fact that the circulation on hot Jupiters occurs in a radiative zone expected
to be stably stratified over most of the planet, vertical overturning can 
occur near the photosphere over timescales of $H_p/W\sim10^5\rm\,sec$.  Because
the interior of a multi-Gyr-old hot Jupiter transports an intrinsic
flux of only $\sim10$--$100\rm\,W\,m^{-2}$ (as compared with typically
$\sim10^5\rm\,W\,m^{-2}$ in the layer where starlight is absorbed and
infrared radiation escapes to space), the net heating $\dot q_{\rm net}$ should
decrease rapidly with depth.  The mean vertical velocities should thus
plummet with depth, leading to very long overturning times in the bottom
part of the radiative zone.

\subsection{Effect of eddies: jet streams and banding}
\label{rhines}

Although turbulence is a challenging problem, several decades of
work in the fields of fluid mechanics and atmosphere/ocean dynamics 
have led to a basic understanding of turbulent flows and how they interact
with planetary rotation.
In a 3D turbulent flow, vorticity stretching and straining drives a fluid's
kinetic energy to smaller and smaller length scales, where it can
finally be removed by viscous dissipation.  This process, the 
so-called {\it forward} energy cascade, occurs when turbulence has 
scales smaller
than a fraction of a pressure scale height ($\sim$$10\,$km for Earth,
$\sim$200$\,$km for hot Jupiters).  On global scales, however,
atmospheres are quasi-two-dimensional, with horizontal flow dimensions
exceeding vertical flow dimensions by typically factors of $\sim$100.
In this quasi-2D regime, vortex stretching is inhibited, and other
nonlinear processes, such as vortex merging, assume a more prominent role,
forcing energy to undergo an {\it inverse} cascade from small
to large scales \citep[pp. 349-361]{vallis-2006}.  
This process has been well-documented
in idealized laboratory and numerical experiments \citep[e.g.,][]
{tabeling-2002} and helps
explain the emergence of large-scale jets and vortices from small-scale
turbulent forcing in planetary atmospheres.

\begin{figure*}
 \epsscale{1.8}
\plotone{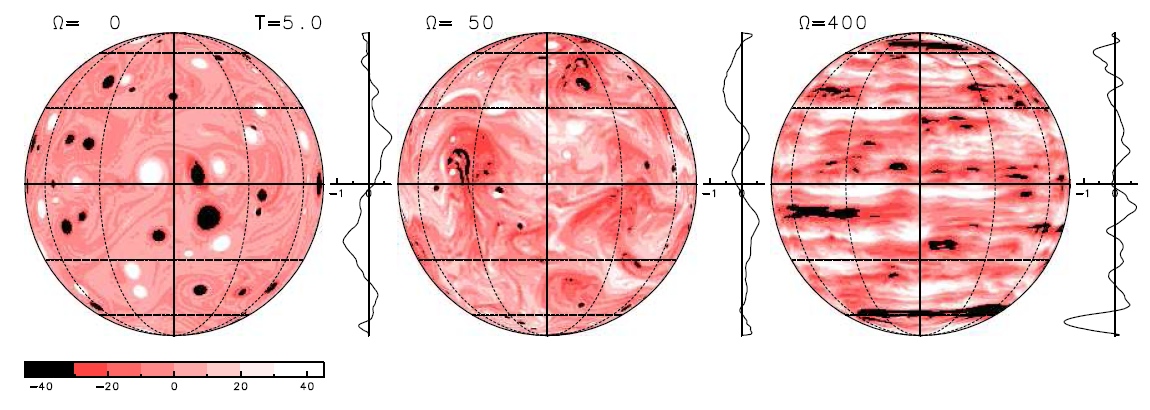}
 \caption{\small Relative vorticity for three 2D non-divergent 
simulations on a sphere initialized from small-scale isotropic turbulence.  
The three simulations are identical except for the planetary rotation rate,
which is zero on the left, intermediate in the middle, and fast on
the right.  Forcing and large-scale friction are zero, but the simulations 
contain a hyperviscosity to maintain numerical stability.  The final
state consists of isotropic turbulence in the non-rotating case but
banded flow in the rotating case, a result of the Rhines effect.  The
spheres are viewed from the equatorial plane, with the rotation axis
oriented upward on the page.  After \citet{yoden-etal-1999},
\citet{hayashi-etal-2000}, \citet{ishioka-etal-1999}, and 
\citet{hayashi-etal-2007}.}
\label{2d-turb}
 \end{figure*}

Where the Coriolis parameter $f$ is approximately constant, this process 
generally produces turbulence that is horizontally isotropic, that is, 
turbulence without any preferred directionality (north-south versus east-west).  
This could explain the existence of some
quasi-circular vortices on Jupiter and Saturn, for example; however,
it fails to explain the existence of numerous jet streams on
Jupiter, Saturn, Uranus, Neptune, and the terrestrial planets.
On the other hand, \citet{rhines-1975} realized that the variation
of $f$ with latitude leads to anisotropy, causing elongation of
structures in the east-west direction relative to the north-south
direction.   This anisotropy can cause the energy to reorganize
into east-west oriented jet streams with a characteristic latitudinal
length scale
\begin{equation}
L_{\beta}=\pi\left({U\over \beta}\right)^{1/2}
\label{rhines-scale}
\end{equation}
where $U$ is a characteristic wind speed and $\beta\equiv df/dy$ is the gradient
of the Coriolis parameter with northward distance $y$.  This length
is called the Rhines scale.  

\begin{figure*}
 \epsscale{1.3}
\plotone{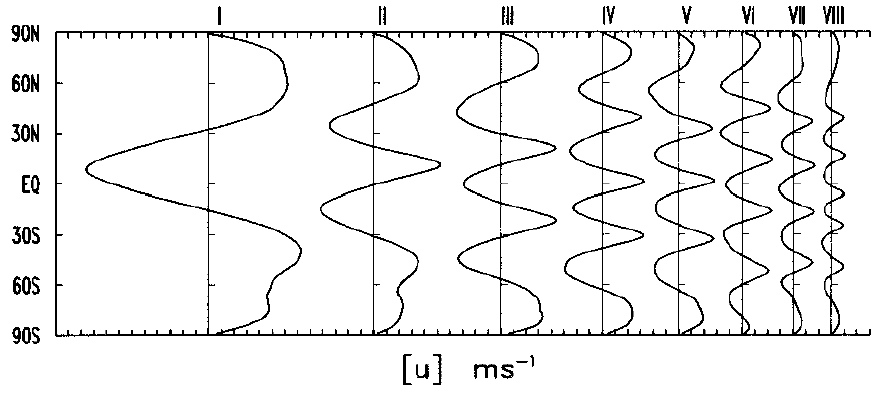}
 \caption{\small  Zonal-mean zonal winds versus latitude for
a series of global, 2D non-divergent simulations performed on a sphere
showing the correlation between jet speed and jet width.  The simulations
are forced by small-scale turbulence and damped by a linear drag.
Each simulation  uses different forcing and friction parameters and 
equilibrates to a different mean wind speed, ranging from fast
on the left to slow on the right.  Simulations with faster jets have
fewer jets and vice versa; 
the jet widths approximately scale with the Rhines length.  From
\citet{huang-robinson-1998}.}
\label{huang-robinson}
 \end{figure*}

Numerous idealized
studies of 2D turbulent flow forced by injection of small-scale turbulence
have demonstrated jet formation by this mechanism [e.g., 
\citet{williams-1978}, \citet{cho-polvani-1996a}, \citet{cho-polvani-1996b}, 
\citet{huang-robinson-1998}, \citet{marcus-etal-2000}; see
\citet{vasavada-showman-2005} for a review]. 
Figure~\ref{2d-turb} illustrates an example.
In 2D non-divergent numerical simulations
of flow energized by small-scale turbulence, the flow remains
isotropic at low rotation rates (Fig.~\ref{2d-turb}, left) but
develops banded structure at high rotation rates (Fig.~\ref{2d-turb},
right).    As shown in Fig.~\ref{huang-robinson}, the
number of jets in such simulations increases as the wind speed decreases,
qualitatively consistent with Eq.~(\ref{rhines-scale}).

The Rhines scale can be interpreted as a transition scale between the regimes 
of turbulence and Rossby waves, which are a large-scale wave solution
to the dynamical equations in the presence of non-zero $\beta$.
Considering the simplest possible case of an unforced 2D non-divergent
fluid (\S\ref{nondivergent}), the vorticity equation reads 
\begin{equation}
{\partial\zeta\over\partial t} + {\bf v}\cdot \nabla\zeta + v\beta=0,
\label{2D-equation}
\end{equation}
where $\zeta$ is the relative vorticity, ${\bf v}$ is the horizontal
velocity vector, and $v$ is the northward velocity component.
The relative vorticity has characteristic scale $U/L$, and so the 
nonlinear term has characteristic scale $U^2/L^2$.  The $\beta$ term has
characteristic scale $\beta U$.  Comparison between these two terms shows
that, for length scales smaller than the Rhines scale, the nonlinear
term dominates, implying the dominance of nonlinear vorticity
advection.  For scales exceeding the Rhines scale, the linear $v\beta$ term
dominates over the nonlinear term, and Rossby waves are the
primary solutions to the equations.\footnote{Adopting
a streamfunction $\psi$ defined by
$u=-\partial\psi/\partial y$ and $v=\partial\psi/\partial x$, the 2D non-divergent
vorticity equation can be written $\partial \nabla^2\psi/\partial t
 + \beta \partial \psi/\partial x=0$ when the nonlinear term is neglected.
In Cartesian geometry with constant $\beta$, this equation has wave 
solutions with a dispersion relationship $\omega=-\beta k^2/(k^2 + l^2)$ 
where $\omega$ is the oscillation frequency and $k$ and $l$ are the 
wavenumbers ($2\pi$ over the wavelength) in the 
eastward and northward directions, respectively.  These are Rossby waves,
which have a westward phase speed and a frequency that depends on the
wave orientation.}

How does jet formation occur at the Rhines scale?  In a 2D fluid
where turbulence is injected at small scales\footnote{In a 2D model,
turbulence behaves in a 2D manner at all scales, but in the atmosphere,
flows only tend to behave 2D when the ratio of widths to heights
$\gg1$ (implying horizontal scales exceeding hundreds of km for Earth).} 
an upscale energy cascade 
occurs, but when the turbulent structures reach sizes comparable 
to the Rhines scale, the transition from nonlinear to quasi-linear 
dynamics prevents the turbulence from easily cascading to scales
larger than $L_{\beta}$.  
At the Rhines
scale, the characteristic Rossby-wave frequency for a typically
oriented Rossby wave, $-\beta/\kappa$, roughly matches the 
turbulent frequency $U\kappa$, where $\kappa$ is the wavenumber
magnitude.  (The Rossby wave frequency results from the dispersion
relation, ignoring the distinction between eastward and northward
wavenumbers; the turbulent frequency is one over an advective
timescale, $L/U$.)  At larger length scales (smaller wavenumbers)
than the Rhines scale, the Rossby-wave oscillation frequency is 
faster than the turbulence frequency and so wave/turbulent interactions
are inefficient at transporting energy into the Rossby-wave regime.
This tends to cause a pile-up of energy
at a wavelength $L_{\beta}$.  As a result, flow structures with a wavelength
of $L_{\beta}$ often contain more energy than flow structures at
any other wavelength.

But why are these dominant structures generally
banded jet streams rather than (say) quasi-circular vortices? The 
reason relates to the anisotropy of Rossby waves.  In a 2D non-divergent
fluid, the Rossby-wave
dispersion relation can be written 
\begin{equation}
\omega=-{\beta k\over k^2 + l^2}=-{\beta\cos\alpha\over \kappa},
\label{rossby}
\end{equation}
where $k$ and $l$ are the wavenumbers in the eastward and northward
directions, $\kappa = \sqrt{k^2 + l^2}$ is the total wavenumber
magnitude, and $\alpha$ is the angle between the Rossby-wave propagation
direction and east.  Equating this frequency to the turbulence frequency
$U\kappa$ leads to an anisotropic Rhines wavenumber
\begin{equation}
k_{\beta}^2 = {\beta\over U}|\cos\alpha|.
\label{rhines2}
\end{equation}
When plotted on the $k$-$l$ wavenumber plane, this relationship traces 
out a dumbbell pattern;
$k_{\beta}$ approaches  $\beta/U$ when $\alpha\approx 0$ or $180^{\circ}$
and approaches zero when $\alpha\approx \pm 90^{\circ}$.  The Rhines
scale $L_{\beta}$ is thus $\sim$$\pi(U/\beta)^{1/2}$ for wavevectors with 
$\alpha\approx 0$ or $180^{\circ}$, but it tends to infinity for 
wavevectors with $\alpha\approx \pm90^{\circ}$.  Because of this
anisotropic barrier,  the inverse cascade can successfully drive 
energy to smaller wavenumbers (larger length scales) along the 
$\alpha = \pm90^{\circ}$ axis than along any other direction. 
Wavevectors with $\alpha\approx \pm90^{\circ}$ correspond to cases 
with $k\ll l$, that is, flow structures whose east-west lengths become 
unbounded compared to their north-south lengths.  The result is 
usually an east-west banded pattern and jet streams whose
latitudinal width is $\sim$$\pi (U/\beta)^{1/2}$.  Although these considerations 
are heuristic, detailed numerical simulations generally confirm that
in 2D flows forced by small-scale isotropic turbulence, the turbulence
reorganizes to produce jets with widths close to $L_{\beta}$ 
\citep[Fig.~\ref{2d-turb}; for a review see][]{vasavada-showman-2005}.

In practice, when small-scale forcing is injected into the 
flow, jet formation does not generally occur as the end product of successive
mergers of continually larger and larger vortices but rather by
a feedback whereby small eddies interact directly with the large-scale
flow and pump up jets \citep[e.g.,][]{nozawa-yoden-1997a, huang-robinson-1998}.
The large-scale background shear associated with jets distorts the eddies,
and the interaction pumps momentum up-gradient into the jet cores, helping to
maintain the jets against friction or other forces.

The above arguments were derived in the context of a 2D non-divergent
model, which is the simplest model where Rossby waves can interact
with turbulence to form jets.  Such a model lacks gravity waves and
has an infinite deformation radius $L_D$ (Eqs.~\ref{sw-deformation-radius}
and \ref{deformation-radius}).  It is possible to extend
the above ideas to one-layer models exhibiting gravity waves and 
a finite deformation radius, such as the shallow-water model.
Doing so suggests that, as long as $\beta$ is strong, jets dominate when 
the deformation radius is large (as in the 2D non-divergent model);
however, the flow becomes
more isotropic (dominated by vortices rather than jets) when
the deformation radius is sufficiently smaller than $(U/\beta)^{1/2}$
\citep[e.g.,][]{okuno-masuda-2003, smith-2004, showman-2007}.

On a spherical
planet, $\beta = 2\Omega a^{-1} \cos\phi$, which implies that a planet
should have a number of jet streams roughly given by
\begin{equation}
N_{\rm jet}\sim \left({2\Omega a\over U}\right)^{1/2}.
\label{number-jets}
\end{equation}

\begin{figure*}
 \epsscale{1.0}
\plotone{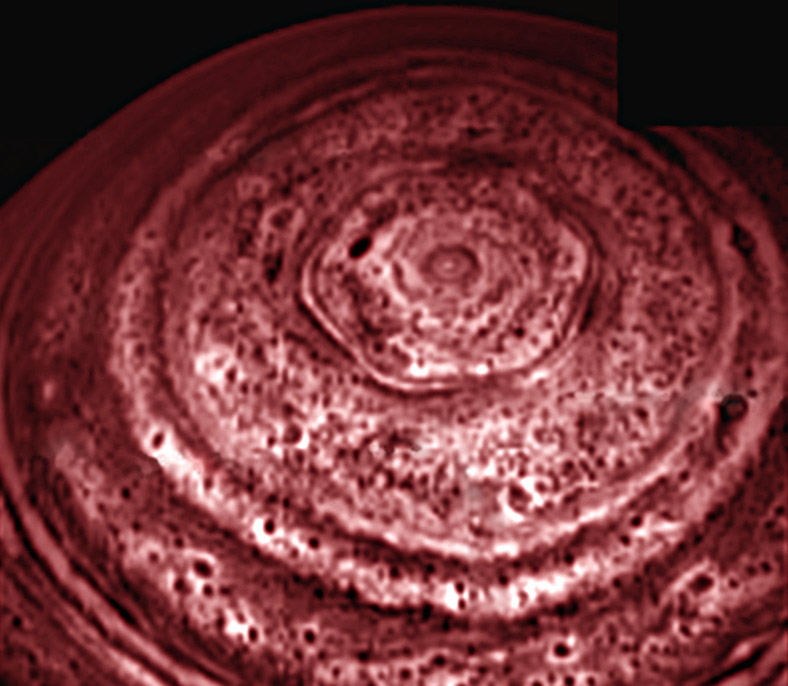}
 \caption{\small Saturn's north polar region as imaged by the Cassini
Visible and Infrared Mapping Spectrometer at 5$\,\mu$m wavelengths.
At this wavelength, scattered sunlight is negligible; bright regions 
are cloud-free regions where thermal emission escapes from the deep 
($\sim3-5\,$bar) atmosphere, whereas dark region are covered by
thick clouds that block this thermal radiation.  Note the scale-dependent
anisotropy: small-scale features tend to be circular, whereas the large-scale
features are banded.  This phenomenon results directly from the Rhines
effect (see text). [Photo credit: NASA/VIMS/Bob Brown/Kevin Baines.]}
\label{vims}
 \end{figure*}

These considerations do a reasonable job of predicting the
latitudinal separations and total number of jets that exist on 
planets in our Solar System.  Given the known wind speeds 
(Table~\ref{planetary-parameters})
and values of $\beta$, the Rhines scale predicts $\sim20$ jet streams 
on Jupiter/Saturn, 4 jet streams on Uranus and Neptune, and 7 jet streams 
on Earth.  This is similar to the observed numbers ($\sim20$ on Jupiter 
and Saturn, 3 on Uranus and Neptune, and 3 to 7 on Earth depending on
how they are defined\footnote{In Earth's troposphere, an instantaneous
snapshot typically shows distinct subtropical and high-latitude jets in each
hemisphere, with weaker flow between these jets and westward flow
at the equator.  The jet latitudes exhibit large excursions in longitude 
and time, however, and when a time- and longitude average is performed,
only a single local maximum in eastward flow exists in each hemisphere,
with westward flow at the equator.}).
Moreover, the scale-dependent anisotropy---with 
quasi-isotropic turbulence at small scales and banded flow
at large scales---is readily apparent on Solar-System planets.  
Fig.~\ref{vims} illustrates an example for Saturn: 
small-scale structures (e.g., the cloud-covered
vortices that manifest as small dark spots in the figure) are relatively
circular, whereas the large-scale structures are banded.

These considerations yield insight into the degree of banding
and size of dominant flow structures to be expected on exoplanets.
Equation~({\ref{number-jets}) suggests that the number of bands scales
as $\Omega^{1/2}$, so rapidly rotating exoplanets will tend to
exhibit numerous strongly banded jets, whereas slowly rotating planets
will exhibit fewer jets with weaker banding.  Moreover, for a given rotation
rate, planets with slower winds should exhibit more bands than planets
with faster winds.  Hot Jupiters
are expected to be tidally locked, implying rotation rates of
typically a few days.  Although wind speeds on hot Jupiters are unknown, 
various numerical models have obtained (or assumed) speeds of 
$\sim$0.5--3$\,\rm km \, sec^{-1}$ \citep{showman-guillot-2002, 
cooper-showman-2005, langton-laughlin-2007, 
dobbs-dixon-lin-2008, menou-rauscher-2009, cho-etal-2003, cho-etal-2008}.  
Inserting these values into Eq.~(\ref{number-jets})
implies $N_{\rm jet}\sim1$--2.  Consistent with these arguments, 
simulations of hot-Jupiter atmospheres forced at small scales
have obtained typically $\sim$1--3 broad jets \citep{cho-etal-2003,
cho-etal-2008, langton-laughlin-2008}.\footnote{Published 3D models of 
hot Jupiters forced by day-night heating contrasts also exhibit only a 
few broad jets \citep[e.g.,][]{cooper-showman-2005, showman-etal-2008a,
showman-etal-2009, dobbs-dixon-lin-2008, menou-rauscher-2009} but for
a different reason.  In this 
case, the jet width may be controlled by the Rossby deformation radius, 
which is close to the planetary radius for many hot Jupiters.  As a result,
the day-night heating contrast injects significant energy into the flow
directly at the planetary scale, and the $\beta$ effect can reorganize
this global-scale energy into a banded jet pattern.  Because the 
forcing and jet scales are comparable, such a flow could lack an inverse
cascade.}
Thus, due to a combination of their slower rotation and presumed faster wind speeds,
hot Jupiters should have only a small number of broad jets---unlike
Jupiter and Saturn.  If the wind speeds are much slower, and/or
rotation rates much greater than assumed here, the number of bands would 
be larger.

\subsection{Role of eddies in 3D atmospheres}
\label{3d-eddies}
Although atmospheres behave in a quasi-2D manner on large
scales due to large horizontal:vertical aspect ratios,
small Rossby numbers, and stable stratification, they are of
course three dimensional, and as a result they can experience
both upscale and downscale energy cascades depending on the
length scales of the turbulence and other factors.  Nevertheless,
the basic mechanism discussed above for the interaction 
of turbulence and the $\beta$ effect still applies and suggests
that even 3D atmospheres can generally exhibit a banded structure with
a characteristic length scale close to $L_{\beta}$.  Consistent
with this idea, numerical simulations show that banding can indeed 
occur in 3D even when $\beta$ is the
only source of horizontal anisotropy \citep[e.g.,][]{sayanagi-etal-2008,
lian-showman-2009}. Nevertheless, on real planets, banding can
also result from anisotropic forcing---such as the fact that
solar heating is primarily a function of latitude rather than
longitude on rapidly rotating planets like Earth and Mars.

A variety of studies show that, in stably stratified, rapidly
rotating atmospheres,
the characteristic vertical length scale of the flow is approximately
$f/N$ times the characteristic horizontal scale \citep[e.g.,][]{charney-1971,
dritschel-etal-1999, reinaud-etal-2003, haynes-2005}.  For typical 
horizontal dimensions of large-scale flows, this often implies vertical 
dimensions of one to several scale heights.

Several processes
can generate turbulent eddies that significantly affect the
large-scale flow.  Convection is particularly important
in giant planet interiors and near the surface of terrestrial
planets. Shear instabilities can occur when the wind shear is
sufficiently large; they transfer energy from the mean flow into
turbulence and reduce the shear of the mean flow. A particularly 
important turbulence-generating process on 
rapidly rotating planets
is {\it baroclinic instability,} which is a dynamical instability driven
by the extraction of potential energy from 
a latitudinal temperature contrast.\footnote{When two adjacent, stably 
stratified air columns have differing temperature profiles, potential 
energy is released when the colder column slides underneath the hotter 
column. For this reason, baroclinic instability, which draws on this
energy source, is sometimes called ``slantwise convection.''}
On non- (or slowly) rotating planets, the presence of a 
latitudinal temperature contrast would simply cause
a direct Hadley-type overturning circulation, which efficiently mutes
the thermal contrasts.  However, on a rapidly rotating planet, 
Hadley circulations cannot penetrate to high latitudes (\S\ref{hadley}),
and, in the absence of instabilities and waves, the atmosphere poleward 
of the Hadley cell would approach 
a radiative-equilibrium temperature structure, with strong latitudinal 
temperature gradients and strong zonal winds in thermal-wind balance 
with the temperature gradients.   In 3D, this radiative-equilibrium
structure can experience baroclinic instabilities, which develop into
three-dimensional eddies that
push cold polar air equatorward and down and push warm low-latitude
air poleward and up.   This process lowers the center of mass of the 
fluid, thereby converting potential energy into kinetic energy and
transporting thermal energy between the equator and poles.  
The fastest instability growth rates occur at length scales comparable
to the Rossby deformation radius (Eq.~\ref{deformation-radius}), 
and the resulting eddies act as a major driver for the mid-latitude 
jet streams on Earth, Mars, and perhaps Jupiter, Saturn, Uranus,
and Neptune.

A formalism for describing the effect of eddies on the mean flow
can be achieved by decomposing the flow into zonal-mean components
and deviations from the zonal mean.  Here, ``eddies'' are defined
as deviations from the zonal-mean flow and can represent the effects
of turbulence, waves, and instabilities.  Denoting zonal means with overbars
and deviations therefrom with primes, we can write $u=\overline{u}+u'$,
$v=\overline{v}+v'$, $\theta=\overline{\theta}+\theta'$, and $\omega=\overline{\omega}+
\omega'$. Substituting these expressions into the 3D primitive equations 
(Eq.~\ref{pe}) and zonally averaging the resulting equations yields
\begin{equation}
{\partial\overline{\theta}\over\partial t}= \overline{\theta q \over T c_p}
 - {\partial(\overline{\theta v})\over\partial y} -
{\partial(\overline{\theta \omega})\over\partial p} 
\label{mean-temp}
\end{equation}
\begin{equation}
{\partial \overline{u}\over\partial t}= f\overline{v} - 
{\partial(\overline{u v})\over\partial y} -
{\partial(\overline{u \omega})\over\partial p} + \overline{\cal F}
- \overline{\cal D}.
\label{mean-wind}
\end{equation}
It can be shown that
\begin{equation}
\overline{u v}= \overline{u}\, \overline{v} + \overline{u'v'}
\end{equation}
and similarly for zonal averages of other quadratic terms.  
Thus, the terms involving derivatives
on the right sides of Eqs.~(\ref{mean-temp})--(\ref{mean-wind})
represent the effect of transport by advection of the {\it mean flow}
(e.g., $\overline{u}\,\overline{v}$) and the effect of transport
by {\it eddies} (e.g., $\overline{u'v'}$).  For example, $\overline{u'v'}$
can be thought of as the northward transport of eastward momentum
by eddies.   If $\overline{u'v'}$ has the same sign as the background
jet shear $\partial\overline{u}/\partial y$, then eddies 
pump momentum up-gradient into the jet cores.  If
they have opposite signs, the eddies transport momentum down-gradient
out of the jet cores. \citep[See][for a brief discussion in the exoplanet
context]{cho-2008}.  Molecular diffusion is a down-gradient process, but
in many cases eddies can transport momentum up-gradient, strengthening 
the jets. 

To illustrate, consider linear Rossby waves, whose dispersion
relation is given by Eq.~(\ref{rossby}) in the case of a 2D non-divergent
flow.  For such waves, it can be shown that the product $\overline{u'v'}$
is opposite in sign to the north-south component of their group velocity
\citep[see][pp.~489-490]{vallis-2006}.  Since the group velocity generally
points away from the wave source, this implies that Rossby waves cause
a flux of eastward momentum {\it toward} the wave source.  The
general result is an eastward eddy acceleration at the latitudes of the
wave source and a westward eddy acceleration at the latitudes where
the waves dissipate or break.  This process is a major mechanism by
which jet streams can form in planetary atmospheres.

The existence of eddy acceleration and heat transport triggers
a mean circulation that has a back-influence on both $\overline{u}$
and $\overline{\theta}$. On a rapidly rotating planet, for 
example, the eddy accelerations and heat transports tend to push
the atmosphere away from geostrophic balance, leading to a mismatch
in the mean north-south pressure gradient and Coriolis forces.
This unbalanced force drives north-south motion, which, through the mass
continuity equation, accompanies vertical motions.  This circulation
induces advection of momentum and entropy, and moreover causes an
east-west Coriolis acceleration $f\overline{v}$, all of which tend
to restore the atmosphere toward geostrophic balance [for
descriptions of this, see \citet[][pp.~100-107]{james-1994} or
\citet[][pp.~313-323]{holton-2004}].  Such mean circulations
tend to have broad horizontal and vertical extents, providing
a mechanism for a localized eddy perturbation to trigger a non-local
response.

Clearly, jets on planets do not continually accelerate, so any jet 
acceleration caused by eddies must (on average)
be resisted by other terms in the equation.
Near the equator (or on slowly rotating planets where Coriolis forces
are small), this jet acceleration is generally balanced by a combination
of friction (e.g., as represented by $\overline{\cal D}$ in 
Eq.~(\ref{mean-wind})) 
and large-scale advection.  However, in the mid- and high-latitudes
of rapidly rotating planets, Coriolis forces are strong, and away
from the surface the eddy accelerations are often balanced by the 
Coriolis force on the mean flow $f\overline{v}$.  

When averaged
vertically, the east-west accelerations due to the Coriolis forces
tend to cancel out\footnote{This occurs because mass continuity requires 
that, in a time average, there be no net north-south mass transport,
hence no net east-west Coriolis acceleration.}, and the {\it vertically
integrated} force balance (on a terrestrial planet, for example) is
between the vertically integrated eddy accelerations and surface 
friction \citep[e.g.,][Chapter 11]
{peixoto-oort-1992}.  This means that regions without active eddy
accelerations must have weak time-averaged surface winds, although the
time-averaged winds may still be strong aloft.  Conversely,
the existence of strong time-averaged winds {\it
at the surface} can only occur at latitudes where eddy accelerations are 
active.  A prime example is the eastward mean surface winds in Earth's 
mid-latitudes (associated with the jet stream), which are enabled 
by the eddy accelerations from mid-latitude baroclinic instabilities.

\section{CIRCULATION REGIMES: GIANT PLANETS}
\label{giants}

\subsection{Thermal structure of giant planets: general considerations}
\label{egp-thermal-structure}

Because of the enormous energy released during planetary accretion,
the interiors of giant planets are hot, and due to
the expected high opacities in their interiors, this energy is thought
to be transported through their interiors into their atmospheres primarily by 
convection rather than radiation \citep[][this volume]
{stevenson-1991, chabrier-baraffe-2000, burrows-etal-2001,
guillot-etal-2004, guillot-2005, fortney-etal-2009}.  
Jupiter, Saturn, and Neptune emit significantly more energy 
than they absorb from the Sun (by factors of 1.7, 1.8, and 2.6, respectively,
implying interior heat fluxes of $\sim$1--$10\rm\,W\,m^{-2}$) showing that 
these planets are still cooling off,
presumably by convection, even 4.6 Gyr after their formation.  Even for
giant exoplanets that lie close to their parent stars, 
the stellar insolation
generally cannot stop the inexorable cooling of the planetary interiors 
\citep[e.g.,][]{guillot-etal-1996, saumon-etal-1996,
burrows-etal-2000, chabrier-etal-2004, fortney-etal-2007}.  
Thus, we expect that in general the interiors of giant planets 
will be convective, and hence the interior 
entropy will be nearly homogenized and the interior temperature and density 
profiles will lie close to an adiabat.\footnote{Compositional gradients 
that force the interior density structures to deviate from a uniform adiabat
could exist in Uranus and some giant exoplanets 
\citep{podolak-etal-1991, chabrier-baraffe-2007}.}

The interiors will not precisely follow an adiabat because
convective heat loss generates descending plumes that are colder than
the background fluid.  We can estimate the deviation from an adiabat as
follows.   Solar-System giant planets are rotationally dominated; as a result,
convective velocities are expected to scale as \citep{stevenson-1979, 
fernando-etal-1991}:
\begin{equation}
w \sim \left({F\alpha g\over \Omega \rho c_p}\right)^{1/2},
\label{mixing-length-velocity}
\end{equation}
where $w$ is the characteristic magnitude of the vertical velocity, 
$F$ is the heat flux transported
by convection, and $\alpha$ and $c_p$ are the thermal expansivity
and specific heat at constant pressure.
To order-of-magnitude, the convected heat flux should be $F\sim \rho w
c_p \, \delta T$, where $\delta T$ is the characteristic magnitude
of the temperature difference between a convective plume and the
environment.  These equations yield
\begin{equation}
\delta T\sim \left({F \Omega \over \rho c_p \alpha g}\right)^{1/2}.
\label{mixing-length-temperature}
\end{equation}
Inserting values for Jupiter's interior ($F\sim10\rm\,W\,m^{-2}$,
$\Omega=1.74\times10^{-4}\rm\,sec^{-1}$, $\rho \sim 1000\rm\,kg
\,m^{-3}$, $\alpha \sim 10^{-5}\rm\,K^{-1}$, $c_p \approx
1.3\times 10^4\rm\,J\,kg^{-1}\,K^{-1}$, and $g\approx 20\rm\,m\,sec^{-2}$),
we obtain $\delta T\sim 10^{-3}\rm\,K$, suggesting fractional density 
perturbations of $\alpha \delta T\sim 10^{-8}$.  Thus, deviations from an
adiabat in the interior should be extremely small.  Even objects
with large interior heat fluxes, such as brown dwarfs or young giant 
planets, will have only modest deviations from an adiabat; for example,
the above equations suggest that an isolated 1000-K Jupiter-like planet 
with a heat flux of $6\times10^4\rm\,W\,m^{-2}$ would experience deviations 
from an adiabat in its interior of only $\sim$$0.1\,$K.

At sufficiently low pressures in the atmosphere, the opacities become small
enough that the outward energy transport transitions from convection to radiation, 
leading to a temperature profile that in radiative equilibrium is
stably stratified and hence suppresses convection.  At a minimum, this
transition (called the {\it radiative-convective boundary})
will occur when the gas becomes optically thin to escaping infrared
radiation (at pressures less than $\sim$0.01--1 bar depending on the opacities).
For Jupiter, Saturn, Uranus, and Neptune, this transition occurs at pressures 
somewhat less than $1\,$bar. However, in the presence of intense
stellar irradiation, the absorbed stellar energy greatly exceeds the 
energy loss from the interior; thus, the mean photospheric temperature only
slightly exceeds the temperature that would exist in thermal equilibrium
with the star.   In this case, cooling of the interior adiabat can only continue
by development of a thick, stably stratified layer that penetrates downward
from the surface and deepens with increasing age.  
Thus, for old, heavily irradiated planets,
the radiative-convective boundary can instead occur at large optical 
depth, at pressures of $\sim$100--1000 bars depending on age 
\citep[see \S\ref{hot-jupiters} and][this volume]{fortney-etal-2009}.

To what extent can large horizontal temperature differences develop in the observable
atmosphere?  This depends greatly on the extent to which the infrared
photosphere\footnote{We define the infrared photosphere as the pressure 
at which the bulk of the planet's infrared radiation escapes to space.}
and radiative-convective boundaries differ in pressure.  On Jupiter, Saturn,
Uranus, and Neptune, the infrared photospheres lie at $\sim$$300\,$mbar, close
to the pressure of the radiative-convective boundary.  Thus, the convective 
interiors---with their exceedingly small lateral temperature 
contrasts---effectively outcrop into the layer where radiation 
streams into space.  
On such a planet, lateral temperature contrasts in the observable atmosphere  
should be small \citep{ingersoll-porco-1978}, and indeed the latitudinal 
temperature contrasts are typically only $\sim$2--$5\,$K in 
the upper tropospheres and lower stratospheres of our Solar-System giant 
planets \citep{ingersoll-1990}.

If instead the radiative-convective boundary lies far deeper than the infrared
photosphere---as occurs on hot Jupiters---then large lateral temperature 
contrasts at the photosphere can potentially develop. Above the 
radiative-convective boundary, convective mixing is inhibited and thus 
cannot force vertical air columns in different regions to lie along a single 
temperature profile.  Laterally varying absorption 
of light from the parent star (or other processes)
can thus generate different vertical temperature gradients in different regions,
potentially leading to significant horizontal temperature contrasts.  In reality,
such lateral temperature contrasts lead to horizontal pressure gradients, which
generate horizontal winds that attempt to reduce the temperature contrasts.  
The resulting temperature contrasts thus depend on a competition between radiation
and dynamics.  These processes will be particularly important when 
strongly uneven external irradiation occurs, as on hot Jupiters.

Several additional processes can produce meteorologically significant temperature 
perturbations on giant planets.  In particular, most giant planets contain trace 
constituents that are gaseous in the high pressure/temperature conditions 
of the interiors but condense in the outermost layers.  This condensation 
releases latent heat and changes the mean molecular weight of the air, both of 
which cause significant density perturbations.  On Jupiter and Saturn, 
ammonia, ammonium hydrosulfide, and water condense at pressures of $\sim$0.5, 
2, and 6 bars, causing temperature increases of $\sim$0.2,
0.1, and $2\,$K, respectively, for solar abundances.  For Jupiter, this 
corresponds to fractional density changes of $\sim$$10^{-3}$ for NH$_3$
and NH$_4$SH and $\sim$$10^{-2}$ for H$_2$O.
At deep levels ($\sim$$2000\,$K, which occurs at pressures of $\sim$$5000\,$bars 
on Jupiter), iron, silicates, and various metal oxides and
hydrides condense, with a total latent heating of $\sim$$1\,$K and a fractional
density perturbation of $\sim$$10^{-3}$.  For Uranus/Neptune, the condensation
structure is similar but also includes methane near the tropopause, with
a latent heating of $\sim$$0.2\,$K and a fractional density change
of $\sim$$3\times10^{-3}$ at solar abundance. Conversion between ortho and para 
forms of the H$_2$ molecule is also important at temperatures $<200\,$K
and produces a fractional density change of $\sim$$10^{-2}$
\citep{gierasch-conrath-1985}.  Note that methane, and plausibly water
and other trace constituents, are enhanced over solar abundances
by factors of $\sim$3, 7, and 20--40 for Jupiter, Saturn, and Uranus/Neptune,
respectively, so the actual density alterations for these planets
likely exceed those listed above by these factors.

The key point is that these
fractional density perturbations exceed those associated with dry convection
in the interior by orders of magnitude; condensation could thus
dominate the meteorology in the region where it occurs.  Such moist
convection could have several effects.  First, if the density contrasts
become organized on large scales, then
significant vertical shear of the horizontal wind can occur 
(via the thermal-wind
equation) that would otherwise not exist in the outermost part of
the convection zone.  Second, it can generate a background temperature
profile that is stably stratified \citep[e.g.][]{stevens-2005}, allowing
wave propagation and various other phenomena.   Third, the buoyancy produced 
by the condensation could act as a powerful driver of the circulation.
For example, on Jupiter and Saturn, condensation leads to powerful 
thunderstorms \citep{little-etal-1999, gierasch-etal-2000} that are a 
leading candidate for driving the global-scale jet streams 
(\S\ref{egp-wind-structure}).  The
cloud formation associated with moist convection can also significantly
alter the temperature structure due to its influence in scattering/absorbing
radiation.

For giant planets hotter than those in our Solar System, the condensation 
sequence shifts to lower pressures and the more volatile species disappear
from the condensation sequence.  Important breakpoints occur
for giant exoplanets with effective temperatures exceeding $\sim$150 and 
$\sim$300--500$\rm\,K$, 
above which ammonia and water, respectively, no longer condense,  thus
removing the effects of their condensation from the meteorology.  The
latter breakpoint will be particularly significant due to the overriding 
dominance of the fractional density change associated with water 
condensation.  Nevertheless, condensation of iron, silicates, and metal 
oxides and hydrides will collectively constitute an important source of 
buoyancy production even for hot giant exoplanets, 
as long as such condensation occurs 
within the convection zone, where its buoyancy dominates over that associated 
with dry convection (Eq.~\ref{mixing-length-temperature}).  For hot Jupiters, 
however, evolution models predict the existence of a quasi-isothermal 
radiative zone extending down to pressures of 
$\sim$100--$1000\,$bars \citep{guillot-2005,
fortney-etal-2009}.  Latent heating occurring within this zone is
less likely to be important, simply because the $\sim$$1\,$K potential 
temperature change associated with silicate condensation is small 
compared to the probable vertical and horizontal potential temperature 
variations associated with the global circulation (\S\ref{hot-jupiters}).
For example, in the vertically isothermal radiative zone of a hot Jupiter, 
a vertical displacement of air by only $\sim$$0.5\,$km would produce a 
temperature perturbation (relative to the background) of $\sim$$1\,$K, 
similar to the magnitude of thermal variation caused by condensation.

\subsection{Circulation of giant planets: general considerations}
\label{egp-wind-structure}
We now turn to the global-scale atmospheric circulation.  As yet, few observational
constraints exist regarding the circulation of exoplanets.
Infrared light curves of several hot Jupiters indirectly suggest the presence
of fast winds able to advect the temperature pattern \citep{knutson-etal-2007b,
cowan-etal-2007, knutson-etal-2009a}, and searches for variability in
brown dwarfs are being made \citep[e.g.,][]{morales-calderon-etal-2006,
goldman-etal-2008}.  But, at present, our local Solar-System giant planets
represent the best proxies to guide our thinking about the circulation on 
a wide class of rapidly rotating giant exoplanets. 
We consider basic issues here and
take up specific models of hot Jupiters in \S\ref{hot-jupiters}.

In the observable atmosphere, the global-scale circulation on Jupiter, Saturn, 
Uranus, and Neptune is dominated by numerous east-west jet 
streams (Fig.~\ref{giantplanets}).  Jupiter and Saturn each exhibit $\sim20$ 
jet streams whereas Uranus and Neptune exhibit three jets each.  Peak 
speeds range from $\sim$$150\m\sec^{-1}$
for Jupiter to over $400\m\sec^{-1}$ for Saturn and Neptune.  
The winds are determined by tracking the motion
of small cloud features, which occur at pressures of $\sim$0.5--4 bars, 
over periods of hours.  The jet streams modulate
the patterns of ascent and descent, leading to a banded cloud pattern
that can be seen in Fig.~\ref{giantplanets}. 

 \begin{figure*}
 \epsscale{1.2}
 \plotone{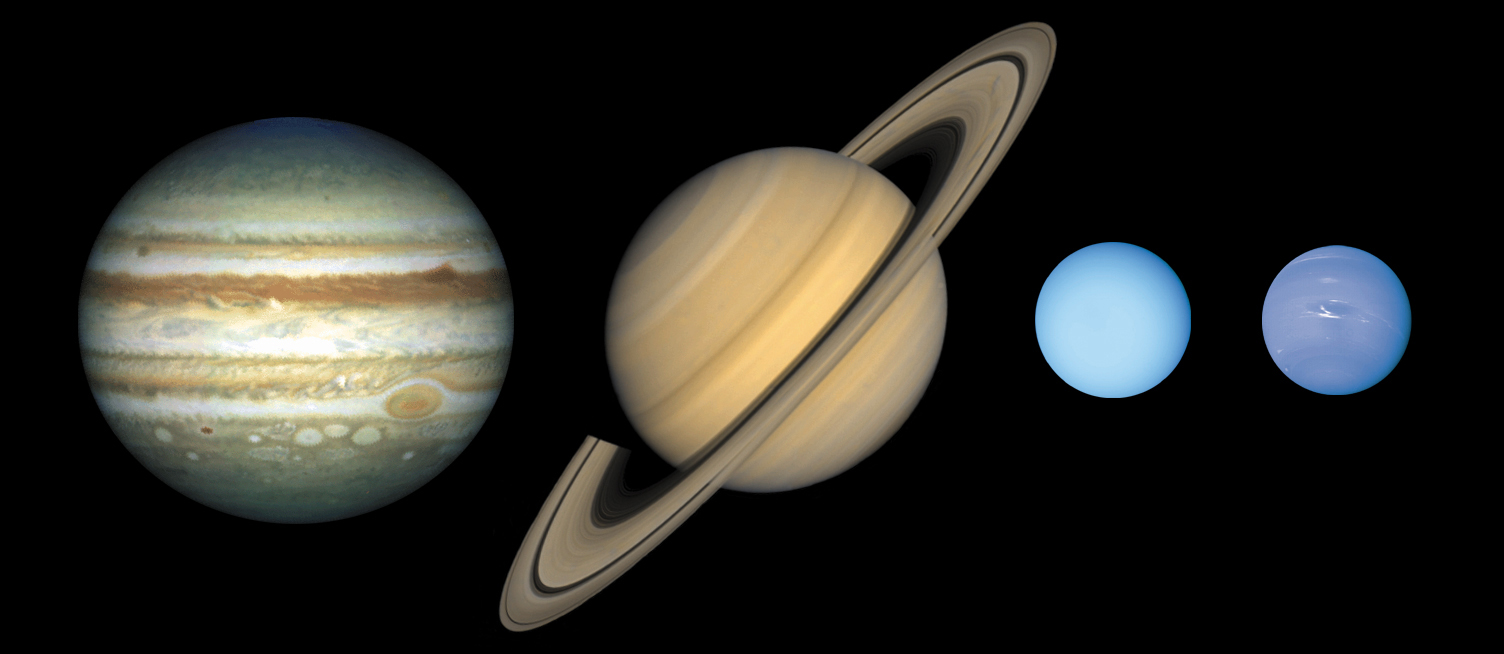}
 \epsscale{1.4}
 \plotone{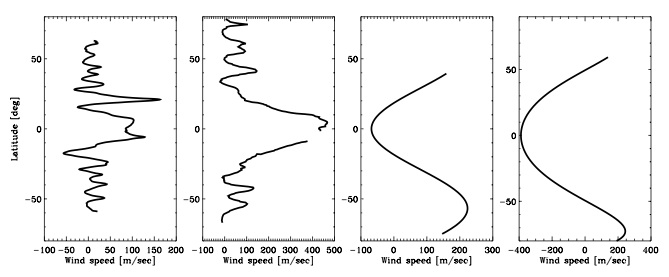}
 \caption{\small {\it Top:} Jupiter, Saturn, Uranus, and Neptune
shown to scale in visible-wavelength images from the Cassini and
Voyager spacecraft.  {\it Bottom:} Zonal-mean zonal wind profiles
obtained from tracking small-scale cloud features over a period
of hours. Jupiter and Saturn have an alternating pattern of $\sim20$
east-west jets, whereas Uranus and Neptune have 3 broad jets.}
\label{giantplanets}
 \end{figure*}

\bigskip
{\it Constraints from balance arguments:}
For giant exoplanets and Solar-System giant planets alike, 
the dynamical link between 
temperatures and winds encapsulated by the thermal-wind
equation allows plausible scenarios for the vertical structure
of the winds to be identified. 
The likelihood that giant planets lose their internal energy by
convection suggests that the interior entropy is nearly homogenized and
hence that the interiors are close to a barotropic state (\S\ref{basic-force-balances}).  
This would lead to the Taylor-Proudman theorem 
(Eqs.~\ref{taylor-proudman}--\ref{taylor-proudman3}), 
which states that the winds are constant on cylinders parallel to
the planetary rotation axis.  This scenario, first suggested by 
\citet{busse-1976} for Jupiter, Saturn, Uranus, and Neptune, postulates
that the jets observed 
in the cloud layer would simply represent the intersection
with the surface of eastward- and westward-moving Taylor columns that 
penetrate throughout the molecular envelope.

How closely will the molecular interior of a giant planet follow the Taylor-Proudman
theorem?  It might not apply if geostrophy does not hold on the length scale of the 
jets (i.e. Rossby number $\gtrsim 1$), but this is unlikely, at least for the rapidly
rotating giant planets of our Solar System \citep{liu-etal-2008}.  However,
for giant exoplanets, 
geostrophy could break down if the interior wind speeds are sufficiently
fast, planetary rotation rate is sufficiently low, the interior heat
flux is sufficiently large (so that convective buoyancy forces exceed 
Coriolis forces $\Omega u$ associated with the jets, for example), or
magnetic effects dominate.  Alternately, even if the interior exhibits 
geostrophic columnar behavior, the interior could exhibit a shear of the 
wind along the Taylor columns if density 
variations on isobars in the interior are sufficiently
large. To illustrate how this might work, 
we cast Eq.~(\ref{3d-vorticity}) in a cylindrical
coordinate system and take the longitudinal component, which gives
\begin{equation}
2\Omega {\partial u\over \partial z_*} = - 
{\nabla \rho\times \nabla p\over \rho^2}\cdot \hat\lambda,
\label{generalized-thermal-wind1}
\end{equation}
where $u$ is the zonal wind, $z_*$ is the coordinate parallel to the rotation
axis (see \S\ref{basic-force-balances}), and $\hat\lambda$ is the unit vector
in the longitudinal direction.  Suppose that horizontal density variations
(on isobars) occur only in latitude.  A geometrical argument shows that the
magnitude of $\nabla\rho\times\nabla p$
is then $|\nabla p|(\partial\rho/\partial y)_p$, where $(\partial \rho/\partial y)_p$
is the partial derivative of density with northward distance, $y$, on a surface
of constant pressure.  The magnitude of the pressure gradient is dominated
by the hydrostatic component, $-\rho g$, and thus we can write
\begin{equation}
2\Omega {\partial u\over\partial z_*} \approx {g\over\rho} 
\left({\partial\rho\over\partial y}\right)_p,
\label{generalized-thermal-wind2}
\end{equation}
which, to order-of-magnitude, implies that the magnitude of 
zonal-wind variation along a Taylor column can be expressed as
\begin{equation}
\Delta u\sim {g\over L \Omega}\left({\delta\rho\over\rho}\right)_p \Delta z_*,
\label{shear-taylor-column}
\end{equation}
where $L$ is the width of the jets in latitude,  $(\delta\rho/\rho)_p$ is 
the fractional density difference (on isobars) across a jet, and $\Delta z_*$ 
is the distance along the Taylor column over which this fractional density 
difference occurs.  In the interior,
convection supplies the fractional density contrasts, and as shown in 
\S\ref{egp-thermal-structure}
these plausibly have extremely small values of $10^{-6}$--$10^{-8}$ 
below the region where condensation can occur.  We can obtain a crude 
order-of-magnitude estimate for the thermal-wind shear 
along a Taylor column by assuming that these fractional density 
differences are organized on the jet scale and coherently extend across the 
full $\sim$$10^4\,$km vertical length of a Taylor column.  Inserting these values 
into Eq.~(\ref{shear-taylor-column}), along with parameters relevant to a 
Jupiter-like planet ($g\approx 20\rm\,m\,sec^{-2}$, $L\approx 5000\,$km, 
$\Omega = 1.74\times10^{-4}\rm\,sec^{-1}$) yields $\Delta u \sim
0.001$--$0.1\rm\,m\,sec^{-1}$.  If this estimate is correct, then Jupiter-like planets
would exhibit only small thermal-wind shear along Taylor columns, and the 
Taylor-Proudman theorem would hold to good approximation in the molecular interior. 

However, the above estimate is crude; if the jet-scale horizontal density contrasts
were much larger than the density contrasts associated with convective plumes,
for example, then the wind shear along Taylor columns could exceed the values
estimated above.  The quantitative validity of Eq.~(\ref{mixing-length-temperature}), 
on which the wind-shear estimates are based, also remains uncertain.  To attack
the problem more rigorously, \citet{kaspi-etal-2009} performed full 3D numerical 
simulations of convection in giant-planet interiors, suggesting that the compressibility 
may play an important role in allowing the generation of thermal-wind shear within
giant planets, especially in the outermost layers of their convection zones.  
Such wind shear is particularly likely to be important for young and/or massive 
planets with large interior heat fluxes.

Although fast winds could exist throughout the molecular envelope,  
Lorentz forces probably act to brake the zonal flows in
the underlying metallic region at pressures exceeding $\sim$1--$3\,$Mbar
\citep{kirk-stevenson-1987, grote-etal-2000a, busse-2002}.  The winds in the deep, 
metallic interior are thus often assumed to be weak.  The transition
between molecular and metallic occurs gradually, and there exists a wide
semi-conducting transition zone over which the Taylor-Proudman theorem approximately holds
yet the electrical conductivity becomes important.
\citet{liu-etal-2008} argued that, if the observed jets on Jupiter and Saturn
penetrated downward as Taylor columns, the Ohmic dissipation would exceed 
the luminosity of Jupiter (Saturn) if the jets extended deeper than
95\% (85\%) of the planetary radius.   
The observed jets are thus probably
shallower than these depths.  However, if the
jets extended partway through the molecular envelope, terminating within
the semi-conducting zone where the Ohmic dissipation occurs, the shear at the 
base of the jets must coexist with lateral 
density contrasts on isobars via the thermal wind relationship 
(Eq.~\ref{generalized-thermal-wind2}).   This is problematic,
because, as previously discussed, sufficiently large sources 
of density contrasts 
are lacking in the deep interior.  \citet{liu-etal-2008} therefore argued that
the jets must be weak {\it throughout} the molecular envelope up to shallow
levels where alternate buoyancy sources become available (e.g., latent heating 
and/or transition to a radiative zone).  

In the outermost layers of a giant planet, density variations associated
with latent heating and/or a transition to a radiative zone allow significant
thermal wind-shear to develop.  On Jupiter and Saturn, this so-called 
``baroclinic'' layer begins with condensation of iron, silicates, and various 
metal oxides and hydrides at pressures of $\sim$$10^4\,$bars
and continues with condensation of water, NH$_4$SH, and
ammonia (at $\sim$10, 2, and 0.5 bars, respectively), finally  
transitioning to the stably stratified stratosphere at $\sim$0.2 bars.  
Early models showed that the observed jet speeds can plausibly result
from thermal-wind shear within this layer assuming the winds in the interior
are zero \citep{ingersoll-cuzzi-1969}.
For example, water condensation could cause a fractional density contrast
on isobars of up to $10^{-2}$, and if these
density differences extend vertically over $\sim$$100\,$km (roughly the altitude
difference between the water-condensation level and the observed cloud deck),
then Eq.~(\ref{shear-taylor-column}) implies $\Delta u\sim30\rm\,m\,sec^{-1}$,
similar to the observed speed of Jupiter's mid-latitude jets 
(Fig.~\ref{giantplanets}).

To summarize, current understanding suggests that giant planets should
exhibit winds within the convective molecular envelope that exhibit columnar
structure parallel to the rotation axis, transitioning in the outermost
layers (pressures less than thousands of bars) to a baroclinic zone where 
horizontal density contrasts associated with latent heating and/or a radiative 
zone allow the zonal winds to vary with altitude.

\bigskip

{\it Constraints from observations:}
Few observations yet exist that constrain the deep wind
structure in Jupiter, Saturn, Uranus, and Neptune.
On Neptune, gravity data from the 1989 {\it Voyager 2} flyby imply that 
the fast jets observed in the cloud layers are confined to the outermost 
few percent or less of the planet's mass,
corresponding to pressures less than $\sim$$10^5\,$bars \citep{hubbard-etal-1991}.
By comparison, only weak constraints currently exist for Jupiter and Saturn.
 The Galileo Probe, which entered Jupiter's atmosphere at a latitude of
$\sim$$7^{\circ}$N in 1995, showed that the equatorial jet penetrates
to at least 22 bars, roughly $150\km$ below the visible cloud deck
\citep{atkinson-etal-1997}, and indirect inferences suggest that the
jets penetrate to at least $\sim$5--$10\,$bars at other latitudes
\citep[e.g.][]{dowling-ingersoll-1989, sanchez-lavega-etal-2008}. 
In 2016, however, NASA's {\it Juno} mission 
will measure Jupiter's gravity field, finally determining whether Jupiter's jets penetrate 
deeply or are confined to pressures as shallow as thousands
of bars or less.  These results will have important implications for
understanding the deep-wind structure in giant exoplanets generally.

\bigskip

{\it Jet-pumping mechanisms:}
Two scenarios exist for the mechanisms to pump the zonal jets on the giant 
planets \citep[for a review see][]{vasavada-showman-2005}.
In the ``shallow forcing'' scenario, the jets are hypothesized
to result from moist convection (e.g., thunderstorms), baroclinic
instabilities, or other turbulence-generating processes in the baroclinic
layer in the outermost region of the planet.  
Two-dimensional and shallow-water models of this process,
where random turbulence is injected as an imposed forcing (or alternately
added as a turbulent initial condition), show
success in producing alternating zonal jets through the Rhines
effect (\S \ref{rhines}). For appropriately chosen forcing and
damping parameters (or for appropriate initial velocities when the
turbulence is added as an initial condition), such models typically exhibit 
jets of approximately the observed speed and spacing \citep{williams-1978,
cho-polvani-1996a, huang-robinson-1998, scott-polvani-2007, 
showman-2007, sukoriansky-etal-2007}.  Most models of this type produce
an equatorial jet that flows westward rather than eastward as on Jupiter 
and Saturn, though robust eastward equatorial flow has been obtained in
a recent shallow-water study \citep{scott-polvani-2008}.
Three-dimensional models of this process can explicitly resolve the turbulent
energy generation and therefore need not inject turbulence
by hand.  These models likewise exhibit multiple zonal jets via the
Rhines effect, in some cases spontaneously producing an eastward 
equatorial jet as on Jupiter and Saturn \citep[e.g.][]{williams-1979,
williams-2003a, lian-showman-2008, lian-showman-2009, schneider-liu-2009}.
Figure~\ref{lian-showman-fig2}, for example, illustrates 3D
simulations from \citet{lian-showman-2009}
where the circulation is driven by latent heating associated 
with condensation of water vapor; the condensation produces turbulent eddies
that interact with the planetary rotation (i.e., non-zero $\beta$) 
to generate jets.  Globes show the zonal wind for a Jupiter 
case (top row), Saturn case (middle), and a case representing Uranus/Neptune 
(bottom row).  The Jupiter and Saturn cases develop $\sim20$ jets, 
including an eastward equatorial jet,
whereas the Uranus/Neptune case develops a 3-jet structure with a broad
westward equatorial flow---qualitatively similar to the observed jet patterns
on these planets (Fig.~\ref{giantplanets}).
Note that shallow {\it forcing} need not imply shallow {\it jets};
because the atmosphere's response to a perturbation is nonlocal, 
deep jets (penetrating many scale heights below the clouds)
can result from jet pumping confined to the cloud layer if the
frictional damping in the interior is sufficiently weak 
\citep{showman-etal-2006, lian-showman-2008}.

\begin{figure*}
 \epsscale{1.0}
 \plotone{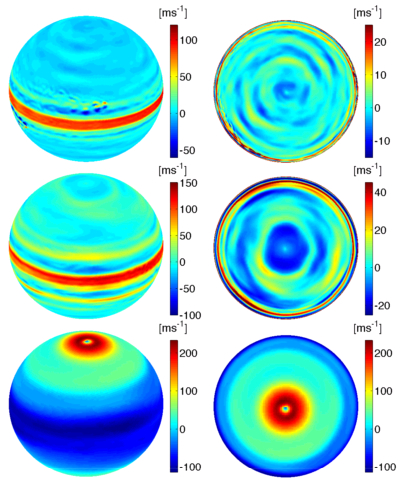}
 \caption{\small Illustration of the effect of latent heating on the circulation
of a rapidly rotating giant planet.  Shows zonal (east-west) wind from 3D
atmospheric simulations 
of Jupiter (top row), Saturn (middle row), and a case representing Uranus/Neptune 
(bottom row) where the circulation is driven by condensation of water vapor,
with an assumed deep abundance of 3, 5, and 30 times solar for the Jupiter, Saturn,
and Uranus/Neptune cases, respectively.  Left column shows an oblique view and
right column shows a view looking down over the north pole.
Note the development of numerous jets (including superrotating equatorial jets
on Jupiter and Saturn), which bear qualitative resemblance to the observed 
jets in Fig.~\ref{giantplanets}.  From \citet{lian-showman-2009}.}
\label{lian-showman-fig2}
 \end{figure*}

In the ``deep forcing'' scenario, convection throughout the
molecular envelope is hypothesized to drive the jets. To date, 
most studies of this process involve 3D numerical simulations of
convection in a self-gravitating spherical shell excluding 
the effects of magnetohydrodynamics \citep[e.g.,][]{aurnou-olson-2001,
christensen-2001, christensen-2002, heimpel-etal-2005}.  
The boundaries are generally
taken as free-slip impermeable spherical surfaces with an outer boundary
at the planetary surface and an inner boundary at a radius of 0.5 
to 0.9 planetary radii. 
Most of these studies assume the mean density, thermal 
expansitivity, and other fluid properties are constant with radius.  
These studies generally produce jets penetrating through the shell
as Taylor columns.  When the shell is thick (inner radius $\le0.7$ of
the outer radius), only $\sim3$--5
jets form \citep{aurnou-olson-2001, christensen-2001, christensen-2002,
aurnou-heimpel-2004}, inconsistent with Jupiter and Saturn.  Only
when the shell has a thickness $\sim10\%$ or less of the planetary
radius do such simulations produce a Jupiter- or Saturn-like profile
with $\sim20$ jets \citep{heimpel-etal-2005}.  However, on giant
planets, the density and thermal expansivity each vary by several
orders of magnitude from the photosphere to the deep interior, and
this might have a major effect on the dynamics.  Only recently has
this effect been included in detailed numerical models of the convection
\citep{glatzmaier-etal-2009, kaspi-etal-2009}, and these studies show
that the compressibility can exert a significant influence on the jet 
structure.  A difficulty in interpreting all the convective simulations 
described here is that, to integrate in reasonable times, they must 
adopt viscosities and heat fluxes orders of magnitude too large.

The fast interior jets in these convection simulations imply the existence of 
strong horizontal pressure contrasts, which in these models are supported
by the impermeable upper and lower boundaries.  On a giant planet,
however, there is no solid surface to support such pressure variations. 
If these fast winds transitioned to weak winds within the metallic region, 
significant lateral density variations would be required 
via the thermal-wind relationship.  As discussed previously, however, there 
exists no obvious source of such density variations (on isobars) 
within the deep interior.  
Resolving this issue will require 3D convection simulations that incorporate 
magnetohydrodynamics and simulate the entire 
molecular$+$metallic interior.  Several groups are currently making such efforts, so
the next decade should see significant advances in this area.

\subsection{Hot Jupiters and Neptunes}
\label{hot-jupiters}

Because of their likelihood of transiting their stars, hot Jupiters
(giant planets within 0.1 AU of their primary star) remain the
best characterized exoplanets and thus far have 
been the focus of most work on exoplanet atmospheric
circulation.  Dayside photometry and/or infrared spectra
now exist for a variety of hot Jupiters, constraining the 
dayside temperature structure.  For some planets, such as HD 189733b, 
this suggests a temperature profile that
decreases with altitude from $\sim0.01$--1 bar \citep{charbonneau-etal-2008,
barman-2008, swain-etal-2009}, 
whereas other hot Jupiters, such as HD 209458b, TrES-2, TrES-4, 
and XO-1b \citep{knutson-etal-2008a, knutson-etal-2009b, machalek-etal-2008}, 
appear to exhibit a thermal inversion layer (i.e., a hot stratosphere) 
where temperatures rise above 2000 K.

Infrared light curves at 8 and/or 
$24\,\mu$m now exist for HD 189733b, Ups And b, HD 209458b, and
several other hot Jupiters \citep{knutson-etal-2007b, knutson-etal-2009a, 
harrington-etal-2006, cowan-etal-2007}, placing constraints
on the day-night temperature distribution of these planets.
Light curves for HD 189733b and HD 209458b exhibit nightside 
brightness temperatures only modestly ($\sim20$--$30$\%) cooler
than the dayside brightness temperatures, suggesting efficient
redistribution of the thermal energy from dayside to nightside.
On the other hand, Ups And b and HD 179949b exhibit larger
day-night phase variations that suggests large (perhaps $>500\,$K)
day-night temperature differences. 

These and other observations provide sufficient constraints on the 
dayside temperature structure and day-night temperature distributions 
to make comparison with detailed atmospheric circulation models 
a useful exercise.  Here we survey the basic dynamical regime and 
recent efforts to model these objects \citep[see also][]
{showman-etal-2008b,cho-2008}.

The dynamical regime on hot Jupiters differs from that on Jupiter and 
Saturn in several important ways.  First, because of the short spindown times
due to their proximity to their stars, hot Jupiters are expected to rotate
nearly synchronously\footnote{or pseudo-synchronously in the case
of hot Jupiters on highly eccentric orbits} with their orbital periods,
implying rotation periods of 1--$5\,$Earth days for most hot Jupiters
discovered to date.  This is 2--12 times slower than Jupiter's 10-hour 
rotation period, implying that, compared to Jupiter, hot Jupiters 
experience significantly weaker Corolis forces for a given wind speed.  
Nevertheless, Coriolis forces can still
play a key role: for global-scale flows (length scales $L\sim10^8\,$m)
and wind speeds of $2\rm\,km\,sec^{-1}$ (see below), the Rossby number for 
a hot Jupiter with a 3-day period is $\sim1$, implying a three-way force
balance between Coriolis, pressure-gradient, and inertial (i.e. advective)
terms in the horizontal equation of motion.  Slower winds would imply smaller
Rossby numbers, implying greater rotational dominance.

Second, hot Jupiters receive enormous energy fluxes 
($\sim10^4$--$10^6\rm\,W\,m^{-2}$ on a global average) from their parent 
stars, in contrast to the $\sim10\rm\,W\,m^{-2}$ received by Jupiter 
and $1\rm\,W\,m^{-2}$ for Neptune.  The resulting high temperatures lead 
to short radiative time constants of days or less at pressures 
$<1\,$bar \citep[][and Eq.~(\ref{tau-rad})]{showman-guillot-2002, 
iro-etal-2005, showman-etal-2008a}, in contrast to Jupiter where 
radiative time constants are years.  Even in the presence of fast winds, 
these short radiative time constants allow the possibility of large 
fractional day-night temperature differences, particularly on the most 
heavily irradiated planets---in contrast to Jupiter where temperatures 
vary horizontally by only a few percent.

Third, over several Gyr of evolution, the intense irradiation received by
hot Jupiters leads to the development of a nearly isothermal radiative 
zone extending to pressures of $\sim100$--1000 bars \citep{guillot-etal-1996, 
guillot-2005, fortney-etal-2009}.  The observable weather in hot 
Jupiters thus occurs not within the convection zone but within the 
stably stratified radiative zone.  Moreover, as described in 
\S\ref{egp-thermal-structure},  this separation between
the photosphere and the radiative-convective boundary may allow large 
horizontal temperature differences to develop and support significant 
thermal-wind shear.  For example, assuming that a lateral temperature contrast 
of $\Delta T\sim 300\,$K extending over several scale heights 
occurs over a lateral distance of $10^8\rm\,m$,
the resulting thermal-wind shear is $\Delta u\sim 2\rm\,km\,sec^{-1}$.  
Thus, balance arguments suggest that the existence of significant 
horizontal temperature contrasts would naturally require fast wind speeds.

The modest rotation rates, large stable stratifications, and possible fast
wind speeds suggest that the Rossby deformation radius and Rhines scale 
are large on hot Jupiters.  On Jupiter, the deformation radius is 
$\sim2000\rm\,$km and the Rhines scale is $\sim10^4\,$km (much smaller 
than the planetary radius of $71,400\,$km), which helps explain why Jupiter 
and Saturn have an abundant population of small vortices  
and numerous zonal jets (Figs.~\ref{vims}--\ref{giantplanets}) 
\citep[][]{williams-1978, cho-polvani-1996a, 
vasavada-showman-2005}.  In contrast, a hot Jupiter with a rotation 
period of 3 Earth days and wind speed of $2\rm\,km\,sec^{-1}$ has a 
deformation radius and Rhines scale comparable to the planetary radius.  
Thus, dynamical structures such as vortices and jets should be more 
global in scale than occurs on Jupiter and Saturn \citep{menou-etal-2003, 
showman-guillot-2002, cho-etal-2003}.  This scaling argument is supported 
by detailed nonlinear numerical simulations of the circulation, which 
generally show the development of only $\sim1$--3 broad jets
\citep{showman-guillot-2002, cho-etal-2003, cho-etal-2008, 
dobbs-dixon-lin-2008, langton-laughlin-2007, langton-laughlin-2008, 
showman-etal-2008a, showman-etal-2009}.

A variety of 2D and 3D models have been used to investigate
the atmospheric circulation of hot Jupiters.  \citet{cho-etal-2003, 
cho-etal-2008} performed global numerical simulations of hot Jupiters
on circular orbits using the one-layer
equivalent barotropic equations (a one-layer model that is mathematically
similar to the shallow-water equations described in \S~\ref{shallow-water}).  
They initialized the
simulations with small-scale balanced turbulence and forced them
with a large-scale deflection of the surface
to provide a crude representation of the pressure effects
of the day-night heating gradient.   However, no explicit heating/cooling
was included.  Together, these effects led
to the production of several broad, meandering jets and drifting
polar vortices (Fig.~\ref{cho-etal-2003}).  
The mean wind speed in the final state is to a large
degree determined by the mean speed of the initial turbulence, which
ranged from $100$--$800\rm\,m\,sec^{-1}$ in their models.
At the large ($\sim$planetary-scale) deformation radius relevant to
hot Jupiters, the equatorial jet in the final state 
can flow either eastward or westward
depending on the initial condition and other details.  The simulations 
exhibit significant time variability that, if present on hot Jupiters, 
would lead to detectable orbit-to-orbit variability in light curves 
and secondary eclipse depths \citep{cho-etal-2003,
rauscher-etal-2007a, rauscher-etal-2008}.

\begin{figure*}
 \epsscale{1.0}
 \plotone{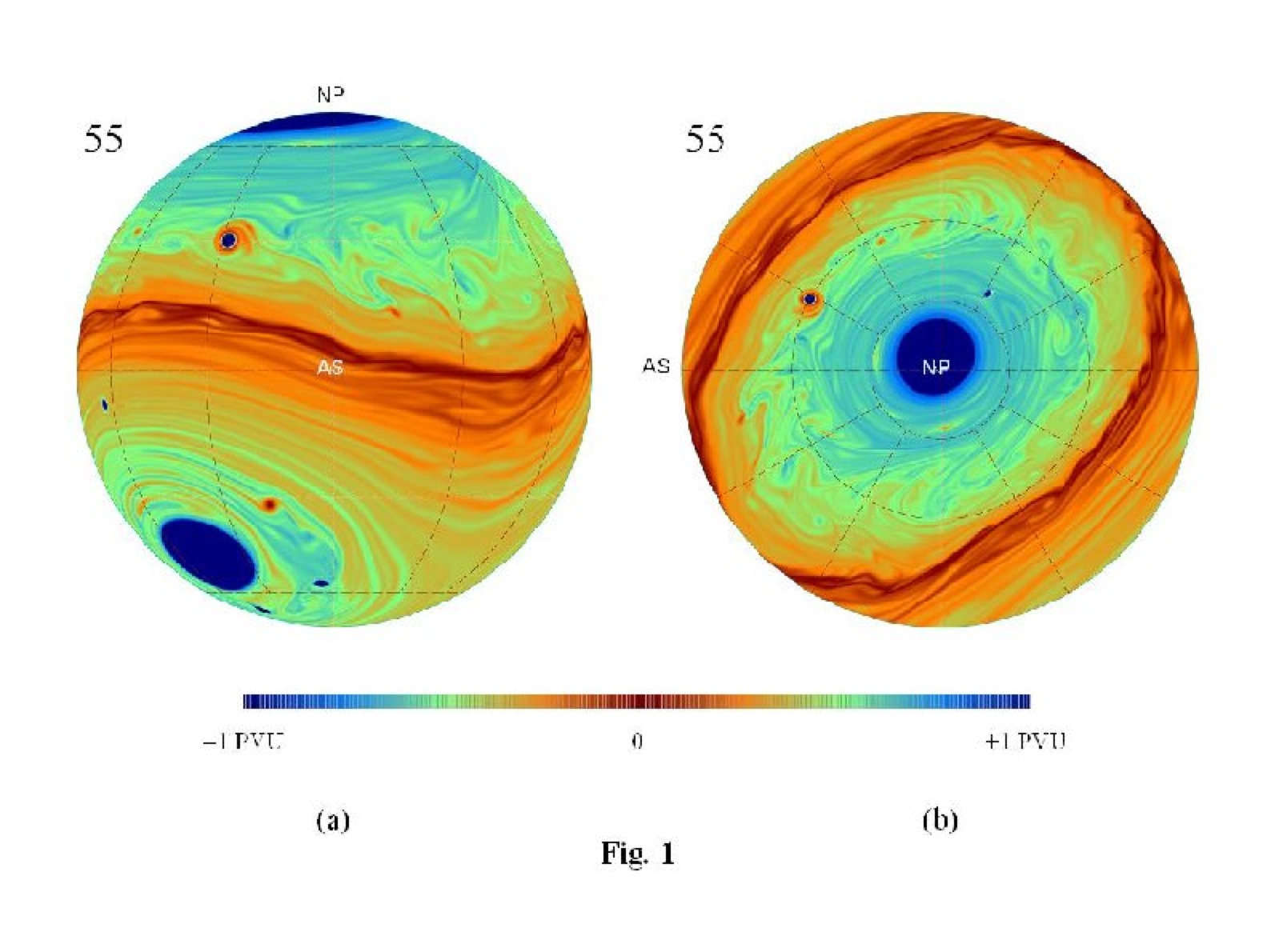}
 \epsscale{1.0}
 \plotone{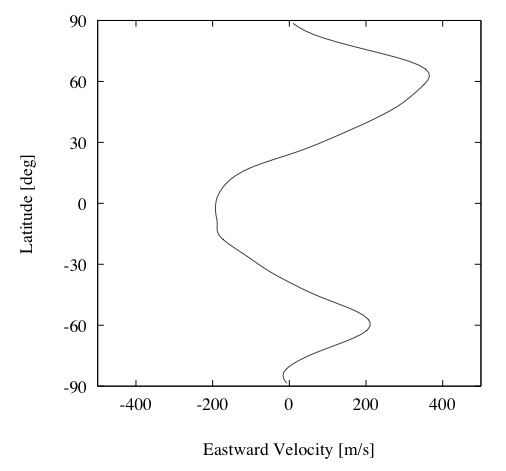}
 \caption{\small One-layer, global equivalent barotropic simulation of 
the hot Jupiter HD 209458b from \citet{cho-etal-2003}. Globes show equatorial 
(left) and polar (right) views of the potential vorticity (see 
\S\ref{equations}); 
note the development of large polar vortices and small-scale turbulent 
structure, which result from a combination of the turbulent initial condition 
and large-scale ``topographic'' forcing intended to qualitatively represent 
the day-night heating contrast.  Plot shows zonal-mean zonal wind
versus latitude, illustrating the 3-jet structure that develops.}
\label{cho-etal-2003}
 \end{figure*}

\citet{langton-laughlin-2007} performed global,
2D simulations of hot Jupiters on circular orbits using
the shallow-water equations with a mass source on
the dayside and mass sink on the nightside to parameterize the 
effects of dayside heating and nightside cooling. 
When the obliquity is assumed zero and the mass sources/sinks
are sufficiently large, their forced flows quickly reach a steady 
state with wind speeds of $\sim1\rm\,km \,sec^{-1}$ and order-unity
lateral variations in the thickness of the shallow-water layer.
On the other hand, \citet{langton-laughlin-2008} numerically solved
the 2D fully compressible equations for the horizontal velocity 
and temperature of hot Jupiters on eccentric orbits and obtained
very turbulent, time-varying flows.  Apparently, in this case,
the large-scale heating patterns produced hemisphere-scale vortices
that were dynamically unstable, leading to the breakdown of these
eddies into small-scale turbulence.

Several authors have also performed 3D numerical simulations
of hot Jupiters.  \citet{showman-guillot-2002}, \citet{cooper-showman-2005,
cooper-showman-2006}, \citet{showman-etal-2008a}, 
and \citet{menou-rauscher-2009} performed global simulations with
the 3D primitive equations where
the dayside heating and nightside cooling was parameterized with
a Newtonian heating/cooling scheme, which relaxes the temperatures
toward a prescribed radiative-equilibrium temperature profile (hot
on the dayside, cold on the nightside) over a prescribed radiative
timescale.  \citet{dobbs-dixon-lin-2008} performed simulations with
the fully compressible equations in a limited-area domain, consisting of 
the equatorial and mid-latitudes but with the poles cut off.  Like
the studies listed above, they also adopted a simplified method for 
forcing their flow, in this case using a radiative diffusion scheme.  
\citet{showman-etal-2009} coupled their global 3D dynamical solver 
to a state-of-the-art, non-gray, cloud-free radiative transfer scheme
with opacities calculated assuming local chemical equilibrium
(Figs.~\ref{showman-etal-2009-hd189733b} and \ref{showman-etal-2009-hd209458b}).
The above models models all generally obtain wind structures with 
$\sim1$--3 broad jets with speeds of $\sim1$--$4\rm\,km\,sec^{-1}$.

\begin{figure*}
 \epsscale{1.2}
 \plotone{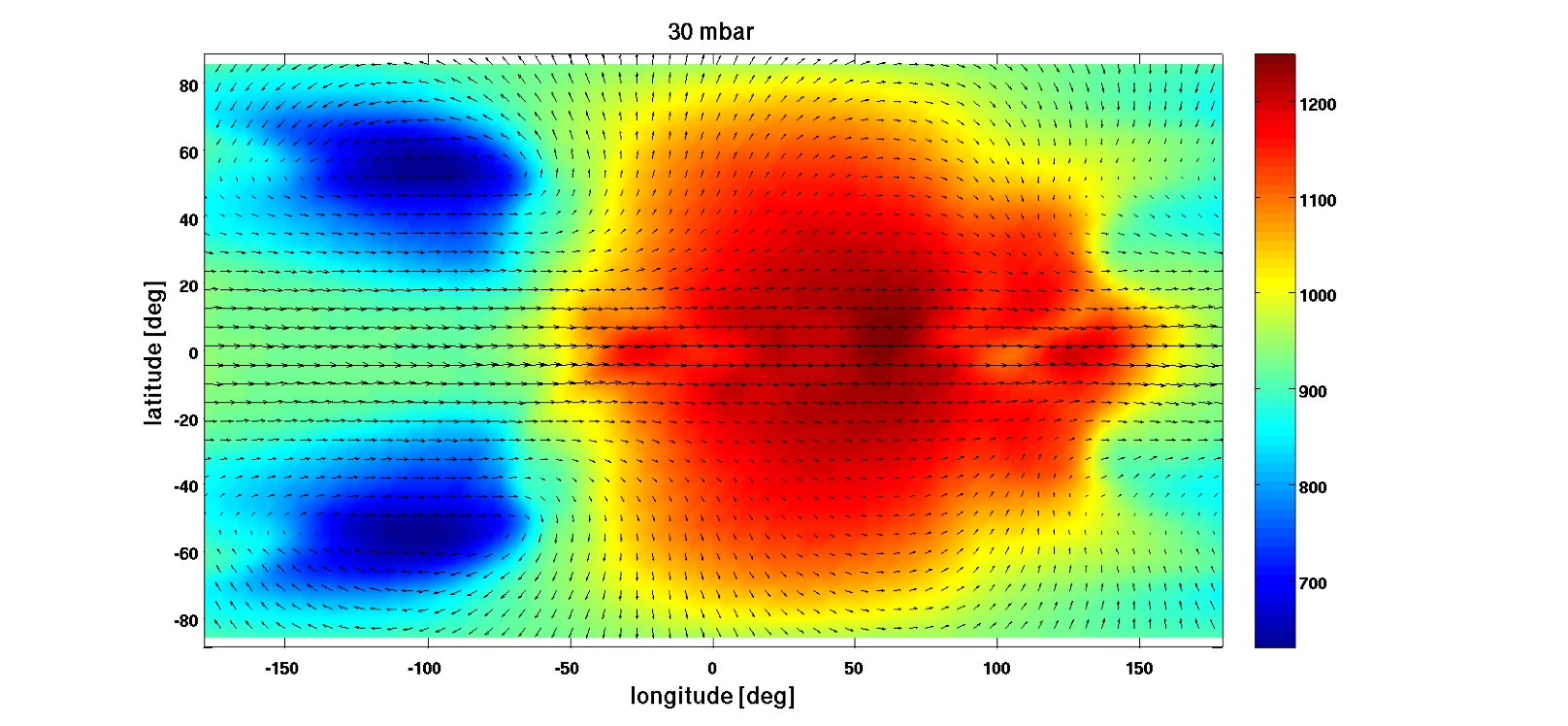}   
 \epsscale{0.8}
 \plotone{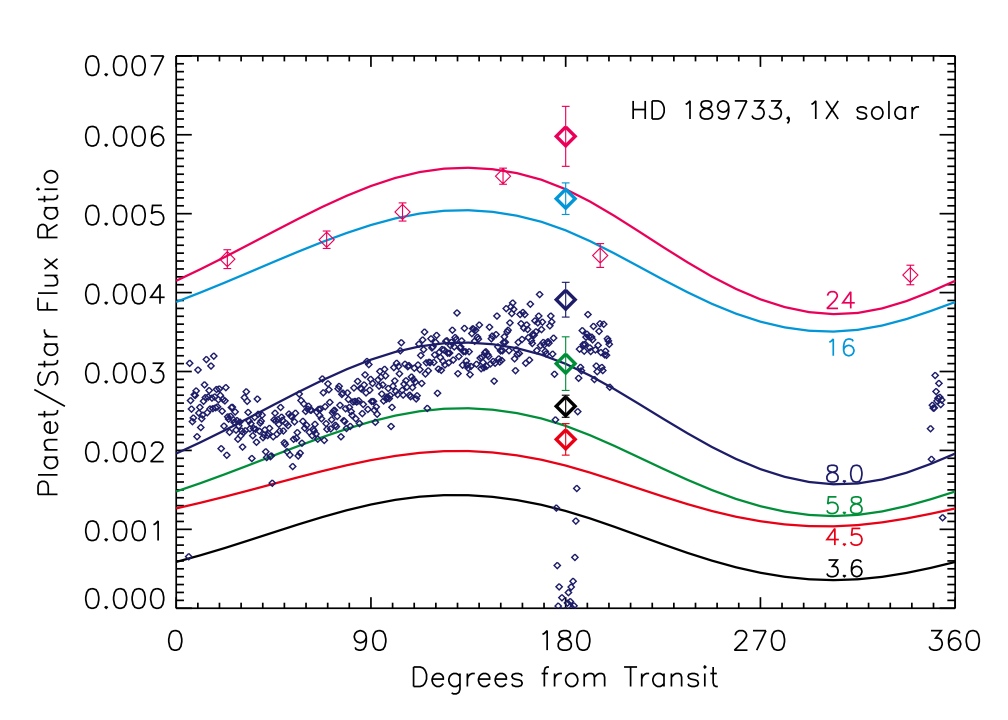}
 \caption{\small Results from a 3D simulation of the hot Jupiter
HD 189733b from \citet{showman-etal-2009}.  The dynamics are coupled
to a realistic representation of cloud-free, non-gray radiative transfer
assuming solar abundances.  {\it Left:} Temperature (colorscale, in K)
and winds (arrows) at the 30 mbar pressure, which is the approximate
level of the mid-IR photosphere.  A strong eastward equatorial jet 
develops that displaces the hottest regions eastward from the substellar
point (which lies at $0^{\circ}$ longitude, $0^{\circ}$ latitude).
{\it Right:} Light curves in {\it Spitzer} bandpasses calculated from the
simulation (curves; labels show wavelength in $\mu$m) in comparison 
to observations (points) from \citet{knutson-etal-2007b, knutson-etal-2009a}, 
\citet{charbonneau-etal-2008}, and \citet{deming-etal-2006}.}
\label{showman-etal-2009-hd189733b}
 \end{figure*}

\begin{figure*}
 \epsscale{1.2}
 \plotone{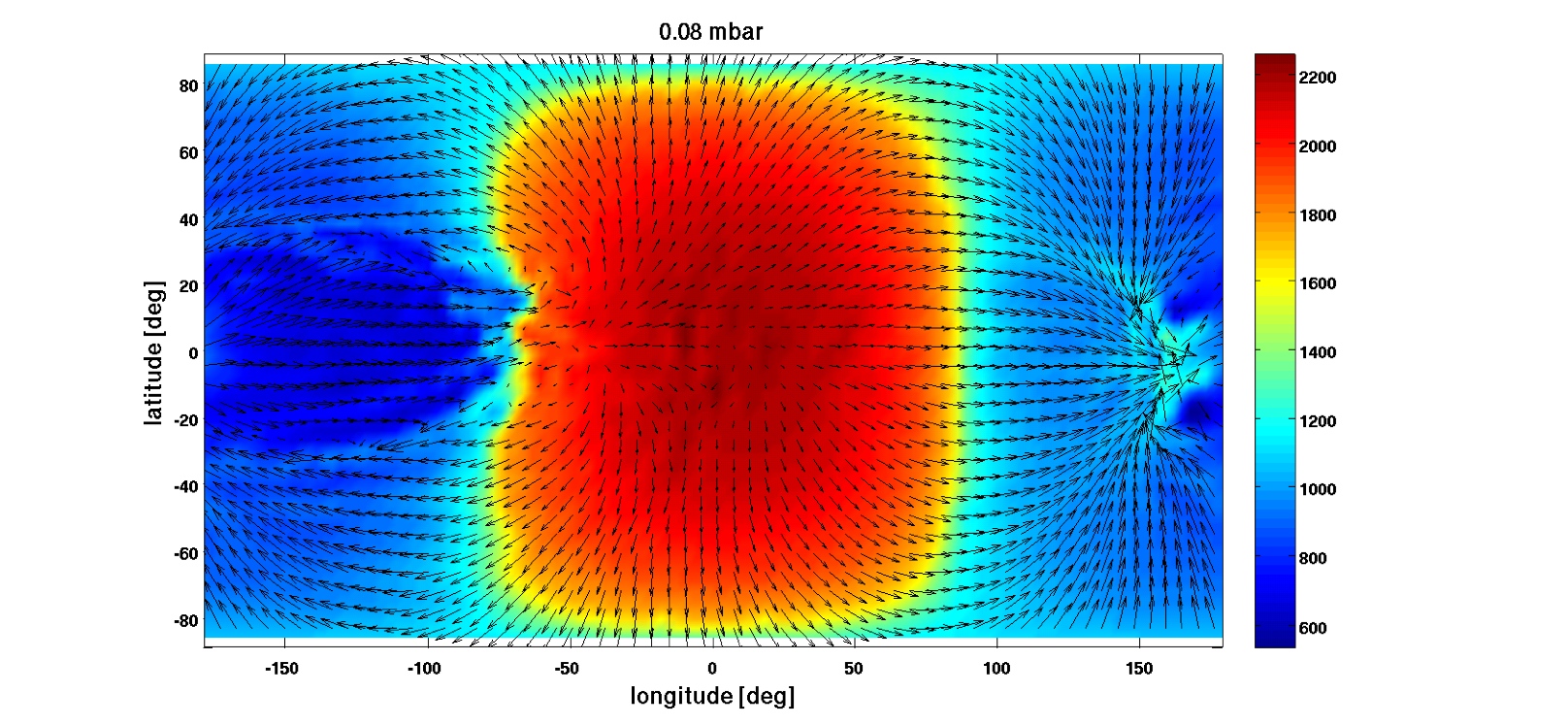}  
 \plotone{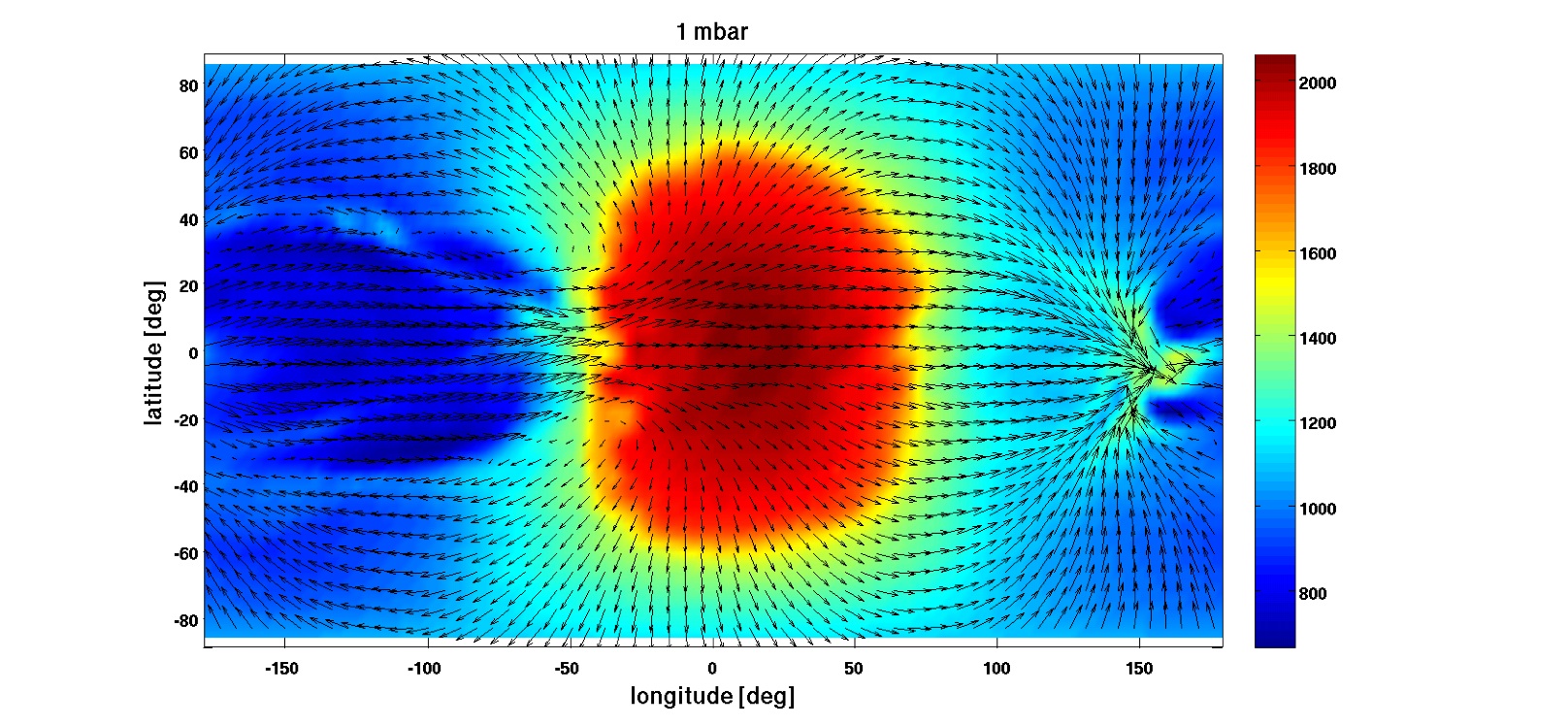}
 \plotone{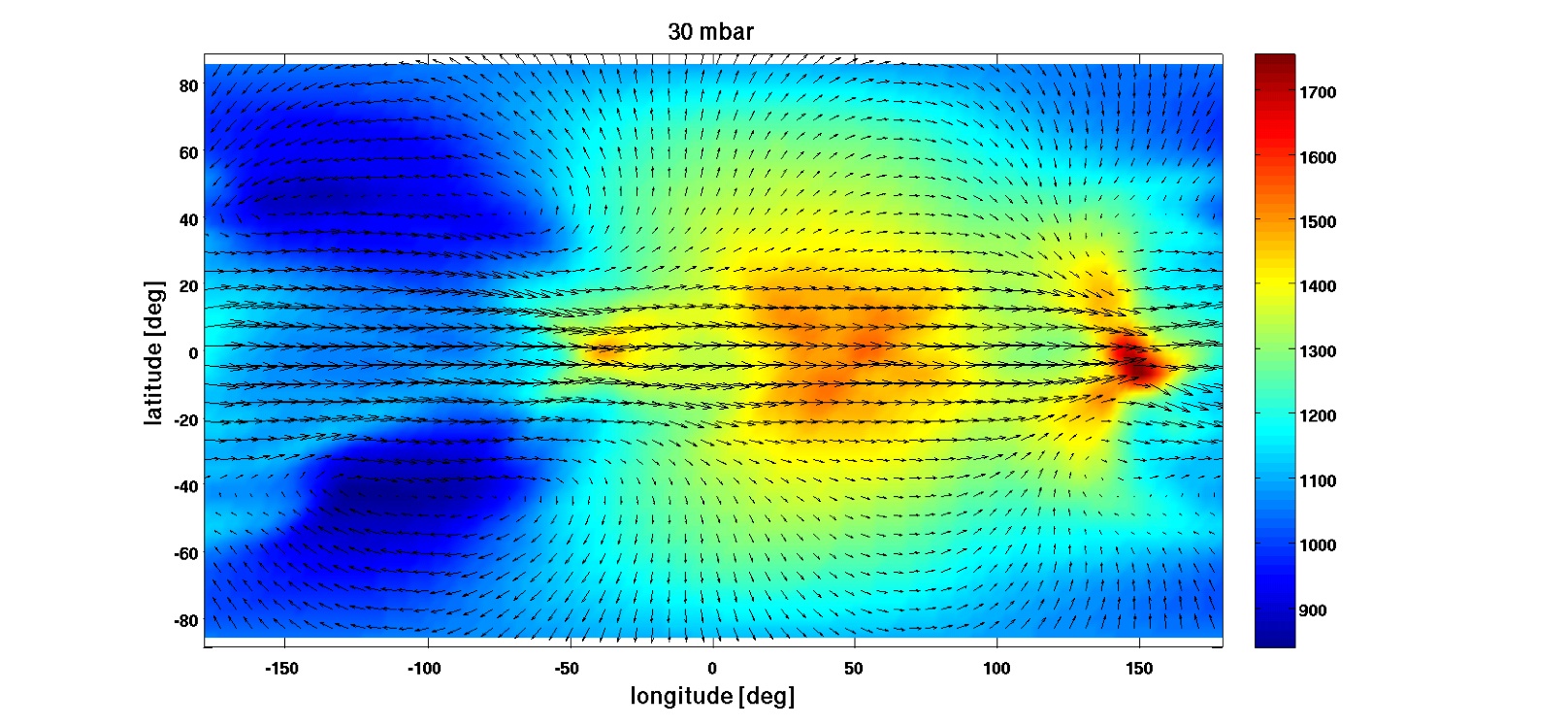}

 \caption{\small 3D simulation of HD 209458b, including realistic
cloud-free, non-gray radiative transfer, illustrating development 
of a dayside stratosphere when opacity from visible-absorbing
species (in this case TiO and VO) are included.  Such stratospheres
are relevant to explaining {\it Spitzer} data of HD 209458b,
TrES-2, TrES-4, XO-1b, and other hot Jupiters. Here, the 
stratosphere (shown in red) begins at pressures
of $\sim10\,$mbar and widens with altitude until it covers most
of the dayside at pressures of $\sim0.1\,$mbar. Panels show temperature 
(colorscale, in K) and winds (arrows) at the 0.08, 1, and 30 mbar levels 
from top to bottom, respectively.  Substellar longitude and latitude
are ($0^{\circ}$, $0^{\circ}$). From \citet{showman-etal-2009}. }
\label{showman-etal-2009-hd209458b}
 \end{figure*}

Despite the diversity in modeling approaches,
the studies described above agree in several key areas. 

\begin{itemize}
\item First, most
of the above studies generally produce peak wind speeds
similar to within a factor of 2--3, in the range of one to several
$\rm km\,sec^{-1}$.\footnote{In intercomparing studies, one must be careful
to distinguish mean versus peak speeds and, in the case of 3D models,
the pressure level at which those speeds are quoted; such quantities
can differ by a factor of several in a single model.}  This similarity
is not a coincidence but results from the force balances that occur
for a global-scale flow in the presence of large fractional
temperature differences.  Consider the longitudinal force
balance at the equator, for example, and suppose the day-night
heating gradient produces a day-night temperature difference
$\Delta T_{\rm horiz}$ that extends vertically over a range of log-pressures
$\Delta \ln p$.  This temperature difference causes a day-night
horizontal pressure-gradient acceleration which, to order-of-magnitude, 
can be written $R\Delta T_{\rm horiz} \Delta\ln p /a$, where $a$ is 
the planetary radius.  At high latitudes, this could
be balanced by the Coriolis force arising from a north-south flow,
but the horizontal Coriolis force is zero at the equator.  At the equator,
such a force instead tends to cause acceleration of the flow in the
east-west direction.  Balancing the pressure-gradient acceleration
by ${\bf v}\cdot\nabla{\bf v}$, which to order-of-magnitude is $U^2/a$
for a global-scale flow, we have
\begin{equation}
U\sim \sqrt{R \Delta T_{\rm horiz} \Delta\ln p}
\label{hot-jupiter-wind-speed1}
\end{equation}
This should be interpreted as the characteristic variation in zonal
wind speed along the equator.  For $R = 3700\rm\,J\,kg^{-1}\,K^{-1}$,
$\Delta T_{\rm horiz}\sim400\rm\,K$, and $\Delta \ln p\sim3$
(appropriate to a temperature difference extending vertically
over three scale heights), this yields $U\sim 2\rm\,km\,sec^{-1}$.

Likewise, consider the latitudinal force balance in the mid-latitudes.
To order-of-magnitude, the latitudinal pressure-gradient acceleration 
is again given by $R\Delta T_{\rm horiz}\Delta\ln p/a$, where here 
$\Delta T_{\rm horiz}$ is the latitudinal temperature contrast 
(e.g., from equator to pole) that extends vertically over $\Delta \ln p$.
If $Ro \gg 1$, this is balanced by the advective acceleration $U^2/a$, 
whereas if $Ro \ll 1$, it would instead be balanced by the Coriolis 
acceleration $f U$, where $f$ is the Coriolis parameter 
(\S\ref{basic-force-balances}).  The former case recovers
Eq.~(\ref{hot-jupiter-wind-speed1}), whereas the latter case
yields
\begin{equation}
U\sim {R \Delta T_{\rm horiz} \Delta\ln p \over f a}
\label{hot-jupiter-wind-speed2}
\end{equation}
Here, $U$ is properly interpreted as the characteristic difference in 
horizontal wind speed vertically across $\Delta\ln p$.  Inserting
$R = 3700\rm\,J\,kg^{-1}\,K^{-1}$,
$\Delta T_{\rm horiz}\sim400\rm\,K$, $\Delta \ln p\sim3$, $f\sim2\times10^{-5}$
(appropriate in mid-latitudes to a hot Jupiter with a 3-day period),
and $a\approx 10^8\rm\,m$, we again obtain $U\sim 2\rm\,km\,sec^{-1}$.
\footnote{The similarity of the numerical estimates from 
Eq.~(\ref{hot-jupiter-wind-speed1}) and (\ref{hot-jupiter-wind-speed2})
results from the fact that we assumed the same horizontal temperature
differences in longitude and latitude and that $Ro\sim1$ for the parameter 
regime explored here.}

While not minimizing the real differences in the numerical results
obtained in the various studies, the above estimates show that the
existence of wind speeds of a few $\rm km\,sec^{-1}$ in the various
numerical studies is a basic outcome of force balance in 
the presence of lateral temperature contrasts of hundreds of K.


\item Second, the various numerical studies all produce a small
number ($\sim1$--4) of broad jets.   This was first pointed out
by \citet{showman-guillot-2002}, \citet{menou-etal-2003}, and 
\citet{cho-etal-2003} and, as described previously, results from 
the fact that the Rossby
deformation radius and Rhines scale are comparable to a planetary
radius for conditions relevant to typical hot Jupiters.  At least
two specific mechanisms seem to be relevant.
When the deformation radius is close to the planetary radius,
the day-night forcing tends to inject energy at horizontal
scales comparable to the planetary radius;
the $\beta$ effect then anisotropizes this
energy into zonal jets whose widths are the order of a planetary radius.
Alternately, if the energy injection occurs primarily
at small horizontal scales, an inverse cascade can reorganize
the energy into planetary-scale jets if the planetary rotation is sufficiently
slow and friction sufficiently weak (allowing fast winds and a large
Rhines scale).  Follow-up studies
are consistent with these general expectations \citep{cooper-showman-2005, 
cho-etal-2008, langton-laughlin-2007, langton-laughlin-2008, 
dobbs-dixon-lin-2008, showman-etal-2008a, showman-etal-2009},
but additional work to clarify the relative roles of $L_D$ and $L_{\beta}$
in setting the jet widths would be beneficial.

\end{itemize}

Despite these similarities, there remain some key differences
in the numerical results obtained by the various groups so far.
These include the following:
\begin{itemize}
\item As yet, little agreement exists on whether hot Jupiters
should exhibit significant temporal variability on the global
scale.  The 2D simulations of \citet{cho-etal-2003, cho-etal-2008}
and \citet{langton-laughlin-2008} produce highly turbulent, time-variable
flows, whereas the shallow-water study of \citet{langton-laughlin-2007} and
most of the 3D studies \citep{showman-guillot-2002, cooper-showman-2005,
dobbs-dixon-lin-2008, showman-etal-2008a, showman-etal-2009} exhibit
relatively steady flow patterns with only modest time variability
for zero-obliquity, zero-eccentricity hot Jupiters.  The Cho et al.
simulations essentially guarantee a time-variable outcome because
of the turbulent initial condition, although breaking Rossby waves and 
dynamical instabilities during the simulations also play a large role;
the \citet{langton-laughlin-2008} simulations were initialized
from rest and naturally
develop turbulence via dynamical instability throughout the course
of the simulation.  

Time variability can result from several mechanisms, including
dynamical (e.g., barotropic or baroclinic) instabilities, large-scale
oscillations, and modification of the jet pattern by atmospheric waves.
If the jet pattern is dynamically stable
(or, more rigorously, if relevant instability growth times are significantly
longer than the characteristic atmospheric heating/cooling times), 
then this mechanism for producing variability would be inhibited;
on the other hand, if the jet pattern is unstable (with
instability growth rates comparable to or less than the heating/cooling times),
the jets will naturally break up and produce a wide range of turbulent
eddies.  In this regard, it is crucial to carefully distinguish between
2D and 3D models, because they exhibit very different stability criteria
\citep[e.g.,][]{dowling-1995a}.
For example, in a 2D, nondivergent model, theoretical and numerical
work shows that jets are guaranteed stable only if
\begin{equation}
{\partial^2 u\over \partial y^2} < \beta
\label{barotropic-stability-criterion}
\end{equation}
a relationship called the {\it barotropic stability criterion}.
Numerical simulations show that 2D, nondivergent fluids initialized
with jets violating Eq.~(\ref{barotropic-stability-criterion}) generally
develop instabilities that rob energy from the jets until the jets
no longer violate the equation.  However, jets in a 3D fluid can in some
cases strongly violate Eq.~(\ref{barotropic-stability-criterion}) 
while remaining stable; Jupiter and Saturn, for example, have several 
jets where $\partial^2 u/\partial y^2$ exceed $\beta$ by a factor of 2--3
\citep{ingersoll-etal-1981, sanchez-lavega-etal-2000}, 
and yet the jet pattern has remained
almost unchanged over multi-decade time scales.  This may simply imply
that the barotropic stability criterion is irrelevant
for a 3D fluid \citep{dowling-1995a, vasavada-showman-2005}. 
Thus, it is {\it a priori}
unsurprising that 2D and 3D models would make different predictions
for time variability. 

However, a recent study by \citet{menou-rauscher-2009} shows that
the issue is not only one of 2D versus 3D.  Making straightforward
extensions to an Earth model, they performed forced 3D shallow simulations 
initialized from rest that developed highly turbulent, 
time variable flows on the global scale, showing that time variability 
from the emergence of horizontal shear instability
may occur in 3D under some conditions relevant to hot Jupiters.   
Like several previous studies,
they forced their flows with a Newtonian heating/cooling scheme that 
generates a day-night temperature difference.  However, 
their forcing set-up differs significantly
from those of previous studies.  They place an impermeable surface at 1 bar
and apply their maximum day/night forcing immediately above the surface;
the day-night radiative-equilibrium temperature difference decreases
with altitude and reaches zero near the top of their model.   This
is the reverse of the set-up used in \citet{cooper-showman-2005,
cooper-showman-2006} and  \citet{showman-etal-2008a, showman-etal-2009} 
where the forcing amplitude peaks near the top of the domain and decreases
with depth, and where the surface is placed significantly deeper than
the region of atmospheric heating/cooling to minimize its interaction with
the circulation in the observable atmosphere. The differences
in the forcing schemes and position of the surface presumably determine
whether or not a given 3D simulation develops global-scale instability
and time variability.  Further work will be necessary 
to identify the specific forcing conditions that lead to time-variable 
or quasi-steady conditions at the global scale.  

It is also worth remembering that 
hot Jupiters themselves exhibit huge diversity; among the known
transiting planets, for example, incident stellar flux varies by
a factor of $\sim150$, gravities range over a factor of $\sim70$,
and expected rotation rates vary over nearly a factor of 10 (e.g.,
Table~\ref{planetary-parameters}).
Given this diversity, it is reasonable to expect that some hot Jupiters
will exhibit time-variable conditions while others will exhibit
relatively steady flow patterns.

\item Another area of difference between models regards the direction 
of the equatorial jet.  The shallow-water and equivalent barotropic
models can produce either eastward or westward equatorial jets, depending
on the initial conditions and other modeling details
\citep{cho-etal-2003, cho-etal-2008,
langton-laughlin-2007, langton-laughlin-2008}.  In contrast, all of the
3D models published to date for synchronously rotating hot Jupiters 
with zero obliquity and eccentricity have produced robust eastward 
equatorial jets
\citep{showman-guillot-2002, cooper-showman-2005, cooper-showman-2006,
showman-etal-2008a, showman-etal-2009, dobbs-dixon-lin-2008, 
menou-rauscher-2009}.  The mechanisms responsible for generating the 
eastward jets in these 3D models remain to be diagnosed in detail but 
presumably involve the global-scale eddies generated by the day-night 
heating contrast.  

\end{itemize}

To date, most work on hot-Jupiter atmospheric circulation has emphasized 
planets on circular orbits.  However, several transiting hot Jupiters and Neptunes
have eccentric orbits, including GJ 436b, HAT-P-2b, HD 17156b, 
and HD 80606b, whose orbital eccentricities are 0.15, 0.5, 0.67,
and 0.93, respectively.  When the eccentricity is large,
the incident stellar flux varies significantly throughout the
orbit---HAT-P-2b, for example, receives $\sim9$ times more
flux at periapse than at apoapse, while for HD 80606b, 
the maximum incident flux exceeds the minimum incident flux by a 
factor of over 800! These variations dwarf those experienced in
our Solar System and constitute an uncharted regime of
dynamical forcing for a planetary atmosphere.  An 8-$\mu$m
light curve of HD 80606b has recently been obtained with the {\it Spitzer
Space Telescope} during a 30-hour interval surrounding periapse passage
\citep{laughlin-etal-2009}, which shows the increase in planetary
flux that presumably accompanies the flash heating as the planet
passes by its star.  2D and 3D numerical simulations suggest that
these planets may exhibit dynamic, time-variable flows and show that 
the planet's emitted infrared flux can peak hours to days after periapse 
passage \citep{langton-laughlin-2008, langton-laughlin-2008b, 
lewis-etal-2009}.

\subsection{Chemistry as a probe of the meteorology}
\label{chemistry}

Chemistry can provide important constraints on the meteorology of 
giant planets.  On Jupiter, for example, gaseous CO, PH$_3$,
GeH$_4$, and AsH$_3$ have been detected in the upper troposphere ($p<5\,$bars)
with mole fractions of $\sim0.8$, 0.2, 0.6, and 0.5--1 ppb, respectively.  
None of these compounds are thermochemically stable in the upper 
troposphere---the chemical equilibrium mole fractions are $<10^{-14}$ 
for AsH$_3$ and $<10^{-20}$ for the other three species 
\citep{fegley-lodders-1994}.  However, the equilibrium abundances of 
all four species rise with depth, exceeding 1 ppb at pressures greater 
than 500, 400, 160, and 20 bars for CO,  PH$_3$, GeH$_4$, and AsH$_3$, 
respectively.  Thus, the high
abundances of these species in the upper troposphere appear
to result from rapid convective mixing from the deep
atmosphere.\footnote{Jupiter also has a high stratospheric CO abundance 
that is inferred to result from exogenous sources \citep{bezard-etal-2002}.}  
Kinetic reaction times are rapid
at depth but plummet with decreasing temperature and pressure,
allowing these species to persist in disequilibrium (i.e., ``quench'')
in the low-pressure and temperature conditions of the upper
troposphere.  Knowledge of the chemical kinetic rate constants allows 
quantitative estimates to be made of the vertical mixing
rate needed to explain the observed abundances.  For all four of
the species listed above, the required vertical mixing rates yield 
order-of-magnitude
matches with the mixing rate expected from thermal convection at
Jupiter's known heat flux \citep[e.g.][]{prinn-barshay-1977, 
fegley-lodders-1994, bezard-etal-2002}.  
Thus, the existence of these disequilibrium species provides
evidence that the deep atmosphere is indeed convective.

Given the difficulty of characterizing the meteorology of exoplanets with
spatially resolved observations, these types of chemical constraints
will likely play an even more important role in our understanding
of exoplanets than is the case for Jupiter.  Several brown dwarfs, including
Gl 229b, Gl 570d, and 2MASS J0415-0935, show evidence for
disequilibrium abundances of CO and/or NH$_3$ \citep[e.g.,][]{noll-etal-1997,
saumon-etal-2000, saumon-etal-2006, saumon-etal-2007}. These
observations can also plausibly 
be explained by vertical mixing.  For these objects, NH$_3$
is quenched in the convection zone, but CO is quenched in the
overlying radiative zone where vertical mixing must result from
some combination of small-scale turbulence (due, for example, to 
the breaking of vertically propagating gravity waves) and
vertical advection due to the large-scale circulation (\S\ref{3d-eddies}).
Current models suggest that such mixing corresponds to an eddy 
diffusivity in the radiative zone of $\sim1$--$10\rm\,m^2\,sec^{-1}$ 
for Gl 229b and 10--$100\rm\,m^2\,sec^{-1}$ for 2MASS J0415-0935 
\citep{saumon-etal-2007}.  These inferences on vertical mixing rate
provide constraints on the atmospheric circulation that would be
difficult to obtain in any other way.

Several hot Jupiters likewise reside in temperature regimes where CH$_4$
is the stable form in the observable atmosphere and CO is the
stable form at depth, or where CH$_4$ is the stable form on
the nightside and CO is the stable form on the dayside.  The abundance
and spatial distribution of CO and CH$_4$ on these objects could
thus provide important clues about the wind speeds and other
aspects of the meteorology.  CH$_4$
has been detected via transmission spectroscopy on HD 189733b
\citep{swain-etal-2008} and CO has been indirectly inferred
from the shape of the dayside emission spectrum \citep{barman-2008},
although further observations and radiative-transfer models are needed
to determine their abundances and spatial variability (if any).
\citet{cooper-showman-2006} coupled a simple model for CO/CH$_4$
interconversion kinetics to their 3D dynamical model of hot Jupiters
to demonstrate that CO and CH$_4$ should be quenched in the observable
atmosphere, where thermochemical CO$\leftrightarrow$CH$_4$
interconversion times are orders of magnitude 
longer than plausible dynamical times.  If so, one would expect CO and
CH$_4$ to have similar abundances on the dayside
and nightside of hot Jupiters, in contrast to the chemical-equilibrium
situation where CH$_4$ would be much more prevalent on the nightside
(note, however, that their models neglect photochemistry,
which could further modify the abundances of both species). 
For temperatures relevant to objects like HD 209458b and HD 189733b, 
\citet{cooper-showman-2006} found that the quenching occurs at pressures
of $\sim1$--$10\,$bars, and that the quenched abundance depends primarily on 
the temperatures and vertical velocities in this pressure range.

Thus, future observational determination of CO and CH$_4$
abundances and comparison with detailed
3D models may allow constraints on the temperatures and vertical velocities
in the deep (1--$10\,$bar) atmospheres of hot Jupiters to be obtained.  
Such constraints probe significantly
deeper than the expected visible and infrared photospheres of hot Jupiters
and hence would nicely complement the characterization of hot-Jupiter thermal
structure via light curves and secondary-eclipse spectra.
Note also that an analogous story involving N$_2$ and NH$_3$ interconversion 
may occur on even cooler objects ($T_{\rm eff}\sim700\,$K ``warm'' Jupiters).

\section{CIRCULATION REGIMES: TERRESTRIAL PLANETS}
\label{terrestrial}

With current groundbased searches and spacecraft
such as NASA's {\it Kepler} mission,
terrestrial planets are likely to be discovered around main-sequence 
stars within the next five years, and basic characterization of their 
atmospheric composition and temperature structure should follow over 
the subsequent decade. At base, we wish to understand not only
whether such planets exist but whether they have atmospheres,
what their atmospheric composition, structure and climate may be, 
and whether they can support life.  
Addressing these issues will require a consideration of relevant
physical and chemical climate feedbacks as well as 
the plausible circulation regimes in these planets' atmospheres.

The study of terrestrial exoplanet atmospheres is just
beginning, and only a handful of papers have specifically investigated
the possible circulation regimes on these objects. 
However, a vast literature has developed to understand
the climate and circulation of Venus, Titan, Mars, and especially 
Earth, and many of the insights developed for understanding these 
planets may be generalizable to exoplanets.  Our goal here is to provide 
conceptual and theoretical guidance on the types of climate and 
circulation processes that exist in terrestrial planet atmospheres
and discuss how those processes vary under diverse planetary conditions.
In \S\ref{climate} we survey basic issues in climate.  The following
sections, \S\ref{regimes}--\S\ref{slowly-rotating}, address basic
circulation regimes on terrestrial planets, emphasizing those
aspects (e.g., processes that determine the horizontal
temperature contrasts) relevant to future exoplanet observations.
\S\ref{unusual-forcing} discusses regimes of exotic forcing
associated with synchronous rotation, large obliquities, and
large orbital eccentricities.

\subsection{Climate}
\label{climate}

Climate can be defined as the mean condition of a planet's atmosphere/ocean
system---the temperatures, pressures, winds, humidities, and cloud
properties---averaged over time intervals longer than the timescale
of typical weather events.  The climate on the terrestrial planets 
results from a wealth of interacting physical, chemical, geological,
and (when relevant) biological effects, 
and even on Earth, understanding the past and present climate has
required a multi-decade interdisciplinary research effort.  In this section,
we provide only a brief sampling of the subject, touching only 
on the most basic physical processes that help to determine a planet's 
global-mean conditions.

The mean temperatures at which a planet radiates to space
depend primarily on the incident stellar flux and the planetary
albedo.  Equating emitted infrared energy (assumed blackbody) with 
absorbed stellar flux yields a planet's global-mean effective
temperature at radiative equilibrium:
\begin{equation}
T_{\rm eff}=\left[{F_* (1-A_B)\over 4\sigma}\right]^{1/4}
\label{teff}
\end{equation}
where $F_*$ is the incident stellar flux, $A_B$ is the planet's
global-mean bond albedo\footnote{The bond albedo is the fraction of 
light incident upon a planet that is scattered back to space, integrated
over all wavelengths and directions.}, 
and $\sigma$ is the Stefan-Boltzmann constant.  
The factor of 4 results from the fact that the planet intercepts a stellar
beam of area $\pi a^2$ but radiates infrared from its
full surface area of $4\pi a^2$.

Given an incident stellar flux, a variety of atmospheric
processes act to determine the planet's surface temperature.
First, the circulation and various atmospheric feedbacks can play 
a key role in determining the planetary albedo.   Bare rock is fairly 
dark, and for terrestrial planets the albedo is largely determined 
by the distribution of clouds and surface ice.  
The Moon, for example, has bond albedo of 0.11, yielding an
effective temperature of $274\,$K.  
In contrast, Venus has a bond albedo of 0.75
because of its global cloud cover, yielding an effective 
temperature of $232\,$K.  With partial cloud and ice cover, Earth 
is an intermediate case, with a bond albedo of 0.31 and an effective 
temperature of $255\,$K.  These examples illustrate that the circulation,
via its effect on clouds and surface ice, can have a major 
influence on mean conditions---Venus' 
effective temperature is less than that of the Earth and Moon
despite receiving nearly double the Solar flux!

Second, $T_{\rm eff}$ is not the surface temperature but the mean 
blackbody temperature at which the planet radiates to space.  The
surface temperatures can significantly exceed $T_{\rm eff}$ through
the {\it greenhouse effect.}  Planets orbiting Sun-like stars receive
most of their energy in the visible, but they tend to radiate their 
energy to space in the infrared.   Most gases tend to be relatively
transparent in the visible, but H$_2$O, CO$_2$, CH$_4$, and
other molecules absorb significantly in the infrared.  On a planet 
like Earth or Venus, a substantial fraction of the
sunlight therefore reaches the planetary surface, but infrared
emission from the surface is mostly absorbed in the atmosphere
and cannot radiate directly to space.   In turn, the 
atmosphere radiates both up (to space) and down (to the surface).
 The surface therefore receives a double whammy of radiation from
both the Sun and the atmosphere; to achieve energy balance, the 
surface temperature becomes elevated relative to the effective temperature.
This is the greenhouse effect.  Contrary to popular descriptions,
the greenhouse effect should {\it not} be thought of as a situation
where the heat is ``trapped'' or ``cannot escape.''   Indeed,
the planet as a whole resides in a near-balance where infrared
emission to space (mostly from the atmosphere when the greenhouse effect
is strong) almost equals absorbed sunlight.

To have a significant greenhouse effect, a planet must have a
massive-enough atmosphere to experience pressure broadening of
the spectral lines.  Mars, for example, has $\sim$~15 times more 
CO$_2$ per area than Earth, yet its greenhouse effect is significantly
weaker because its surface pressure is only 6 mbar.

The climate is influenced by numerous feedbacks that can
affect the mean state and how the atmosphere responds to
perburbations.  Some of the more important feedbacks are
as follows:

\begin{itemize}

\item {\it Thermal feedback:}  Increases or decreases in the temperature
at the infrared photosphere lead to enhanced or reduced radiation to
space, respectively.  This is a negative feeback that allows planets
to reach a stable equilibrium with absorbed starlight.

\item {\it Ice-albedo feedback:}  Surface ice can form on planets exhibiting
trace gases that can condense to solid form.  Because of the brightness
of ice and snow, an increase in snow/ice coverage decreases the absorbed
starlight, promoting colder conditions and growth of even more ice.  
Conversely, melting of surface ice increases the absorbed starlight,
promoting warmer conditions and continued melting.  This is a 
positive feedback.

Simple models of the ice-albedo feedback show that, over a range
of $F_*$ values (depending on the strength of the greenhouse effect), 
an Earth-like planet can exhibit two stable equilibrium states for a 
given value of $F_*$: a warm, ice-poor state with a low albedo 
and a cold, ice-covered state with a high albedo \citep[e.g.,][]
{north-etal-1981}. Geologic evidence suggests that $\sim0.6$--2.4 Gyr
ago Earth experienced several multi-Myr-long glaciations 
with global or near-global ice cover \citep[dubbed ``snowball Earth'' 
events;][]{hoffman-schrag-2002}, suggesting that Earth has flipped 
back and forth between these equilibria.  
The susceptibility of a planet to entering such a snowball
state is highly sensitive to the strength of latitudinal heat
transport \citep{spiegel-etal-2008}, thereby linking this feedback
to the global circulation.

\item {\it Condensable-greenhouse-gas feedback:} 
When an atmospheric constituent is a greenhouse gas that also exists 
in condensed form on the surface, the atmosphere can
experience a positive feedback that affects the temperature and
the distribution of this constituent between the atmosphere and surface.  
An increase in surface temperature increases the constituent's
saturation vapor pressure, hence the atmospheric abundance 
and therefore the greenhouse effect.  A decrease in surface 
temperature decreases the saturation vapor pressure,
reducing the atmospheric abundance and therefore
the greenhouse effect.  Thermal perturbations are therefore amplified.
For Earth, this process occurs with water vapor (Earth's
most important greenhouse gas) and is called the ``water-vapor feedback''
in the Earth climate literature \citep{held-soden-2000}.  
The water-vapor feedback will play a major role in 
determining how Earth responds to anthropogenic increases in 
carbon dioxide over the next century.

In some circumstances, this positive feedback is so strong that
it can trigger a runaway that shifts the atmosphere into a drastically
different state.  In the case of warming, this would constitute a
{\it runaway greenhouse} that leads to the complete evaporation/sublimation 
of the condensable constituent from the surface.  
If early Venus had oceans, for example, they might have experienced
runaway evaporation, leading to a monstrous early water vapor
atmosphere \citep{ingersoll-1969, kasting-1988} that would have had
major effects on subsequent planetary evolution.  

In the case of cooling, such a runaway would remove most of the
condensable constituent from the atmosphere, and if the relevant
gas dominates the atmosphere, this could lead to 
{\it atmospheric collapse}.  For example, if a Venus-like
planet (with its $\sim500$-K greenhouse effect) were moved sufficiently
far from the Sun, CO$_2$ condensation would initiate, potentially
collapsing the atmosphere to a Mars-like state with most of the 
CO$_2$ condensed on the ground, a cold, thin CO$_2$ atmosphere 
in vapor-pressure
equilibrium with the surface ice, and minimal greenhouse effect.
The CO$_2$ cloud formation that precedes such a collapse 
is often used to define the outer edge of
the classical habitable zone for terrestrial planets whose greenhouse
effect comes primarily from CO$_2$ \citep{kasting-etal-1993}.

Because CO$_2$ condensation naturally initiates in the coldest
regions (the poles for a rapidly rotating planet; the nightside
for a synchronous rotator), the conditions under which atmospheric collapse
initiates depend on the atmospheric circulation.  Weak equator-to-pole 
(or day-night) heat transport leads to colder polar (or nightside)
temperatures for a given solar flux, promoting atmospheric collapse.  
These dynamical effects have yet to be fully included in climate models.

\item {\it Clouds:} Cloud coverage increases the albedo,
lessening the absorption of starlight and promoting cooler conditions.
On the other hand, because the tropospheric temperatures generally
decrease with altitude, cloud tops are typically cooler than the ground and
therefore radiate less infrared energy to space, promoting warmer
conditions (an effect that depends sensitively on cloud altitude and 
latitude).  These effects compete with each other.  By determining
the detailed properties of clouds (e.g., fractional coverage, 
latitudes, altitudes, and size-particle distributions), the atmospheric
circulation therefore plays a major role in determining the mean
surface temperature.

To the extent that cloud properties 
depend on atmospheric temperature, clouds can act as a feedback that
affects the mean state and amplifies or reduces a thermal perturbation 
to the climate system.  However, because the sensitivity of cloud
properties to the global circulation and mean climate is extremely
difficult to predict, the net sign of this feedback (positive
or negative) remains unknown even for Earth. For
exoplanets, clouds could plausibly act as a positive feedback in
some cases and negative feedback in others. 

\item {\it Long-term atmosphere/geology feedbacks:} On geological
timescales, terrestrial planets can experience significant exchange
of material between the interior and atmosphere. An example
that may be particularly relevant for Earth-like (i.e. ocean and 
continent-bearing) exoplanets is the carbonate-silicate cycle,
which can potentially buffer atmospheric CO$_2$ in a temperature-dependent
way that tends to stabilize the atmospheric temperature against
variations in solar flux \citep[see, e.g.][]{kasting-catling-2003}.
Such feedbacks, while beyond the scope of this chapter, may
be critical in determining the mean composition and temperature
of terrestrial exoplanets.
\end{itemize}

In addition to the major feedbacks outlined above, there exist
dozens of additional interacting physical, chemical, and biological
feedbacks that can influence the mean climate and its sensitivity to
perturbations.  \citet{hartmann-etal-2003} provide a 
thorough assessment for the Earth's current climate; although 
the details will be different for other planets, this gives a flavor
for the complexity that can be expected.

\subsection{Global circulation regimes}
\label{regimes}

For atmospheres with radiative time constants greatly exceeding
the planet's solar day, the atmosphere cannot respond rapidly
to day-night variations in stellar heating and instead
responds primarily to the daily-mean insolation, which
is a function of latitude.  This situation applies to
most planets in the Solar System, including the (lower) atmospheres
of Venus, Earth, Titan, Jupiter, Saturn, Uranus, and Neptune.\footnote{Mars
is a transitional case, with a radiative time constant $\sim2$--3
times its 24.6-hour day.}  The resulting patterns of  
temperature and winds vary little with longitude
in comparision to their variation in latitude and height.  In
such an atmosphere, the primary task of the atmospheric circulation
is to transport thermal energy not from day to night but between the 
equator and the poles.\footnote{At low obliquity ($<54^{\circ}$), yearly
averaged starlight is absorbed primarily at low latitudes, so the 
circulation transports thermal energy poleward, but at high obliquity 
($>54^{\circ}$), yearly averaged starlight is absorbed primarily at the 
poles \citep{ward-1974}, and the yearly averaged energy transport
by the circulation is equatorward.  At high obliquity, strong seasonal
cycles will also occur.}  On the other hand, when the atmosphere's
radiative time constant is much shorter than the solar day, the
day-night heating gradient will be paramount and day-night temperature
variations at the equator could potentially rival temperature variations
between the equator and poles. 

In the next several subsections 
we survey our understanding of global atmospheric circulations
for these two regimes.  Because the first situation applies to most 
Solar-System atmospheres, it has received the vast majority of work
and remains better understood.  This regime, which we discuss in 
\S\S\ref{hadley}--\ref{slowly-rotating}, will apply to planets
whose orbits are sufficiently far from their stars that the planets
have not despun into a synchronously rotating state; it can
also apply to atmosphere-bearing moons of hot Jupiters even at
small distances from their stars.  The latter regime, surveyed
in \S\ref{unusual-forcing}, prevails
for planets exhibiting small enough orbital eccentricities and semi-major
axes to become synchronously locked to their stars.  This situation
probably applies to most currently known transiting giant exoplanets
and may also apply to terrestrial planets in the habitable zones of
M dwarfs, which are the subject of current observational searches.
Despite its current relevance, this novel forcing regime has come under
investigation only in the past decade and remains incompletely understood 
due to a lack of Solar-System analogs.

\subsection{Axisymmetric flows: Hadley cells}
\label{hadley}

Perhaps the simplest possible
idealization of a  circulation that transports heat from equator
to poles is an axisymmetric circulation---that is, 
a circulation that is independent of longitude---where
hot air rises at the equator, moves poleward aloft, cools, sinks at high
latitudes, and returns equatorward at depth (near the surface
on a terrestrial planet).  Such a circulation is termed a {\it Hadley
cell}, and was first envisioned by Hadley in 1735 to explain 
Earth's trade winds.  Most planetary atmospheres 
in our Solar System, including those of Venus, Earth, Mars, Titan, and 
possibly the giant planets, exhibit Hadley circulations.  

Hadley circulations
on real planets are of course not truly axisymmetric; on the terrestrial 
planets, longitudinal variations in topography and thermal properties (e.g.,
associated with continent-ocean contrasts) induce 
asymmetry in longitude.  Nevertheless, the fundamental idea is
that the longitudinal variations are not {\it crucial} for
driving the circulation.  This differs from the circulation in
midlatitudes, whose longitudinally averaged properties are fundamentally
controlled by the existence of non-axisymmetric baroclinic eddies that
are inherently three-dimensional (see \S\ref{3d-eddies}).

Planetary rotation generally prevents Hadley circulations from
extending all the way to the poles.  Because of planetary rotation,
equatorial air contains considerable angular momentum about the
planetary rotation axis; to conserve angular momentum, 
equatorial air would accelerate to unrealistically high speeds 
as it approached the pole, a phenomenon which is dynamically inhibited.
To illustrate, the specific angular momentum about the rotation axis
on a spherical planet is $M=(\Omega a \cos\phi
+ u)a \cos\phi$, where the first and second terms represent
angular momentum due to planetary rotation and winds, respectively
(recall that $\Omega$, $a$, and $\phi$ are planetary rotation rate,
planetary radius, and latitude).
If $u=0$ at the equator, then $M=\Omega a^2$, and an angular-momentum
conserving circulation would then exhibit winds of
\begin{equation}
u=\Omega a {\sin^2\phi\over \cos\phi}  
\label{u_M}
\end{equation}
Given Earth's radius and rotation rate, 
this equation implies zonal-wind speeds of $134\rm\,m\,sec^{-1}$ at 
$30^{\circ}$ latitude, $700\rm\,m\,sec^{-1}$ at $60^{\circ}$ latitude, and
$2.7\rm\,km\,sec^{-1}$ at $80^{\circ}$ latitude.  Such high-latitude 
wind speeds are unrealistically high and would furthermore be violently 
unstable to 3D instabilities.  On Earth, the actual Hadley circulations 
extend to $\sim30^{\circ}$ latitude.

The Hadley circulation exerts strong control over the wind structure,
latitudinal temperature contrast, and climate.  Hadley circulations
transport thermal energy by the most efficient means possible, namely
straightforward advection of air from one latitude to another.
As a result, the latitudinal temperature
contrast across a Hadley circulation tends to be modest;
the equator-to-pole temperature contrast on a planet will therefore
depend strongly on the width of the Hadley cell.  Moreover, on planets
with condensable gases, Hadley cells exert control over 
the patterns of cloudiness and rainfall.  On Earth, the rising branch 
of the Hadley circulation leads to cloud formation and abundant rainfall 
near the equator, helping to explain for example the prevalence of 
tropical rainforests in Southeast Asia/Indonesia, Brazil, and
central Africa.\footnote{Regional circulations, such as monsoons, also
contribute.}  On the other hand, because condensation and rainout
dehydrates the rising air, the descending branch of the Hadley cell is
relatively dry, which explains the abundances of arid climates on Earth
at 20--$30^{\circ}$ latitude, including the deserts of the
African Sahara, South Africa, Australia, central Asia, and the
southwestern United States. The Hadley cell can also influence
the mean cloudiness, hence albedo and thereby the mean surface 
temperature.  Venus's slow rotation rate leads to a global equator-to-pole
Hadley cell, with a broad ascending branch in low latitudes.
Coupled with the presence of trace condensable gases, this widespread
ascent contributes to a near-global cloud layer that helps generate Venus' high
bond albedo of 0.75.  Different Hadley cell patterns would presumably
cause different cloudiness patterns, different albedos, and therefore
different global-mean surface temperatures.  On exoplanets, Hadley circulations
will also likewise help control latitudinal heat fluxes, equator-to-pole
temperature contrasts, and climate.

\begin{figure*}
 \epsscale{1.0}
\plotone{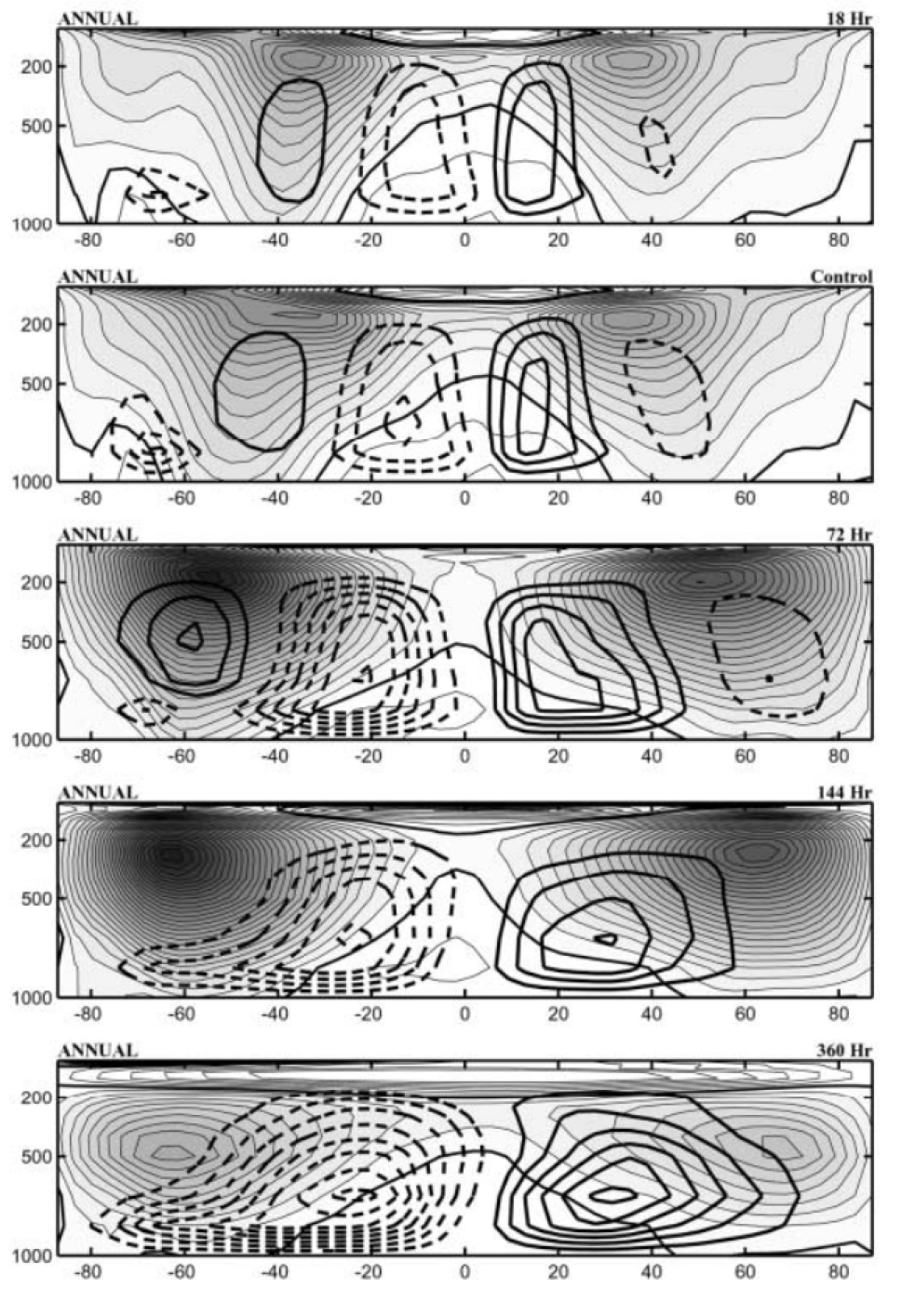}
 \caption{\small {\bf Left:} Zonal-mean circulation versus latitude 
in degrees (abscissa)
and pressure in mbar (ordinate) in a series of Earth-based GCM experiments
from \citet{navarra-boccaletti-2002} 
where the rotation period is varied from 18 hours (top) to 360 hours 
(bottom).  Greyscale and thin grey contours depict zonal-mean zonal wind,
$\overline{u}$,
and thick black contours denote streamfunction of the circulation
in the latitude-height plane (the so-called ``meridional circulation''), 
with solid being clockwise and dashed being counterclockwise.  
The meridional circulation flows parallel to the streamfunction
contours, with greater mass flux when contours are more closely spaced.
The two cells closest to the equator
correspond to the Hadley cell.  As rotation period increases,
the jets move poleward and the Hadley cell widens, becoming
nearly global at the longest rotation periods.}
\label{navarra-boccaletti}
 \end{figure*}

A variety of studies have been carried out using fully nonlinear,
global 3D numerical circulation models to determine the sensitivity
of the Hadley cell to the planetary rotation rate and other
parameters \citep[e.g.][]{hunt-1979, williams-holloway-1982,
williams-1988a, williams-1988b, delgenio-suozzo-1987,
navarra-boccaletti-2002, walker-schneider-2005, walker-schneider-2006}.
These studies show that as the rotation rate is decreased 
the width of the Hadley cell increases, the equator-to-pole 
heat flux increases, and the equator-to-pole temperature contrast decreases.
Figs.~\ref{navarra-boccaletti}--\ref{delgenio-suozzo-fig3} illustrates examples
from \citet{navarra-boccaletti-2002} and \citet{delgenio-suozzo-1987}. 
For Earth parameters, the circulation exhibits mid-latitude eastward 
jet streams that peak in the upper troposphere ($\sim 200\rm\,mbar$ pressure), 
with weaker wind at the equator (Fig.~\ref{navarra-boccaletti}).  
The Hadley cells extend from the equator to the 
equatorward flanks of the mid-latitude jets.  As the rotation rate decreases,
the Hadley cells widen and the jets shift poleward.  At first, the jet 
speeds increase with decreasing rotation rate, which results from the 
fact that as the Hadley cells extend poleward (i.e., closer to the rotation 
axis) the air can spin-up faster (cf Eq.~\ref{u_M}).  Eventually, once the 
Hadley cells extend almost to the pole (at rotation periods exceeding 
$\sim5$--10 days for Earth radius, gravity, and vertical thermal structure), 
further decreases in rotation rate reduce the mid-latitude jet speed.

Perhaps more interestingly for exoplanet observations, these
changes in the Hadley cell significantly influence the planetary temperature 
structure.  This is illustrated in Fig.~\ref{delgenio-suozzo-fig3} from 
a series of simulations by \citet{delgenio-suozzo-1987}.  Because the 
Hadley cells transport heat extremely efficiently, the temperature 
remains fairly constant across the width of the Hadley cells.  Poleward 
of the Hadley cells, however, the heat is transported in latitude by 
baroclinic instabilities (\S\S\ref{3d-eddies} and \ref{baroclinic-zone}), 
which are less efficient, so a large latitudinal temperature 
gradient exists within this so-called ``baroclinic zone.'' 
The equator-to-pole temperature contrast depends strongly on the width of the 
Hadley cell.

\begin{figure*}
 \epsscale{1.0}
\plotone{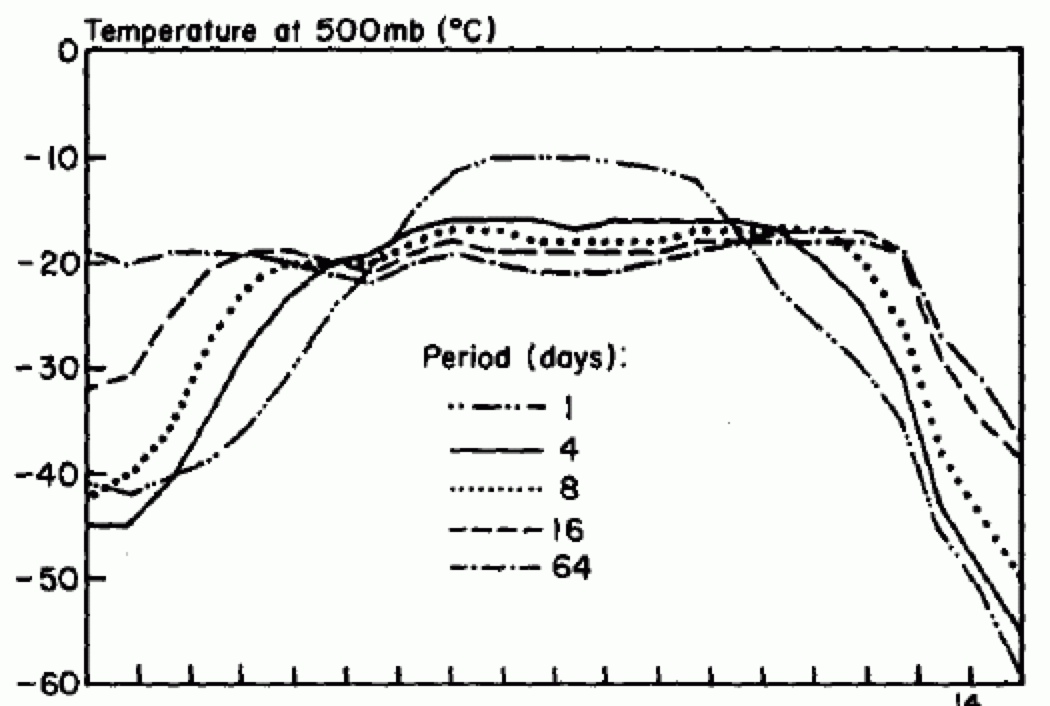}
 \caption{\small Zonally averaged temperature in $^{\circ}$C 
versus latitude at 500 mbar 
(in the mid-troposphere) for a sequence of Earth-like GCM runs that
vary the planetary rotation period between 1 and 64 Earth days (labelled
in the graph). Latitude runs from $90^{\circ}$ on the left
to $-90^{\circ}$ on the right, with the equator at the center of the 
horizontal axis.  The flat region near the equator in each run results
from the Hadley cell, which transports thermal energy extremely
efficiently and leads to a nearly isothermal equatorial temperature 
structure.  The region of steep temperature gradients at high latitudes
is the baroclinic zone, where the temperature structure and latitudinal
heat transport are controlled by eddies resulting from baroclinic
instability.  Note that the width of the Hadley cell increases, and the 
equator-to-pole temperature difference decreases, as the planetary rotation 
period is increased.  From \citet{delgenio-suozzo-1987}.
}
\label{delgenio-suozzo-fig3}
 \end{figure*}

Despite the value of the 3D circulation models described above, 
the complexity of these models tends to obscure the physical mechanisms 
governing the Hadley circulation's strength and latitudinal extent and 
cannot easily be extrapolated to different planetary parameters.  
A conceptual theory for the Hadley cell, due to \citet{held-hou-1980}, provides
considerable insight into Hadley cell dynamics and allows estimates
of how, for example, the width of the Hadley cell should scale
with planetary size and rotation rate
[see reviews in \citet[][pp. 80-92]{james-1994}, 
\citet[][pp.~457-466]{vallis-2006}, and \citet{schneider-2006}].
Stripped to its basics, the scheme
envisions an axisymmetric two-layer model, where the lower layer
represents the equatorward flow near the surface and the upper layer
represents the poleward flow in the upper troposphere. For simplicity,
\citet{held-hou-1980} adopted a basic-state density that is constant
with altitude.   Absorption
of sunlight and loss of heat to space generate a latitudinal temperature
contrast that drives the circulation; for concreteness, let us parameterize
the radiation as a relaxation toward a radiative-equilibrium potential
temperature profile that varies with latitude as $\theta_{\rm rad}=\theta_0 - 
\Delta\theta_{\rm rad}\sin^2\phi$, where $\theta_0$ is the 
radiative-equilibrium potential temperature at the equator and 
$\Delta\theta_{\rm rad}$ is the equator-to-pole difference in 
radiative-equilibrium potential temperature.  If
we make the small-angle approximation for simplicity (valid for 
a Hadley cell that is confined to low latitudes), we can express
this as $\theta_{\rm rad}=\theta_0 - \Delta\theta_{\rm rad}\phi^2$.  

In the lower layer, we assume that friction against the ground keeps the
wind speeds low; in the upper layer, assumed to occur at an 
altitude $H$, the flow conserves 
angular momentum.  The upper layer flow is then specified by
Eq.~(\ref{u_M}), which is just $u=\Omega a \phi^2$ in the small-angle
limit.  We expect that the upper-layer wind will be in 
thermal-wind balance with the latitudinal temperature
contrast\footnote{This form differs slightly from Eq.~(\ref{thermal-wind})
because Eq.~(\ref{hh-thermal-wind}) adopts a constant basic-state density
(the so-called ``Boussinesq'' approximation) whereas Eq.~(\ref{thermal-wind}) 
adopts the compressible ideal-gas equation of state.}:
\begin{equation}
f{\partial u\over\partial z}= f{u\over H}= -{g\over\theta_0}
{\partial\theta\over\partial y}
\label{hh-thermal-wind}
\end{equation}
where $\partial u/\partial z$ is simply given by $u/H$ in this 
two-layer model.  Inserting $u=\Omega a \phi^2$ into 
Eq.~(\ref{hh-thermal-wind}), approximating the Coriolis parameter as
$f=\beta y$ (where $\beta$ is treated as constant), and integrating, 
we obtain a temperature that varies with latitude as
\begin{equation}
\theta=\theta_{\rm equator} - {\Omega^2\theta_0\over 2 g a^2 H}y^4
\label{hh-theta}
\end{equation}
where $\theta_{\rm equator}$ is a constant to be determined.

At this point, we introduce two constraints.  First, \citet{held-hou-1980} 
assumed the circulation is energetically closed, i.e. that no net exchange
of mass or thermal energy occurs between the Hadley cell and higher latitude 
circulations.  Given an energy equation with radiation parameterized
using Newtonian cooling, $d\theta/dt=(\theta_{\rm rad} - \theta)/
\tau_{\rm rad}$, where $\tau_{\rm rad}$ is a radiative time constant, 
the assumption that the circulation is steady and closed requires that
\begin{equation}
\int_0^{\phi_H} \theta\, dy = \int_0^{\phi_H} \theta_{\rm rad} dy
\label{hh-energy}
\end{equation}
where we are integrating from the equator to the poleward edge
of the Hadley cell, at latitude $\phi_H$.  Second, temperature
must be continuous with latitude at the poleward edge of the Hadley
cell.  In the axisymmetric model, baroclinic instabilities are suppressed, 
and the regions poleward of the Hadley cells reside in a state of radiative 
equilibrium.  Thus, $\theta$ must equal $\theta_{\rm rad}$ at the poleward 
edge of the cell.  Inserting our expressions for $\theta$ and 
$\theta_{\rm rad}$ into these two constraints yields a system of two equations 
for $\phi_H$ and $\theta_{\rm equator}$.  The solution yields a
Hadley cell with a latitudinal half-width of
\begin{equation}
\phi_H = \left({5\Delta\theta_{\rm rad} g H\over 
3\Omega^2 a^2 \theta_0}\right)^{1/2}
\label{hh-hadley-width}
\end{equation}
in radians. This solution suggests that the width of the Hadley cell scales
as the square root of the fractional equator-to-pole radiative-equilibrium
temperature difference, the square root of the gravity, the
square root of the height of the cell, and inversely with
the rotation rate.  Inserting Earth annual-mean values 
($\Delta \theta_{\rm rad} \approx 
70\,\rm K$, $\theta=260\,$K, $g=9.8\rm\,m\,sec^{-2}$, $H=15\,$km, 
$a=6400\,$km, and $\Omega =7.2\times10^{-5}\rm\,sec^{-1}$) 
yields $\sim30^{\circ}$.

Redoing this analysis without the small-angle approximation leads to
a transcendental equation for $\phi_H$ \citep[see Eq.~(17) in][]
{held-hou-1980}, which can be solved numerically.
Figure~\ref{delgenio-suozzo-fig2} ({\it solid curve}) illustrates 
the solution.  As expected, $\phi_H$ ranges from $0^{\circ}$ as 
$\Omega\to\infty$ to 
$90^{\circ}$ as $\Omega\to0$, and, for planets of Earth radius
with Hadley circulations $\sim10\,\rm km$ tall, bridges these extremes
between rotation periods of $\sim0.5$ and 20 days.  Although deviations
exist, the agreement between the simple Held-Hou model and the 3D GCM 
simulations are surprisingly good given the simplicity of the Held-Hou
model.

\begin{figure*}
 \epsscale{1.0}
\plotone{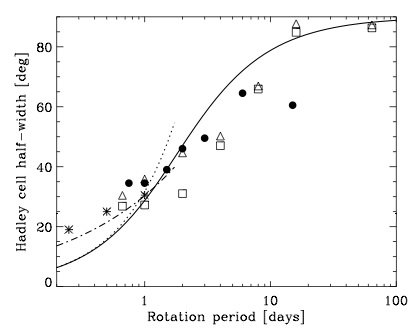}
 \caption{\small Latitudinal width of the Hadley cell from a sequence
of Earth-like GCM runs that vary the planetary rotation period
(symbols) and comparison to the \citet{held-hou-1980} theory (solid curve),
the small-angle approximation to the Held-Hou theory from
Eq.~(\ref{hh-hadley-width}) (dotted curve), and the \citet{held-2000}
theory from Eq.~(\ref{eddies-hadley-width}) (dashed-dotted curve).
GCM results are from \citet{delgenio-suozzo-1987} (squares
and triangles, depicting northern and southern hemispheres, respectively),
\citet{navarra-boccaletti-2002} (filled circles), and 
\citet{korty-schneider-2008} (asterisks; only cases adopting 
$\theta_{\rm rad} \approx 70\,$K and their parameter $\Gamma=0.7$ are shown).
Parameters adopted for the curves are $g=9.8\rm\,m\,sec^{-2}$, $H=15\rm\,km$,
$\Delta\theta_{\rm rad}=70\rm\,K$, $a=6400\rm\,km$, 
$\theta_0=260\rm\,K$, and $\Delta\theta_v=30\rm\,K$.  In the GCM
studies, different authors define the width of the Hadley cell in 
different ways, so some degree of scatter is inevitable.
}
\label{delgenio-suozzo-fig2}
 \end{figure*}

Importantly, the Held-Hou model demonstrates 
that latitudinal confinement
of the Hadley cell occurs even in an axisymmetric atmosphere.
Thus, the cell's latitudinal confinement does not require
(for example) three-dimensional baroclinic or barotropic 
instabilities associated with the jet at the poleward branch 
of the cell.  Instead, the confinement results from energetics:
the twin constraints of angular-momentum conservation in the 
upper branch and 
thermal-wind balance specify the latitudinal temperature profile
in the Hadley circulation (Eq.~\ref{hh-theta}).  This generates
equatorial temperatures colder than (and subtropical temperatures
warmer than) the radiative equilibrium, implying radiative heating
at the equator and cooling in the subtropics.  This properly allows
the circulation to transport thermal energy poleward.  If the cell
extended globally on a rapidly rotating planet, however, the circulation
would additionally produce high-latitude temperatures {\it colder} 
than the radiative equilibrium temperature, which in steady state would 
require radiative {\it heating} at high latitudes.   This is thermodynamically 
impossible  given the specified latitudinal dependence of $\theta_{\rm rad}$.  
The highest latitude to which the cell can extend without 
encountering this problem is simply given by Eq.~(\ref{hh-energy}).

The model can be generalized to consider a more realistic
treatment of radiation than the simplified Newtonian cooling/heating
scheme employed by \citet{held-hou-1980}.  \citet{caballero-etal-2008}
reworked the scheme using a two-stream, non-grey representation of
the radiative transfer with parameters appropriate for Earth and Mars.  
This leads to a prediction for the width of the Hadley cell that differs 
from Eq.~(\ref{hh-hadley-width}) by a numerical constant of order unity.

Although the prediction of Held-Hou-type models for $\phi_H$
provides important insight, several failures of these models exist.  
First,  the model underpredicts
the strength of the Earth's Hadley cell (e.g., as characterized
by the magnitude of the north-south wind) by about an order of magnitude.
This seems to result from the lack of turbulent eddies in axisymmetric
models; several studies have shown that turbulent three-dimensional
eddies exert stresses that act to strengthen the Hadley cells
beyond the predictions of axisymmetric models
\citep[e.g.][]{kim-lee-2001b, walker-schneider-2005, walker-schneider-2006,
schneider-2006}.\footnote{\citet{held-hou-1980}'s original model 
neglected the seasonal cycle, and it has been suggested that generalization of
the Held-Hou axisymmetric model to include seasonal effects could
alleviate this failing \citep{lindzen-hou-1988}.  Although this 
improves the agreement with Earth's observed {\it annual-mean} Hadley-cell
strength, 
it predicts large solstice/equinox oscillations in Hadley-cell strength
that are lacking in the observed Hadley circulation \citep{dima-wallace-2003}.}
Second, the Hadley cells on Earth and probably Mars 
are not energetically closed; rather, mid-latitude baroclinic eddies 
transport thermal energy out of the Hadley cell into the polar regions.  
Third, the poleward-moving upper tropospheric branches of the Hadley cells 
do not conserve angular momentum---although the zonal wind does become
eastward as one moves poleward across the cell, for Earth this increase 
is a factor of $\sim2$--3 less than predicted by Eq.~(\ref{u_M}).
Overcoming these failings requires the inclusion of three-dimensional 
eddies.

Several studies have shown that turbulent
eddies in the mid- to high-latitudes---which are neglected in
the Held-Hou and other axisymmetric models---can affect the width
of the Hadley circulation and alter the parameter 
dependences suggested by Eq.~(\ref{hh-hadley-width}) \citep[e.g.,][]
{delgenio-suozzo-1987, walker-schneider-2005, walker-schneider-2006}.
Turbulence can produce an acceleration/deceleration of the zonal-mean
zonal wind, which breaks the angular-momentum conservation constraint
in the upper-level wind, causing $u$ to deviate from Eq.~(\ref{u_M}).  
With a different $u(\phi)$ profile, the latitudinal dependence
of temperature will change (via Eq.~\ref{hh-thermal-wind}), 
and hence so will the latitudinal
extent of the Hadley cell required to satisfy Eq.~(\ref{hh-energy}).
Indeed, within the context of axisymmetric models, the addition
of strong drag into the upper-layer flow (parameterizing turbulent mixing
with the slower-moving surface air, for example) can lead 
Eq.~(\ref{hh-energy}) to predict that the Hadley
cell should extend to the poles even for Earth's rotation rate
\citep[e.g.,][]{farrell-1990}.

It could thus be the case that the width of the Hadley cell is 
strongly controlled by eddies.  For example, in the
midlatitudes of Earth and Mars, baroclinic eddies generally accelerate 
the zonal flow eastward in the upper troposphere; in steady state, 
this is generally counteracted
by a westward Coriolis acceleration, which requires an {\it equatorward}
upper tropospheric flow---backwards from the flow direction in
the Hadley cell.  Such eddy effects can thereby terminate
the Hadley cell, forcing its confinement to low latitudes.  
Based on this idea, \citet{held-2000} suggested that the Hadley cell 
width is determined by the latitude beyond which the troposphere first becomes 
baroclinically unstable (requiring isentrope slopes to exceed a 
latitude-dependent critical value). Adopting the horizontal thermal
gradient implied by the angular-momentum conserving wind (Eq.~\ref{hh-theta}), 
making the small-angle approximation, and utilizing
a common two-layer model of baroclinic
instability, this yields \citep{held-2000}
\begin{equation}
\phi_H \approx \left({g H \Delta\theta_v\over \Omega^2 a^2 \theta_0}
\right)^{1/4}
\label{eddies-hadley-width}
\end{equation}
in radians, where $\Delta \theta_v$ is the vertical difference
in potential temperature from the surface to the top of the Hadley
cell.  Note that the predicted dependence of $\phi_H$
on planetary radius, gravity, rotation
rate, and height of the Hadley cell is weaker than predicted by
the Held-Hou model.  Earth-based GCM simulations suggest that
Eq.~(\ref{eddies-hadley-width}) may provide a better representation
of the parameter dependences \citep{frierson-etal-2007, lu-etal-2007,
korty-schneider-2008}.  Nevertheless, even discrepancies with
Eq.~(\ref{eddies-hadley-width}) are expected since the actual
zonal wind does not follow the angular-momentum conserving profile
(implying that the actual thermal gradient will deviate from 
Eq.~\ref{hh-theta}).
Substantially more work is needed to generalize these ideas
to the full range of conditions relevant for exoplanets.

\subsection{High-latitude circulations: the baroclinic zones}
\label{baroclinic-zone}

For terrestrial planets heated at the equator and cooled at the poles,
several studies suggest that the equator-to-pole heat engine
can reside in either of two regimes depending 
on rotation rate and other parameters \citep{delgenio-suozzo-1987}.
When the rotation period is long, the Hadley cells extend nearly
to the poles, dominate the equator-to-pole heat transport, and 
thereby determine the equator-to-pole temperature gradient.
Baroclinic instabilities are suppressed because of the small latitudinal
thermal gradient and large Rossby deformation radius (at which
baroclinic instabilities have maximal growth rates), which exceeds
the planetary size for slow rotators such as Titan and Venus.
On the other hand, for rapidly rotating planets like Earth and 
Mars, the Hadley cells are confined to low latitudes;  in
the absence of eddies the mid- and high-latitudes would relax into a 
radiative-equilibrium state, leading to minimal equator-to-pole
heat transport and a large equator-to-pole temperature difference.  
This structure is baroclinically unstable, however, and the resulting 
baroclinic eddies provide the dominant mechanism for transporting thermal
energy from the poleward edges of the Hadley cells ($\sim30^{\circ}$
latitude for Earth) to the poles.  This baroclinic heat transport
significantly reduces the equator-to-pole temperature contrast in
the mid- and high-latitudes.  For gravity, planetary radius, and heating rates
relevant for Earth, the breakpoint between these regimes occurs
at a rotation period of $\sim5$--10 days \citep{delgenio-suozzo-1987}.

On rapidly rotating terrestrial planets like Earth and Mars, then, the 
equatorial and high-latitude circulations fundamentally differ: 
the equatorial Hadley cells, while strongly affected by eddies, do not 
{\it require} eddies to exist, nor to transport heat poleward.  
In contrast, the circulation and heat transport at high latitudes 
(poleward of $\sim30^{\circ}$ latitude on Earth and Mars) fundamentally
depends on the existence of eddies. At high latitudes, interactions 
of eddies with the mean flow controls the latitudinal 
temperature gradient, latitudinal heat transport, and structure of 
the jet streams.  This range of latitudes is called the 
{\it baroclinic zone.}

The extent to which baroclinic eddies can reduce the equator-to-pole
temperature gradient has important implications for the mean climate.
Everything else being equal, an Earth-like planet with colder poles
will develop more extensive polar ice and be more susceptible to
an ice-albedo feedback that triggers a globally glaciated ``snowball''
state \citep{spiegel-etal-2008}.  Likewise, a predominantly CO$_2$
atmosphere can become susceptible to atmospheric collapse if the
polar temperatures become sufficiently cold for CO$_2$ condensation.

There is thus a desire to understand the extent to which baroclinic
eddies can transport heat poleward.  General circulation models (GCMs) 
attack this problem by spatially and temporally resolving the
full life of every baroclinic eddy and their effect on the mean
state, but this is computationally intensive and often sheds little
light on the underlying sensitivity of the process to rotation
rate and other parameters.  

Two simplified approaches have been advanced that illuminate this issue.
Although GCMs for terrestrial exoplanets will surely be needed in the future,
simplified approaches can guide our understanding of such 
GCM results and provide testable hypotheses regarding the dependence
of heat transport on planetary parameters.  They also may provide 
guidance for parameterizations of the latitudinal heat transport 
in simpler energy-balance climate models that do not explicitly
attempt to resolve the full dynamics.  We devote the rest of 
this section to discussing these simplified approaches.

The first simplified approach postulates that baroclinic eddies relax
the midlatitude thermal structure into a state that is neutrally
stable to baroclinic instabilities \citep{stone-1978}, a process called
{\it baroclinic adjustment}.  This idea
is analogous to the concept of convective adjustment: when radiation
or other processes drive the vertical temperature gradient steeper
than an adiabat ($dT/dz < -g/c_p$ for an ideal gas, where $z$ is height), 
convection ensues and drives the temperature profile toward the adiabat,
which is the neutrally stable state for convection.  
If the convective overturn timescales are much
shorter than the radiative timescales, then convection overwhelms
the ability of radiation to destabilize the environment, and the temperature
profile then deviates only slightly from an adiabat over a wide range of
convective heat fluxes (see, e.g., \S\ref{egp-thermal-structure}). 
In a similar way, the concept of baroclinic adjustment postulates that
when timescales for baroclinic eddy growth are much shorter than the 
timescale for radiation to create a large equator-to-pole temperature 
contrast, the eddies will transport thermal energy poleward at just the rate 
needed to maintain a profile that is neutral to baroclinic instabilities.

   This idea
was originally developed for a simplified two-layer system, for 
which baroclinic instability first initiates when the slope of
isentropes exceeds $\sim H/a$, where $H$ is a pressure scale height
and $a$ is the planetary radius.  Comparison of this critical
isentrope slope with Earth observations shows impressive agreement 
poleward of 30-$40^{\circ}$ latitude,
where baroclinic instabilities are expected to be active \citep{stone-1978}.
A substantial literature has subsequently developed to explore
the idea further \citep[for a review see][]{zurita-gotor-lindzen-2006}.
The resulting latitudinal temperature gradient is then approximately
$H/a$ times the vertical potential temperature gradient.  Interestingly,
this theory suggests that otherwise identical planets with different 
Brunt-V\"ais\"al\"a frequencies (due to differing opacities, 
vertical heat transport by large-scale eddies, or role for latent heating,
all of which can affect the stratification) would exhibit very different 
midlatitude 
temperature gradients---the planet with smaller stable stratification (i.e.,
smaller Brunt-V\"ais\"al\"a frequency; Eq.~\ref{bvf}) exhibiting
a smaller latitudinal temperature gradient.  Moreover, although Earth's
ocean transports substantial heat between the equator and poles, the
theory also suggests that the ocean may only exert a modest influence on the 
latitudinal temperature gradients in the {\it atmosphere} 
\citep{lindzen-farrell-1980}: in the absence of oceanic heat transport, the
atmospheric eddies would simply take up the slack to maintain the atmosphere
in the baroclinically neutral state. (There could of course be an
indirect effect on latitudinal temperature gradients if removing/adding 
the oceans altered the tropospheric stratification.)  Finally, the 
theory suggests that the latitudinal temperature gradient in the 
baroclinic zone does not depend on planetary rotation rate except indirectly 
via the influence of rotation rate on static stability.  It is worth
emphasizing, however, that even if the latitudinal temperature gradient 
{\it in the baroclinic zone} were constant with rotation rate,
the total equator-to-pole temperature difference would still decrease
with decreasing rotation rate because, with decreasing rotation rate, 
the Hadley cell occupies a greater latitude
range and the baroclinically adjusted region would be compressed toward
the poles (see Fig.\ref{delgenio-suozzo-fig3}). 

Despite \citet{stone-1978}'s encouraging results, several complicating 
factors exist.  First, the time-scale separation between radiation
and dynamics is much less obvious for baroclinic adjustment than
for convective adjustment.  In an Earth or Mars-like context, the convective
instability timescales are $\sim1$ hour, the baroclinic instability timescales
are days, and the radiation timescale is $\sim20\,$days (Earth) or 
$\sim 2\,$days (Mars).  The ability of baroclinic eddies to adjust the 
environment to a baroclinically neutral state may thus be marginal, 
especially for planets with short radiative time constants (such as Mars).  
Second, the neutrally stable state in the two-layer model---corresponding
to an isentrope slope of $\sim H/a$---is an artefact of the vertical
discretization in that model.  When a multi-layer model is used, 
the critical isentrope slope for initiating baroclinic instability decreases 
as the number of layers increases, and it approaches zero in a vertically 
continuous model.  One might then wonder why a baroclinic zone can support
{\it any} latitudinal temperature contrast (as it obviously does on
Earth and Mars).  The reason is that the
instability growth timescales are long at small isentropic slopes
\citep{lindzen-farrell-1980b}; the instabilities only develop a substantial 
ability to affect the mean state when isentrope slopes become steep.
 In practice, then, the baroclinic eddies
are perhaps only able to relax the atmosphere to a state with
isentrope slopes of $\sim H/a$ rather than something significantly
shallower.

The second simplified approach to understanding equator-to-pole
temperature contrasts on terrestrial planets seeks to describe
the poleward heat transport by baroclinic eddies as a diffusive process
\citep{held-1999}.
This approach is based on the idea that baroclinic eddies
have maximal growth rates at scales close to the Rossby radius of 
deformation, $L_D$ (see Eq.~\ref{deformation-radius}), 
and if the baroclinically unstable 
zone has a latitudinal width substantially exceeding $L_D$ then 
there will be scale separation between the mean flow and the 
eddies.\footnote{As pointed out by \citet{held-1999}, this differs from many 
instability problems in fluid mechanics, such as shear instability 
in pipe flow or convection in a fluid driven by a heat flux between 
two plates, where maximal growth rates occur at length scales comparable 
to the domain size and no eddy/mean-flow scale separation occurs.}
The possible existence of a scale separation in the baroclinic instability
problem implies that representing the heat transport as a function
of {\it local} mean-flow quantities (such as the mean latitudinal
temperature gradient) is a reasonable prospect. Still, caution
is warranted, since an inverse energy cascade (\S\ref{rhines}) could
potentially transfer the energy to scales larger than $L_D$, thereby
weakening the scale separation.

The diffusive approach is typically cast in the context of a 
1D energy-balance model that seeks to determine the variation with
latitude of the zonally averaged surface temperature.  This approach has a long
history for Earth climate studies \citep[see review in][]{north-etal-1981},
Solar-System planets \citep[e.g.][]{hoffert-etal-1981}, and
even in preliminary studies of the climates and habitability of
terrestrial exoplanets \citep{williams-kasting-1997,
williams-pollard-2003, spiegel-etal-2008, spiegel-etal-2009}.
In its simplest form, the governing equation reads
\begin{equation}
c{\partial T\over\partial t}=\nabla\cdot (c D\nabla T) +
S(\phi) - I
\label{ebm}
\end{equation}
where $T$ is surface temperature, $D$ ($\m^2\sec^{-1}$) is
the diffusivity associated with heat transport by baroclinic eddies,
$S(\phi)$ is the absorbed stellar flux as a function of latitude,
and $I$ is the emitted thermal flux, which is a function of 
temperature.  In the simplest possible case, $I$ is represented
as a linear function of temperature, $I=A + B T$ \citep{north-etal-1981},
where $A$ and $B$ are positive constants.  In Eq.~(\ref{ebm}), 
$c$ is a heat capacity  (with units $\rm J\,m^{-2}\,K^{-1}$)
that represents the atmosphere and oceans (if any).
The steady-state solution in the absence of transport ($D=0$) is
$T=[S(\phi)-A]/B$.  On the other hand, when transport dominates
($D\to\infty$), the solution yields constant $T$.

The term $S(\phi)$ includes the effect of latitudinally varying 
albedo (e.g., due to ice cover), but if albedo is constant with
latitude, then $S$ represents the latitudinally varying insolation.
To schematically illustrate the effect of heat transport on the 
equator-to-pole temperature gradient, parameterize $S$ as a constant
plus a term proportional to the Legendre polynomial $P_2(\cos\phi)$.
In this case, and if $D$ and $c$ are constant, the equation has a steady 
analytic solution with an equator-to-pole temperature difference 
given by \citep{held-1999}
\begin{equation}
\Delta T_{\rm eq-pole} = {\Delta T_{\rm rad}\over 1 + 6{c D\over B a^2}}
\label{ebm-solution}
\end{equation}
where $a$ is the planetary radius and $\Delta T_{\rm rad}$ is
the equator-to-pole difference in radiative-equilibrium temperature.  
The equation implies that 
atmospheric heat transport significantly influences the
equator-to-pole temperature difference when the diffusivity
exceeds $\sim B a^2/(6 c)$.  For Earth, $B\approx 2\rm\,W\,m^{-2}\,K^{-1}$,
$a=6400\rm\,km$, and $c\approx 10^7\rm\,J\,m^{-2}\,K^{-1}$, suggesting
that atmospheric transport becomes important when $D \gtrsim 10^6\rm\,m^2\,
sec^{-1}$.

The question comes down to how to determine the diffusivity, $D$.
In Earth models, $D$ is typically chosen by tuning the models to
match the current climate, and then that value of $D$ is used
to explore the regimes of other climates \citep[e.g.][]{north-1975}.
However, that approach sidesteps the underlying physics and prevents
an extrapolation to other planetary environments.  

Motivated by an interest in understanding the feedback between baroclinic 
instabilities and the mean state, substantial effort has been devoted to 
determining the dependence of the diffusivity on control parameters such as 
rotation rate and latitudinal thermal gradient \citep[for a review see]
[]{held-1999}. This work can help guide efforts to understand the
temperature distribution on rapidly rotating exoplanets.  We expect
that the diffusivity will scale as 
\begin{equation}
D\approx u_{\rm eddy} L_{\rm eddy}
\label{diffusivity1}
\end{equation}
where $u_{\rm eddy}$ and $L_{\rm eddy}$ are the characteristic velocities
and horizontal sizes of the heat-transporting eddies.
Because baroclinic eddies will be in near-geostrophic balance 
(Eq.~\ref{geostrophy}) on 
a rapidly rotating planet, the characteristic eddy
velocities, $u_{\rm eddy}$, will relate to characteristic eddy 
potential-temperature perturbations $\theta'_{\rm eddy}$ via the
thermal-wind relation, giving
$u_{\rm eddy} \sim g \theta'_{\rm eddy} h_{\rm eddy}/
(f L_{\rm eddy})$, where $h_{\rm eddy}$ is the characteristic
vertical thickness of the eddies.  Under the assumption that the 
ratio of vertical to horizontal scales is $h_{\rm eddy}/L_{\rm eddy}\sim f/N$
\citep{charney-1971, haynes-2005}, this yields $u_{\rm eddy} \approx
g \theta'_{\rm eddy}/(\theta_0 N)$ (where $N$ is Brunt-V\"ais\"al\"a frequency),
which simply states that eddies with larger thermal perturbations will
also have larger velocity perturbations.  Under the assumption that
the thermal perturbations scale as $\theta'_{\rm eddy}\approx L_{\rm
eddy} \partial \overline{\theta}/\partial y$, we then have
\begin{equation}
D\approx L_{\rm eddy}^2 {g \over N\theta_0}{\partial\overline{\theta}
\over \partial y}.
\label{diffusivity2}
\end{equation}
Greater thermal gradients and eddy-length scales lead to greater
diffusivity and, importantly, the diffusivity depends on the {\it square}
of the eddy size.

\begin{deluxetable}{lcrllllcrl}
\tabletypesize{\small}
\tablecaption{Proposed diffusivities for high-latitude heat transport on terrestial planets\label{diffusivities}}
\tablewidth{0pt}
\tablehead  
{Scheme &$L_{\rm eddy}$          &$D$ \\   
        &  &    &}
\startdata
Stone (1972) &$L_D$  &${H^2 N g\over f^2\theta_0}{\partial
\bar\theta\over \partial y}$  \\ \\
Green (1970) &$L_{\rm zone}$ &$L_{\rm zone}^2 {g\over N \theta_0}
{\partial\bar\theta\over\partial y}$   \\ \\
\citet{held-larichev-1996}    &$L_{\rm \beta}$  &${g^3\over N^3 \beta^2
\theta_0^3}\left({\partial\bar\theta\over\partial y}\right)^3$\\ \\

\citet{barry-etal-2002}  &$L_{\beta}$  &$\left({e a \dot q_{\rm net} \over 
\theta_0}{\partial\overline{\theta}\over\partial y}\right)^{3/5} 
\left({2\over \beta}\right)^{4/5}$
\enddata
\end{deluxetable}

Several proposals for the relevant eddy size have been put forward, which
we summarize in Table~\ref{diffusivities}.
Based on the idea that baroclinic instabilities have the greatest growth
rates for lengths comparable to the deformation radius, \citet{stone-1972}
suggested that $L_{\rm eddy}$ is the deformation radius, $NH/f$.
On the other hand, baroclinic eddies could energize an inverse cascade,
potentially causing the dominant heat-transporting eddies to have 
sizes exceeding $L_D$.  In the limit of this process \citep{green-1970}, 
the eddies would reach the width of the baroclinic zone, $L_{\rm zone}$ 
(potentially close to a planetary radius for a planet with a narrow 
Hadley cell). This simply 
leads to Eq.~(\ref{diffusivity2}) with $L_{\rm eddy}=L_{\rm zone}$.
In contrast, \citet{held-larichev-1996} argued that the inverse cascade
would produce an eddy scale not of order $L_{\rm zone}$ but instead
of order the Rhines scale, $(u_{\rm eddy}/\beta)^{1/2}$.  
Finally, \citet{barry-etal-2002} used heat-engine arguments to 
propose that
\begin{equation}
D \approx \left({e a q \over \theta_0}{\partial\overline{\theta}
\over\partial y}\right)^{3/5} \left({2\over \beta}\right)^{4/5}
\label{diffusivity-barry}
\end{equation}
where $\dot q_{\rm net}$ is the net radiative heating/cooling per mass that the
eddy fluxes are balancing, $a$ is the planetary radius, and $e$ is 
a constant of order unity. Equation~\ref{diffusivity-barry} is 
probably most robust for the dependences on heating rate $\dot q_{\rm net}$ 
and thermal gradient $\partial\overline{\theta}
/\partial y$, which they varied by factors of $\sim200$ and 
6, respectively; planetary radius and rotation rates were varied
by only $\sim70\%$, so the dependences on those parameters should
be considered tentative. 

As can be seen in Table~\ref{diffusivities},
these proposals have divergent implications for the dependence of
diffusivity on background parameters.   \citet{stone-1972}'s diffusivity
is proportional to the latitudinal temperature gradient and, because
of the variation of $L_D$ with rotation rate, inversely proportional to
the square of the planetary rotation rate.   \citet{green-1970}'s diffusivity
likewise scales with the latitudinal temperature gradient; it contains
no explicit dependence on the planetary rotation rate, but a 
rotation-rate dependence could enter because the Hadley cell shrinks
and $L_{\rm zone}$ increases with increasing rotation rate.
\citet{held-larichev-1996}'s diffusivity has the same rotation-rate
dependence as that of \citet{stone-1972}, but it has a much stronger
depenence on gravity and temperature gradient [$g^3$ and
$(\partial\bar\theta/\partial y)^3$].  Moreover, it increases
with decreasing static stability as $N^{-3}$, unlike Stone's
diffusivity that scales with $N$.\footnote{The $N^{-3}$ dependence
in \citet{held-larichev-1996}'s 
gives the impression that the diffusivity becomes unbounded as the
atmospheric vertical temperature profile becomes neutrally stable
(i.e. as $N\to0$), but this is misleading.
In reality, one expects the latitudinal temperature gradient and
potential energy available for driving baroclinic instabilities to
decrease with decreasing $N$.  Noting that the slope of isentropes 
is $m_{\theta}\equiv(\partial \theta/\partial y)
(\partial\theta/\partial z)^{-1}$, one can re-express 
\citet{held-larichev-1996}'s diffusivity as scaling with
$m_{\theta} N^3$.  Thus, at constant isentrope slope,
the diffusivity properly drops to zero as the vertical temperature
profile becomes neutrally stable.} Finally, \citet{barry-etal-2002}'s
diffusivity suggests a somewhat weaker dependence, scaling with
as $(\partial\bar\theta/\partial y)^{3/5}$ and $\Omega^{-3/5}$.
Note that the rotation-rate dependences described above should be
treated with caution, because altering the rotation rate or gravity
could change the circulation in a way that alters the Brunt-V\"ais\"al\"a 
frequency, leading to additional changes in the diffusivity.

Additional work is needed to determine which of these schemes
(if any) is valid over a wide range of parameters relevant to
terrestrial exoplanets.  Most of the work performed to test the schemes of
\citet{green-1970}, \citet{stone-1972}, \citet{held-larichev-1996}
and others has adopted simplified two-layer quasi-geostrophic
models\footnote{That is, models where geostrophic balance is imposed
as an external constraint.} under idealized assumptions such as planar geometry
with constant Coriolis parameter (a so-called ``f-plane'').  Because 
variation of $f$ with latitude alters the properties of baroclinic 
instability, schemes developed for constant $f$ may not 
translate directly into a planetary context.  More recently,
two-layer planar models with non-zero $\beta$ have been explored 
by \citet{thompson-young-2007} and \citet{zurita-gotor-vallis-2009},
while full three-dimensional GCM calculations on a sphere (with
$\sim30$ layers) investigating the latitudinal heat flux
have been performed by \citet{barry-etal-2002} and 
\citet{schneider-walker-2008} under conditions relevant to Earth.
Additional simulations in the spirit
of \citet{barry-etal-2002} and \citet{schneider-walker-2008}, 
considering a wider range of planetary
and atmospheric parameters, can clarify the true sensitivities 
of the heat transport rates in three-dimensional atmospheres.

\subsection{Slowly rotating regime}
\label{slowly-rotating}

At slow rotation rates, the GCM simulations shown in 
Figs.~\ref{navarra-boccaletti}--\ref{delgenio-suozzo-fig3}
develop near-global Hadley cells with high-latitude jets,
but the wind remains weak at the equator.   In contrast,
Titan and Venus (with rotation periods of 16 and 243 days, respectively)
have robust ($\sim$$100\rm\,m\,sec^{-1}$) superrotating 
winds in their equatorial upper tropospheres.  Because the 
equator is the region of
the planet lying farthest from the rotation axis, such
a flow contains a local maximum of angular momentum at
the equator.  Axisymmetric Hadley circulations (\S\ref{hadley})
cannot produce such superrotation; rather, up-gradient
transport of momentum by eddies is required.  This process
remains poorly understood.  One class of models suggests
that this transport occurs from high latitudes; for example,
the high-latitude jets that result from the Hadley circulation
(which can be seen in Fig.~\ref{navarra-boccaletti}) experience
a large-scale shear instability that pumps eddy momentum
toward the equator, generating the equatorial superrotation 
\citep{delgenio-zhou-1996}.  Alternatively, thermal tide or wave interaction 
could induce momentum transports that generate the equatorial superrotation
\citep[e.g.,][]{fels-lindzen-1974}.  Recently, a variety of simplified 
Venus and Titan GCMs have been developed that show encouraging
progress in capturing the superrotation \citep[e.g.,][]
{yamamoto-takahashi-2006, richardson-etal-2007, lee-etal-2007, 
herrnstein-dowling-2007}.  

The Venus/Titan superrotation problem may have important implications 
for understanding the circulation of synchronously locked
exoplanets.  All published 3D circulation models of synchronously 
locked hot Jupiters (\S\ref{hot-jupiters}), and even the few
published studies of sychrounously locked terrestrial planets
\citep[][see \S\ref{unusual-forcing}]{joshi-etal-1997, joshi-2003},
develop robust equatorial superrotation.  The day-night heating
pattern associated with synchronous locking should generate thermal
tides and, in analogy with Venus and Titan, these could be relevant
in driving the superrotation in the 3D exoplanet models.  Simplified
Earth-based two-layer calculations that include longitudinally
varying heating, and which also develop equatorial superrotation,
seem to support this possibility \citep{suarez-duffy-1992,
saravanan-1993}.

\subsection{Unusual forcing regimes}
\label{unusual-forcing}

Our discussion of circulation regimes on terrestrial planets has so far
largely focused on annual-mean forcing conditions. On Earth,
seasonal variations represent relatively modest perturbations around
the annual-mean climate due to Earth's $23.5^{\circ}$ obliquity. On
longer timescales, the secular evolution of Earth's orbital
elements is thought to be responsible for paleoclimatic trends,
such as ice ages, according to Milankovitch's interpretation 
\citep[e.g.][]{kawamura-etal-2007}.
While it may be surprising to refer to Earth's
seasons or ice ages as minor events from a human perspective, it is
clear that they constitute rather mild versions of the more diverse
astronomical forcing regimes expected to occur on extrasolar worlds.

In principle, terrestrial exoplanets could possess large obliquities
($i \to 90 \deg$) and eccentricities ($e \to 1$). This would result in
forcing conditions with substantial seasonal variations around
the annual mean, such as order-of magnitude variations in the global
insolation over the orbital period at large eccentricities. Even the
annual mean climate can be dramatically affected under unusual forcing
conditions, for instance at large obliquities when the poles receive
more annual-mean insolation than the planetary equator (for $i > 54
^{\circ}$). Yet another unusual forcing regime occurs if the terrestrial
planet is tidally locked to its parent star and thus possesses
permanent day and night sides much like the hot Jupiters discussed 
in \S\ref{hot-jupiters}. Relatively few studies of
circulation regimes under such unusual forcing conditions have been
carried out to date.

\citet{williams-kasting-1997} and \citet{spiegel-etal-2009} investigated 
the climate of oblique Earth-like planets with simple, diffusive
energy-balance models of the type described by Eq.~(\ref{ebm}). While
seasonal variations were found to be severe at high obliquity,
\citet{williams-kasting-1997} concluded that highly oblique planets could
nevertheless remain regionally habitable, especially if they possessed
thick CO$_2$-enriched atmospheres with large thermal inertia relative
to Earth, as may indeed be expected for terrestrial planets in the
outermost regions of a system's habitable zone. \citet{spiegel-etal-2009}
confirmed these results and highlighted the risks that global
glaciation events and partial atmospheric CO$_2$ collapse constitute
for highly oblique terrestrial exoplanets. \citet{williams-pollard-2003}
presented a much more detailed set of three-dimensional climate and
circulation models for Earth-like planets at various
obliquities. Despite large seasonal variations at extreme obliquities,
\citet{williams-pollard-2003} found no evidence for any runaway
greenhouse effect or global glaciation event in their climate models
and thus concluded that Earth-like planets should generally be
hospitable to life even at high obliquity.

\citet{williams-pollard-2002} also studied the climate on eccentric
versions of the Earth with a combination of detailed three-dimensional
climate models and simpler energy-balance models. They showed that the
strong variations in insolation occurring at large eccentricities can
be rather efficiently buffered by the large thermal inertia of the
atmosphere$+$ocean climate system. These authors thus argued
that the annually averaged insolation, which is an increasing function
of $e$, may be the most meaningful quantity to describe in simple
terms the climate of eccentric, thermally blanketed Earth-like
planets. It is presumably the case that terrestrial planets with
reduced thermal inertia from lower atmospheric/oceanic masses would be
more strongly affected by such variations in insolation.

It should be emphasized that, despite the existing work on oblique or
eccentric versions of the Earth, a fundamental understanding of
circulation regimes under such unusual forcing conditions is still
largely missing. For instance, variations in the Hadley circulation
regime or the baroclinic transport efficiency, expected as forcing
conditions vary along the orbit, have not been thoroughly
explored. Similarly, the combined effects of a substantial obliquity
and eccentricity have been ignored. As the prospects for finding and
characterizing exotic versions of the Earth improve, these issues 
should become the subject of increasing scrutiny.

One of the best observational prospects for the next decade is the
discovery of tidally locked terrestrial planets around nearby M-dwarfs
\citep[e.g.,][]{irwin-etal-2009, seager-etal-2008}. 
In anticipation of such discoveries, \citet{joshi-etal-1997} 
presented a careful investigation of the circulation regime
on this class of unusually forced planets, with permanent day and night
sides. The specific focus of their work was to determine the 
conditions under which CO$_2$ atmospheric collapse could occur on the
cold night sides of such planets and act as a trap for this important
greenhouse gas, even when circulation and heat transport are
present. Using a simplified general circulation model, these authors
explored the problem's parameter space for various global planetary
and atmospheric attributes and included a study of the potential risks
posed by violent flares from the stellar host. The unusual circulation
regime that emerged from this work was composed of a direct circulation
cell at surface levels (i.e. equatorial day to night transport with
polar return) and a superrotating wind higher up in the
atmosphere. While detrimental atmospheric collapse did occur under
some combinations of planetary attributes (e.g., for thin
atmospheres), the authors concluded that efficient atmospheric heat
transport was sufficient to prevent collapse under a variety of
plausible scenarios. More recently, \citet{joshi-2003} revisited some of
these results with a much more detailed climate model with explicit
treatments of non-gray radiative transfer and a hydrological cycle. It is
likely that this interesting circulation regime will be reconsidered
in the next few years as the discovery of such exotic worlds is on our
horizon.

\section{RECENT HIGHLIGHTS}
\label{highlights}

Since the first discovery of giant exoplanets around 
Sun-like stars via the Doppler velocity technique \citep{mayor-queloz-1995, 
marcy-butler-1996}, exoplanet research has
experienced a spectacular series of observational breakthroughs.  
The first discovery of a transiting hot Jupiter \citep{charbonneau-etal-2000,
henry-etal-2000} opened the door to a wide range of clever techniques
for characterizing such transiting planets.  The drop in flux that
occurs during secondary eclipse, when the planet passes behind its star,
led to direct infrared detections of the 
thermal flux from the planet's dayside \citep{deming-etal-2005a,
charbonneau-etal-2005}. This was quickly followed by the detection
of day/night temperature variations \citep{harrington-etal-2006}, detailed
phase curve observations \citep[e.g.][]{knutson-etal-2007b, knutson-etal-2009a,
cowan-etal-2007}, infrared spectral and photometric measurements 
\citep{grillmair-etal-2007, grillmair-etal-2008,
richardson-etal-2007, charbonneau-etal-2008,
knutson-etal-2008a} and a variety of transit spectroscopic
constraints \citep{Tinetti-etal-2007b, swain-etal-2008, barman-2008}.
Collectively, these observations help to constrain the composition,
albedo, three-dimensional (3D) temperature structure, and hence
the atmospheric circulation regime of hot Jupiters.

The reality of atmospheric circulation on these objects is suggested by 
exquisite light curves of HD 189733b showing modest day-night infrared 
brightness temperature variations and displacement of the longitudes of 
minimum/maximum flux from the antistellar/substellar points, presumably 
the result of an efficient circulation able to distort the temperature 
pattern \citep{knutson-etal-2007b, knutson-etal-2009a}. Dayside
photometry also hints at the existence of an atmospheric circulation 
on this planet \citep{barman-2008}.  Nevertheless, other hot Jupiters,
such as Ups And b and HD 179949b, apparently exhibit large day-night 
temperature variations with no discernable displacement of the hot 
regions from the substellar point \citep{harrington-etal-2006,
cowan-etal-2007}.  Secondary-eclipse photometry suggests that some hot Jupiters
have dayside temperatures that decrease with altitude, while
other hot Jupiters appear to exhibit dayside thermal inversion layers where
temperatures increase with altitude.  Some authors have suggested
that these two issues are linked \citep{fortney-etal-2008}, although
subtantial additional observations are needed for a robust assessment.
Understanding the dependence
of infrared phase variations and dayside temperature profiles on
planetary orbital and physical properties 
remains an ongoing observational challenge for the coming decade.

In parallel, this observational vanguard has
triggered a growing body of theoretical and modeling studies to
investigate plausible atmospheric circulation patterns on hot Jupiters.
The models agree on some issues, such as that synchronously rotating
hot Jupiters on $\sim3$-day orbits should exhibit fast winds with
only a few broad zonal jets.  On the other hand, the models disagree
on other issues, such as the details of the 3D flow patterns and the 
extent to which significant global-scale time variability is likely.  
The effect of stellar flux, planetary rotation rate, obliquity,
atmospheric composition, and orbital eccentricity on the circulation 
have barely been explored.   Further theoretical 
work and comparison with observations should help move our
understanding onto a firmer foundation.

\section{FUTURE PROSPECTS}
\label{future}

Over the next decade, the observational
characterization of hot Jupiters and Neptunes will continue apace, 
aided especially by the warm {\it Spitzer} mission and subsequently {\it JWST}.
Equally exciting, NASA's {\it Kepler} mission and groundbased 
surveys \citep[e.g., MEarth; see][]{irwin-etal-2009} may soon lead
to the discovery of not only additional super-Earths but 
Earth-sized terrestrial exoplanets, including ``hot Earths'' as well as
terrestrial planets cool enough to lie within the habitable zones of their
stars.  If sufficiently
favorable systems are discovered, follow-up by {\it JWST} 
may allow basic characterization of their spectra and light curves
(albeit at significant resources per planet),
providing constraints on their atmospheric properties including
their composition, climate, circulation, and the extent to which
they may be habitable.  Finally, over the next decade,
significant opportunities exist for
expanding our understanding of basic dynamical processes 
by widening the parameters adopted in circulation models beyond
those typically explored in the study of Solar-System planets.

To move the observational characterization of exoplanet atmospheric
circulation into the next
generation, we recommend full-orbit light curves for a variety
of hot Jupiters and Neptunes (and eventually terrestrial planets).
Target objects should span a range of incident 
stellar fluxes, planetary masses, and orbital periods and eccentricities
(among other parameters) to characterize planetary diversity.  
To constrain the planetary thermal-energy budgets, light curves 
should sample the blackbody peaks at several
wavelengths, including both absorption bands and low-opacity
windows as expected from theoretical models; this will provide
constraints on the day-night temperature contrasts---and hence
day-night heat transport---at a range of atmospheric pressure levels.
Dayside infrared spectra obtained from secondary-eclipse measurements,
while not necessarily diagnostic of the atmospheric circulation in isolation,
will provide crucial constraints when combined with infrared light curves.
By providing constraints on atmospheric composition, transit spectroscopy
will enable much better interpretation of light curves and secondary-eclipse
spectra. Secondary eclipses (or full light curves) that are observed repeatedly
will provide powerful constraints on global-scale atmospheric variability.
{\it Kepler} may prove particularly useful in this regard, with its
ability to observe hundreds of secondary eclipses of hot Jupiters 
\citep[such as HAT-P-7b;][]{borucki-etal-2009} in its field of view.
Such an extensive temporal dataset could provide information on the
frequency spectrum of atmospheric variability (if any), which would
help to identify the mechanisms of variability as well as the
background state of the mean circulation.  Other novel techniques,
such as the possibility of separately measuring transit spectra on
the leading and trailing limbs and the possibility of detecting
the Doppler shifts associated with planetary rotation and/or
winds, should be considered.   

Specific research questions for the next decade include but are
not limited to the following:
\begin{itemize}

\item{How warm are the nightsides of hot Jupiters?  How does
this depend on the stellar flux, atmospheric composition, 
and other parameters?}

\item{What are the dynamical mechanisms for shifting the hot
and cold points away from the substellar and antistellar
points, respectively?  Why does HD 189733b appear to have both
its hottest and coldest points on the same hemisphere?}

\item{What is the observational and physical relationship (if any) between 
the amplitude of the day-night temperature contrast and existence
or absence of a dayside temperature inversion that has been
inferred on some hot Jupiters?}

\item{How does the expected atmospheric circulation regime
depend on planetary size, rotation rate, gravity, obliquity,
atmospheric composition, and incident stellar flux?}

\item{What are the important dissipation mechanisms in the
atmospheres of hot Jupiters? What are the relative roles
of turbulence, shocks, and magnetohydrodynamic processes
in frictionally braking the flows?}

\item{How coupled are the atmospheric flows on hot Jupiters
to convection in the planetary interior?  How are momentum, heat, and
constituents transported across the deep radiative zone in
hot Jupiters?}

\item{Does the atmospheric circulation influence the long-term evolution
of hot Jupiters?  Can downward transport of atmospheric energy into the
interior help explain the radii of some anomalously large
hot Jupiters?}

\item{What is the mechanism for equatorial superrotation that
occurs in most 3D models of synchronously rotating hot Jupiters?
Does such superrotation exist on real hot Jupiters?}

\item{Are the atmospheres of hot Jupiters strongly time variable on
the global scale, and if so, what is the mechanism for the
variability?}

\item{How does the circulation regime of planets in highly
eccentric orbits differ from that of planets in circular orbits?}

\item{To what extent can obliquity, rotation rate,
atmospheric mass, and atmospheric circulation properties be inferred
from disk-integrated spectra or light curves?}

\item{To what extent, if any, does the atmospheric circulation
influence the escape of an atmosphere to space?}

\item{Do the deep molecular envelopes of giant planets differentially
rotate?}

\item{What controls the wind speeds on giant planets?  Why are
Neptune's winds faster than Jupiter's, and what can this teach
us about strongly irradiated planets?}

\item{What can observations of chemical disequilibrium
species tell us about the atmospheric circulation of giant planets
and brown dwarfs?}

\item{What is the role of clouds in affecting the circulation,
climate, evolution, and observable properties of exoplanets?}

\item{For terrestrial planets, how does the atmospheric circulation
affect the boundaries of the classical habitable zone?}

\item{What is the role of the atmospheric circulation in affecting
the ice-albedo feedback, atmospheric collapse, runaway greenhouse,
and other climate feedbacks?}

\item{Can adequate scaling theories be developed to predict
day-night or equator-to-pole temperature contrasts as a function
of planetary and atmospheric parameters?}

\item{What is the dynamical role of a surface (i.e. the ground)
in the atmospheric circulation?  For super-Earths, is there a critical 
atmospheric mass beyond which the surface becomes unimportant and
the circulation behaves like that of a giant planet?}

\end{itemize}

Ultimately, unraveling the atmospheric circulation and climate
of exoplanets from global-scale observations will be
a difficult yet exciting challenge.  Degeneracies of interpretation 
will undoubtedly exist (e.g., multiple circulation patterns explaining 
a given light curve), and forward progress will require not only
high-quality observations but a careful exploration of a hierarchy 
of models so that the nature of these degeneracies can be understood.  
Despite the challenge, the potential payoff will be the unleashing of
planetary meteorology beyond the confines of our Solar System,
leading to not only an improved understanding of basic circulation
mechanisms (and how they may vary with planetary rotation
rate, gravity, incident stellar flux, and other parameters)
but a glimpse of the actual atmospheric circulations, climate, 
and habitability of planets orbiting other stars in our neighborhood
of the Milky Way.

\bigskip
\textbf{ Acknowledgments.} This paper was supported by NASA Origins
grant NNX08AF27G to APS, NASA grants NNG04GN82G and STFC PP/E001858/1 
to JYKC, and NASA contract NNG06GF55G to KM.
\\

\bigskip
\parskip=0pt
{\small
\baselineskip=11pt

\bibliographystyle{ametsoc}

\begin{thebibliography}{195}
\expandafter\ifx\csname natexlab\endcsname\relax\def\natexlab#1{#1}\fi

\bibitem[{{Atkinson} et~al.(1997){\it {Atkinson}, {Ingersoll}, and
  {Seiff}\/}}]{atkinson-etal-1997}
{Atkinson}, D.~H., A.~P. {Ingersoll}, and A.~{Seiff}, 1997:
\newblock {Deep zonal winds on Jupiter: Update of Doppler tracking the Galileo
  probe from the orbiter}.
\newblock {\it Nature\/}, {\bf 388}, 649--650.

\bibitem[{{Aurnou} and {Heimpel}(2004){\it {Aurnou} and
  {Heimpel}\/}}]{aurnou-heimpel-2004}
{Aurnou}, J.~M., and M.~H. {Heimpel}, 2004:
\newblock {Zonal jets in rotating convection with mixed mechanical boundary
  conditions}.
\newblock {\it Icarus\/}, {\bf 169}, 492--498.

\bibitem[{{Aurnou} and {Olson}(2001){\it {Aurnou} and
  {Olson}\/}}]{aurnou-olson-2001}
{Aurnou}, J.~M., and P.~L. {Olson}, 2001:
\newblock {Strong zonal winds from thermal convection in a rotating spherical
  shell}.
\newblock {\it Geophys. Res. Lett.\/}, {\bf 28}, 2557--2560.

\bibitem[{{Barman}(2008){\it {Barman}\/}}]{barman-2008}
{Barman}, T.~S., 2008:
\newblock {On the Presence of Water and Global Circulation in the Transiting
  Planet HD 189733b}.
\newblock {\it \apjl\/}, {\bf 676}, L61--L64.

\bibitem[{{Barry} et~al.(2002){\it {Barry}, {Craig}, and
  {Thuburn}\/}}]{barry-etal-2002}
{Barry}, L., G.~C. {Craig}, and J.~{Thuburn}, 2002:
\newblock {Poleward heat transport by the atmospheric heat engine}.
\newblock {\it Nature\/}, {\bf 415}, 774--777.

\bibitem[{{Batchelor}(1967){\it {Batchelor}\/}}]{batchelor-1967}
{Batchelor}, G.~K., 1967:
\newblock {\it An Introduction to Fluid Dynamics\/}.
\newblock Cambridge University Press.

\bibitem[{{B{\'e}zard} et~al.(2002){\it {B{\'e}zard}, {Lellouch}, {Strobel},
  {Maillard}, and {Drossart}\/}}]{bezard-etal-2002}
{B{\'e}zard}, B., E.~{Lellouch}, D.~{Strobel}, J.-P. {Maillard}, and
  P.~{Drossart}, 2002:
\newblock {Carbon Monoxide on Jupiter: Evidence for Both Internal and External
  Sources}.
\newblock {\it Icarus\/}, {\bf 159}, 95--111.

\bibitem[{{Borucki} and {colleagues}(2009){\it {Borucki} and
  {colleagues}\/}}]{borucki-etal-2009}
{Borucki}, W.~J., and {colleagues}, 2009:
\newblock {Kepler's optical phase curve of the exoplanet HAT-P-7b}.
\newblock {\it Nature\/}, {\bf 325}, 709.

\bibitem[{{Burrows} et~al.(2000){\it {Burrows}, {Guillot}, {Hubbard}, {Marley},
  {Saumon}, {Lunine}, and {Sudarsky}\/}}]{burrows-etal-2000}
{Burrows}, A., T.~{Guillot}, W.~B. {Hubbard}, M.~S. {Marley}, D.~{Saumon},
  J.~I. {Lunine}, and D.~{Sudarsky}, 2000:
\newblock {On the Radii of Close-in Giant Planets}.
\newblock {\it \apjl\/}, {\bf 534}, L97--L100.

\bibitem[{{Burrows} et~al.(2001){\it {Burrows}, {Hubbard}, {Lunine}, and
  {Liebert}\/}}]{burrows-etal-2001}
{Burrows}, A., W.~B. {Hubbard}, J.~I. {Lunine}, and J.~{Liebert}, 2001:
\newblock {The theory of brown dwarfs and extrasolar giant planets}.
\newblock {\it Reviews of Modern Physics\/}, {\bf 73}, 719--765.

\bibitem[{{Busse}(1976){\it {Busse}\/}}]{busse-1976}
{Busse}, F.~H., 1976:
\newblock {A simple model of convection in the Jovian atmosphere}.
\newblock {\it Icarus\/}, {\bf 29}, 255--260.

\bibitem[{{Busse}(2002){\it {Busse}\/}}]{busse-2002}
{Busse}, F.~H., 2002:
\newblock {Convective flows in rapidly rotating spheres and their dynamo
  action}.
\newblock {\it Physics of Fluids\/}, {\bf 14}, 1301--1314.

\bibitem[{{Caballero} et~al.(2008){\it {Caballero}, {Pierrehumbert}, and
  {Mitchell}\/}}]{caballero-etal-2008}
{Caballero}, R., R.~T. {Pierrehumbert}, and J.~L. {Mitchell}, 2008:
\newblock {Axisymmetric, nearly inviscid circulations in non-condensing
  radiative-convective atmospheres}.
\newblock {\it Quarterly Journal of the Royal Meteorological Society\/}, {\bf
  134}, 1269--1285.

\bibitem[{{Chabrier} and {Baraffe}(2000){\it {Chabrier} and
  {Baraffe}\/}}]{chabrier-baraffe-2000}
{Chabrier}, G., and I.~{Baraffe}, 2000:
\newblock {Theory of Low-Mass Stars and Substellar Objects}.
\newblock {\it \araa\/}, {\bf 38}, 337--377.

\bibitem[{{Chabrier} and {Baraffe}(2007){\it {Chabrier} and
  {Baraffe}\/}}]{chabrier-baraffe-2007}
{Chabrier}, G., and I.~{Baraffe}, 2007:
\newblock {Heat Transport in Giant (Exo)planets: A New Perspective}.
\newblock {\it \apjl\/}, {\bf 661}, L81--L84.

\bibitem[{{Chabrier} et~al.(2004){\it {Chabrier}, {Barman}, {Baraffe},
  {Allard}, and {Hauschildt}\/}}]{chabrier-etal-2004}
{Chabrier}, G., T.~{Barman}, I.~{Baraffe}, F.~{Allard}, and P.~H. {Hauschildt},
  2004:
\newblock {The Evolution of Irradiated Planets: Application to Transits}.
\newblock {\it \apjl\/}, {\bf 603}, L53--L56.

\bibitem[{{Charbonneau} et~al.(2000){\it {Charbonneau}, {Brown}, {Latham}, and
  {Mayor}\/}}]{charbonneau-etal-2000}
{Charbonneau}, D., T.~M. {Brown}, D.~W. {Latham}, and M.~{Mayor}, 2000:
\newblock {Detection of Planetary Transits Across a Sun-like Star}.
\newblock {\it \apjl\/}, {\bf 529}, L45--L48.

\bibitem[{{Charbonneau} et~al.(2008){\it {Charbonneau}, {Knutson}, {Barman},
  {Allen}, {Mayor}, {Megeath}, {Queloz}, and {Udry}\/}}]{charbonneau-etal-2008}
{Charbonneau}, D., H.~A. {Knutson}, T.~{Barman}, L.~E. {Allen}, M.~{Mayor},
  S.~T. {Megeath}, D.~{Queloz}, and S.~{Udry}, 2008:
\newblock {The Broadband Infrared Emission Spectrum of the Exoplanet HD
  189733b}.
\newblock {\it \apj\/}, {\bf 686}, 1341--1348.

\bibitem[{{Charbonneau} et~al.(2005){\it {Charbonneau},
  et~al.\/}}]{charbonneau-etal-2005}
{Charbonneau}, D., et~al., 2005:
\newblock {Detection of Thermal Emission from an Extrasolar Planet}.
\newblock {\it \apj\/}, {\bf 626}, 523--529.

\bibitem[{{Charney}(1971){\it {Charney}\/}}]{charney-1971}
{Charney}, J.~G., 1971:
\newblock {Geostrophic turbulence}.
\newblock {\it Journal of Atmospheric Sciences\/}, {\bf 28}, 1087--1095.

\bibitem[{{Cho}(2008){\it {Cho}\/}}]{cho-2008}
{Cho}, J.~Y.-K., 2008:
\newblock {Atmospheric dynamics of tidally synchronized extrasolar planets}.
\newblock {\it Royal Society of London Philosophical Transactions Series A\/},
  {\bf 366}, 4477--4488.

\bibitem[{{Cho} and {Polvani}(1996{\natexlab{a}}){\it {Cho} and
  {Polvani}\/}}]{cho-polvani-1996a}
{Cho}, J.~Y.-K., and L.~M. {Polvani}, 1996{\natexlab{a}}:
\newblock The morphogenesis of bands and zonal winds in the atmospheres on the
  giant outer planets.
\newblock {\it Science\/}, {\bf 8}(1), 1--12.

\bibitem[{{Cho} and {Polvani}(1996{\natexlab{b}}){\it {Cho} and
  {Polvani}\/}}]{cho-polvani-1996b}
{Cho}, J.~Y.-K., and L.~M. {Polvani}, 1996{\natexlab{b}}:
\newblock {The emergence of jets and vortices in freely evolving, shallow-water
  turbulence on a sphere}.
\newblock {\it Physics of Fluids\/}, {\bf 8}, 1531--1552.

\bibitem[{{Cho} et~al.(2003){\it {Cho}, {Menou}, {Hansen}, and
  {Seager}\/}}]{cho-etal-2003}
{Cho}, J.~Y.-K., K.~{Menou}, B.~M.~S. {Hansen}, and S.~{Seager}, 2003:
\newblock {The Changing Face of the Extrasolar Giant Planet HD 209458b}.
\newblock {\it \apjl\/}, {\bf 587}, L117--L120.

\bibitem[{{Cho} et~al.(2008){\it {Cho}, {Menou}, {Hansen}, and
  {Seager}\/}}]{cho-etal-2008}
{Cho}, J.~Y.-K., K.~{Menou}, B.~M.~S. {Hansen}, and S.~{Seager}, 2008:
\newblock {Atmospheric Circulation of Close-in Extrasolar Giant Planets. I.
  Global, Barotropic, Adiabatic Simulations}.
\newblock {\it \apj\/}, {\bf 675}, 817--845.

\bibitem[{{Christensen}(2001){\it {Christensen}\/}}]{christensen-2001}
{Christensen}, U.~R., 2001:
\newblock {Zonal flow driven by deep convection in the major planets}.
\newblock {\it Geophys. Res. Lett.\/}, {\bf 28}, 2553--2556.

\bibitem[{{Christensen}(2002){\it {Christensen}\/}}]{christensen-2002}
{Christensen}, U.~R., 2002:
\newblock {Zonal flow driven by strongly supercritical convection in rotating
  spherical shells}.
\newblock {\it Journal of Fluid Mechanics\/}, {\bf 470}, 115--133.

\bibitem[{{Cooper} and {Showman}(2005){\it {Cooper} and
  {Showman}\/}}]{cooper-showman-2005}
{Cooper}, C.~S., and A.~P. {Showman}, 2005:
\newblock {Dynamic Meteorology at the Photosphere of HD 209458b}.
\newblock {\it \apjl\/}, {\bf 629}, L45--L48.

\bibitem[{{Cooper} and {Showman}(2006){\it {Cooper} and
  {Showman}\/}}]{cooper-showman-2006}
{Cooper}, C.~S., and A.~P. {Showman}, 2006:
\newblock {Dynamics and Disequilibrium Carbon Chemistry in Hot Jupiter
  Atmospheres, with Application to HD 209458b}.
\newblock {\it \apj\/}, {\bf 649}, 1048--1063.

\bibitem[{{Cowan} et~al.(2007){\it {Cowan}, {Agol}, and
  {Charbonneau}\/}}]{cowan-etal-2007}
{Cowan}, N.~B., E.~{Agol}, and D.~{Charbonneau}, 2007:
\newblock {Hot nights on extrasolar planets: mid-infrared phase variations of
  hot Jupiters}.
\newblock {\it \mnras\/}, {\bf 379}, 641--646.

\bibitem[{{Del Genio} and {Suozzo}(1987){\it {Del Genio} and
  {Suozzo}\/}}]{delgenio-suozzo-1987}
{Del Genio}, A.~D., and R.~J. {Suozzo}, 1987:
\newblock {A comparative study of rapidly and slowly rotating dynamical regimes
  in a terrestrial general circulation model}.
\newblock {\it Journal of Atmospheric Sciences\/}, {\bf 44}, 973--986.

\bibitem[{{Del Genio} and {Zhou}(1996){\it {Del Genio} and
  {Zhou}\/}}]{delgenio-zhou-1996}
{Del Genio}, A.~D., and W.~{Zhou}, 1996:
\newblock {Simulations of Superrotation on Slowly Rotating Planets: Sensitivity
  to Rotation and Initial Condition}.
\newblock {\it Icarus\/}, {\bf 120}, 332--343.

\bibitem[{{Deming} et~al.(2005){\it {Deming}, {Seager}, {Richardson}, and
  {Harrington}\/}}]{deming-etal-2005a}
{Deming}, D., S.~{Seager}, L.~J. {Richardson}, and J.~{Harrington}, 2005:
\newblock {Infrared radiation from an extrasolar planet}.
\newblock {\it \nat\/}, {\bf 434}, 740--743.

\bibitem[{{Deming} et~al.(2006){\it {Deming}, {Harrington}, {Seager}, and
  {Richardson}\/}}]{deming-etal-2006}
{Deming}, D., J.~{Harrington}, S.~{Seager}, and L.~J. {Richardson}, 2006:
\newblock {Strong Infrared Emission from the Extrasolar Planet HD 189733b}.
\newblock {\it \apj\/}, {\bf 644}, 560--564.

\bibitem[{{Dima} and {Wallace}(2003){\it {Dima} and
  {Wallace}\/}}]{dima-wallace-2003}
{Dima}, I.~M., and J.~M. {Wallace}, 2003:
\newblock {On the Seasonality of the Hadley Cell.}
\newblock {\it Journal of Atmospheric Sciences\/}, {\bf 60}, 1522--1527.

\bibitem[{{Dobbs-Dixon} and {Lin}(2008){\it {Dobbs-Dixon} and
  {Lin}\/}}]{dobbs-dixon-lin-2008}
{Dobbs-Dixon}, I., and D.~N.~C. {Lin}, 2008:
\newblock {Atmospheric Dynamics of Short-Period Extrasolar Gas Giant Planets.
  I. Dependence of Nightside Temperature on Opacity}.
\newblock {\it \apj\/}, {\bf 673}, 513--525.

\bibitem[{{Dowling}(1995){\it {Dowling}\/}}]{dowling-1995a}
{Dowling}, T.~E., 1995:
\newblock {Dynamics of jovian atmospheres}.
\newblock {\it Annual Review of Fluid Mechanics\/}, {\bf 27}, 293--334.

\bibitem[{{Dowling} and {Ingersoll}(1989){\it {Dowling} and
  {Ingersoll}\/}}]{dowling-ingersoll-1989}
{Dowling}, T.~E., and A.~P. {Ingersoll}, 1989:
\newblock {Jupiter's Great Red Spot as a shallow water system}.
\newblock {\it Journal of Atmospheric Sciences\/}, {\bf 46}, 3256--3278.

\bibitem[{{Dritschel} et~al.(1999){\it {Dritschel}, {de La Torre Ju{\'a}rez},
  and {Ambaum}\/}}]{dritschel-etal-1999}
{Dritschel}, D.~G., M.~{de La Torre Ju{\'a}rez}, and M.~H.~P. {Ambaum}, 1999:
\newblock {The three-dimensional vortical nature of atmospheric and oceanic
  turbulent flows}.
\newblock {\it Physics of Fluids\/}, {\bf 11}, 1512--1520.

\bibitem[{{Farrell}(1990){\it {Farrell}\/}}]{farrell-1990}
{Farrell}, B.~F., 1990:
\newblock {Equable Climate Dynamics.}
\newblock {\it Journal of Atmospheric Sciences\/}, {\bf 47}, 2986--2995.

\bibitem[{{Fegley} and {Lodders}(1994){\it {Fegley} and
  {Lodders}\/}}]{fegley-lodders-1994}
{Fegley}, B.~J., and K.~{Lodders}, 1994:
\newblock {Chemical models of the deep atmospheres of Jupiter and Saturn}.
\newblock {\it Icarus\/}, {\bf 110}, 117--154.

\bibitem[{{Fels} and {Lindzen}(1974){\it {Fels} and
  {Lindzen}\/}}]{fels-lindzen-1974}
{Fels}, S.~B., and R.~S. {Lindzen}, 1974:
\newblock {The Interaction of Thermally Excited Gravity Waves with Mean Flows}.
\newblock {\it Geophysical and Astrophysical Fluid Dynamics\/}, {\bf 6},
  149--191.

\bibitem[{{Fernando} et~al.(1991){\it {Fernando}, {Chen}, and
  {Boyer}\/}}]{fernando-etal-1991}
{Fernando}, H.~J.~S., R.-R. {Chen}, and D.~L. {Boyer}, 1991:
\newblock {Effects of rotation on convective turbulence}.
\newblock {\it Journal of Fluid Mechanics\/}, {\bf 228}, 513--547.

\bibitem[{{Fortney} et~al.(2007){\it {Fortney}, {Marley}, and
  {Barnes}\/}}]{fortney-etal-2007}
{Fortney}, J.~J., M.~S. {Marley}, and J.~W. {Barnes}, 2007:
\newblock {Planetary Radii across Five Orders of Magnitude in Mass and Stellar
  Insolation: Application to Transits}.
\newblock {\it \apj\/}, {\bf 659}, 1661--1672.

\bibitem[{{Fortney} et~al.(2008){\it {Fortney}, {Lodders}, {Marley}, and
  {Freedman}\/}}]{fortney-etal-2008}
{Fortney}, J.~J., K.~{Lodders}, M.~S. {Marley}, and R.~S. {Freedman}, 2008:
\newblock {A Unified Theory for the Atmospheres of the Hot and Very Hot
  Jupiters: Two Classes of Irradiated Atmospheres}.
\newblock {\it \apj\/}, {\bf 678}, 1419--1435.

\bibitem[{{Fortney} et~al.(2009){\it {Fortney}, {Militzer}, and
  {Baraffe}\/}}]{fortney-etal-2009}
{Fortney}, J.~J., B.~{Militzer}, and I.~{Baraffe}, 2009:
\newblock {Giant planets: theory of interiors and evolution}.
\newblock  {\it Exoplanets\/}, S.~{Seager}, Ed., Univ. Arizona Press.

\bibitem[{{Frierson} et~al.(2007){\it {Frierson}, {Lu}, and
  {Chen}\/}}]{frierson-etal-2007}
{Frierson}, D.~M.~W., J.~{Lu}, and G.~{Chen}, 2007:
\newblock {Width of the Hadley cell in simple and comprehensive general
  circulation models}.
\newblock {\it Geophys. Res. Lett.\/}, {\bf 34}, L18,804.

\bibitem[{{Gierasch} and {Conrath}(1985){\it {Gierasch} and
  {Conrath}\/}}]{gierasch-conrath-1985}
{Gierasch}, P.~J., and B.~J. {Conrath}, 1985:
\newblock {Energy conversion processes in the outer planets}.
\newblock  {\it Recent Advances in Planetary Meteorology\/}, G.~E. {Hunt}, Ed.,
  Cambridge Univ. Press, New York, pp. 121--146.

\bibitem[{{Gierasch} et~al.(1997){\it {Gierasch},
  et~al.\/}}]{gierasch-etal-1997}
{Gierasch}, P.~J., et~al., 1997:
\newblock {The General Circulation of the Venus Atmosphere: an Assessment}.
\newblock  {\it Venus II: Geology, Geophysics, Atmosphere, and Solar Wind
  Environment\/}, S.~W. {Bougher}, D.~M. {Hunten}, and R.~J. {Philips}, Eds.,
  pp. 459--500.

\bibitem[{{Gierasch} et~al.(2000){\it {Gierasch},
  et~al.\/}}]{gierasch-etal-2000}
{Gierasch}, P.~J., et~al., 2000:
\newblock {Observation of moist convection in Jupiter's atmosphere}.
\newblock {\it Nature\/}, {\bf 403}, 628--630.

\bibitem[{{Glatzmaier} et~al.(2009){\it {Glatzmaier}, {Evonuk}, and
  {Rogers}\/}}]{glatzmaier-etal-2009}
{Glatzmaier}, G.~A., M.~{Evonuk}, and T.~M. {Rogers}, 2009:
\newblock {Differential rotation in giant planets maintained by
  density-stratified turbulent convection}.
\newblock {\it Geophys. Astrophy. Fluid Dyn.\/}, {\bf 103}, 31--51.

\bibitem[{{Goldman} et~al.(2008){\it {Goldman}, et~al.\/}}]{goldman-etal-2008}
{Goldman}, B., et~al., 2008:
\newblock {CLOUDS search for variability in brown dwarf atmospheres. Infrared
  spectroscopic time series of L/T transition brown dwarfs}.
\newblock {\it \aap\/}, {\bf 487}, 277--292.

\bibitem[{{Goodman}(2009){\it {Goodman}\/}}]{goodman-2009}
{Goodman}, J., 2009:
\newblock {Thermodynamics of Atmospheric Circulation on Hot Jupiters}.
\newblock {\it \apj\/}, {\bf 693}, 1645--1649.

\bibitem[{{Green}(1970){\it {Green}\/}}]{green-1970}
{Green}, J.~S.~A., 1970:
\newblock {Transfer properties of the large-scale eddies and the general
  circulation of the atmosphere}.
\newblock {\it Quarterly Journal of the Royal Meteorological Society\/}, {\bf
  96}, 157--185.

\bibitem[{{Grillmair} et~al.(2007){\it {Grillmair}, {Charbonneau}, {Burrows},
  {Armus}, {Stauffer}, {Meadows}, {Van Cleve}, and
  {Levine}\/}}]{grillmair-etal-2007}
{Grillmair}, C.~J., D.~{Charbonneau}, A.~{Burrows}, L.~{Armus}, J.~{Stauffer},
  V.~{Meadows}, J.~{Van Cleve}, and D.~{Levine}, 2007:
\newblock {A Spitzer Spectrum of the Exoplanet HD 189733b}.
\newblock {\it \apjl\/}, {\bf 658}, L115--L118.

\bibitem[{{Grillmair} et~al.(2008){\it {Grillmair}, {Burrows}, {Charbonneau},
  {Armus}, {Stauffer}, {Meadows}, {van Cleve}, {von Braun}, and
  {Levine}\/}}]{grillmair-etal-2008}
{Grillmair}, C.~J., A.~{Burrows}, D.~{Charbonneau}, L.~{Armus}, J.~{Stauffer},
  V.~{Meadows}, J.~{van Cleve}, K.~{von Braun}, and D.~{Levine}, 2008:
\newblock {Strong water absorption in the dayside emission spectrum of the
  planet HD189733b}.
\newblock {\it \nat\/}, {\bf 456}, 767--769.

\bibitem[{{Grote} et~al.(2000){\it {Grote}, {Busse}, and
  {Tilgner}\/}}]{grote-etal-2000a}
{Grote}, E., F.~H. {Busse}, and A.~{Tilgner}, 2000:
\newblock {Regular and chaotic spherical dynamos}.
\newblock {\it Physics of the Earth and Planetary Interiors\/}, {\bf 117},
  259--272.

\bibitem[{{Guillot}(2005){\it {Guillot}\/}}]{guillot-2005}
{Guillot}, T., 2005:
\newblock {The interiors of giant planets: Models and outstanding questions}.
\newblock {\it Annual Review of Earth and Planetary Sciences\/}, {\bf 33},
  493--530.

\bibitem[{{Guillot} et~al.(1996){\it {Guillot}, {Burrows}, {Hubbard}, {Lunine},
  and {Saumon}\/}}]{guillot-etal-1996}
{Guillot}, T., A.~{Burrows}, W.~B. {Hubbard}, J.~I. {Lunine}, and D.~{Saumon},
  1996:
\newblock {Giant Planets at Small Orbital Distances}.
\newblock {\it \apjl\/}, {\bf 459}, L35--L38.

\bibitem[{{Guillot} et~al.(2004){\it {Guillot}, {Stevenson}, {Hubbard}, and
  {Saumon}\/}}]{guillot-etal-2004}
{Guillot}, T., D.~J. {Stevenson}, W.~B. {Hubbard}, and D.~{Saumon}, 2004:
\newblock {The interior of Jupiter}.
\newblock  {\it Jupiter: The Planet, Satellites and Magnetosphere\/},
  F.~{Bagenal}, T.~E. {Dowling}, and W.~B. {McKinnon}, Eds., Cambridge Univ.
  Press, pp. 35--57.

\bibitem[{{Hadley}(1735){\it {Hadley}\/}}]{hadley-1735}
{Hadley}, G., 1735:
\newblock {Concerning the cause of the general trade-winds}.
\newblock {\it Phil. Trans.\/}, {\bf 39}, 58--62.

\bibitem[{{Harrington} et~al.(2006){\it {Harrington}, {Hansen}, {Luszcz},
  {Seager}, {Deming}, {Menou}, {Cho}, and
  {Richardson}\/}}]{harrington-etal-2006}
{Harrington}, J., B.~M. {Hansen}, S.~H. {Luszcz}, S.~{Seager}, D.~{Deming},
  K.~{Menou}, J.~Y.-K. {Cho}, and L.~J. {Richardson}, 2006:
\newblock {The Phase-Dependent Infrared Brightness of the Extrasolar Planet
  $\upsilon$ Andromedae b}.
\newblock {\it Science\/}, {\bf 314}, 623--626.

\bibitem[{{Hartmann} and {coauthors}(2003){\it {Hartmann} and
  {coauthors}\/}}]{hartmann-etal-2003}
{Hartmann}, D.~L., and {coauthors}, 2003:
\newblock {\it Understanding Climate Change Feedbacks\/}.
\newblock National Research Council of the National Academy of Sciences, The
  National Academies Press, Washington, D.C.

\bibitem[{{Hayashi} et~al.(2000){\it {Hayashi}, {Ishioka}, {Yamada}, and
  {Yoden}\/}}]{hayashi-etal-2000}
{Hayashi}, Y.-Y., K.~{Ishioka}, M.~{Yamada}, and S.~{Yoden}, 2000:
\newblock {Emergence of circumpolar vortex in two dimensional turbulence on a
  rotating sphere}.
\newblock  {\it Proceedings of the IUTAM Symposium on Developments in
  Geophysical Turbulence (Fluid Mechanics and its Applications V. 58\/}, R.~M.
  Kerr and Y.~Kimura, Eds., Kluwer Academic Pub., pp. 179--192.

\bibitem[{{Hayashi} et~al.(2007){\it {Hayashi}, {Nishizawa}, {Takehiro},
  {Yamada}, {Ishioka}, and {Yoden}\/}}]{hayashi-etal-2007}
{Hayashi}, Y.-Y., S.~{Nishizawa}, S.-I. {Takehiro}, M.~{Yamada}, K.~{Ishioka},
  and S.~{Yoden}, 2007:
\newblock {Rossby Waves and Jets in Two-Dimensional Decaying Turbulence on a
  Rotating Sphere}.
\newblock {\it Journal of Atmospheric Sciences\/}, {\bf 64}, 4246--4269.

\bibitem[{{Haynes}(2005){\it {Haynes}\/}}]{haynes-2005}
{Haynes}, P., 2005:
\newblock {Stratospheric Dynamics}.
\newblock {\it Annual Review of Fluid Mechanics\/}, {\bf 37}, 263--293.

\bibitem[{{Heimpel} et~al.(2005){\it {Heimpel}, {Aurnou}, and
  {Wicht}\/}}]{heimpel-etal-2005}
{Heimpel}, M., J.~{Aurnou}, and J.~{Wicht}, 2005:
\newblock {Simulation of equatorial and high-latitude jets on Jupiter in a deep
  convection model}.
\newblock {\it \nat\/}, {\bf 438}, 193--196.

\bibitem[{{Held}(2005){\it {Held}\/}}]{held-2005}
{Held}, I., 2005:
\newblock {The gap between simulation and understanding in climate modeling}.
\newblock {\it Bull. Amer. Meteorological Soc.\/}, {\bf 86}, 1609--1614.

\bibitem[{{Held}(1999){\it {Held}\/}}]{held-1999}
{Held}, I.~M., 1999:
\newblock {The macroturbulence of the troposphere}.
\newblock {\it Tellus\/}, {\bf 51A-B}, 59--70.

\bibitem[{{Held}(2000){\it {Held}\/}}]{held-2000}
{Held}, I.~M., 2000:
\newblock {The general circulation of the atmosphere}.
\newblock {\it Paper presented at 2000 Woods Hole Oceanographic Institute
  Geophysical Fluid Dynamics Program, Woods Hole Oceanographic Institute, Woods
  Hole, MA (available at http://www.whoi.edu/page.do?pid=13076)\/}.

\bibitem[{{Held} and {Hou}(1980){\it {Held} and {Hou}\/}}]{held-hou-1980}
{Held}, I.~M., and A.~Y. {Hou}, 1980:
\newblock {Nonlinear Axially Symmetric Circulations in a Nearly Inviscid
  Atmosphere.}
\newblock {\it Journal of Atmospheric Sciences\/}, {\bf 37}, 515--533.

\bibitem[{{Held} and {Larichev}(1996){\it {Held} and
  {Larichev}\/}}]{held-larichev-1996}
{Held}, I.~M., and V.~D. {Larichev}, 1996:
\newblock {A scaling theory for horizontally homogeneous, baroclinically
  unstable flow on a beta plane}.
\newblock {\it Journal of Atmospheric Sciences\/}, {\bf 53}, 946--952.

\bibitem[{{Held} and {Soden}(2000){\it {Held} and {Soden}\/}}]{held-soden-2000}
{Held}, I.~M., and B.~J. {Soden}, 2000:
\newblock {Water vapor feedback anad global warming}.
\newblock {\it Annu. Rev. Energy Environ.\/}, {\bf 25}, 441--475.

\bibitem[{{Henry} et~al.(2000){\it {Henry}, {Marcy}, {Butler}, and
  {Vogt}\/}}]{henry-etal-2000}
{Henry}, G.~W., G.~W. {Marcy}, R.~P. {Butler}, and S.~S. {Vogt}, 2000:
\newblock {A Transiting ``51 Peg-like'' Planet}.
\newblock {\it \apjl\/}, {\bf 529}, L41--L44.

\bibitem[{{Herrnstein} and {Dowling}(2007){\it {Herrnstein} and
  {Dowling}\/}}]{herrnstein-dowling-2007}
{Herrnstein}, A., and T.~E. {Dowling}, 2007:
\newblock {Effects of topography on the spin-up of a Venus atmospheric model}.
\newblock {\it Journal of Geophysical Research (Planets)\/}, {\bf 112}({E11}).

\bibitem[{{Hoffert} et~al.(1981){\it {Hoffert}, {Callegari}, {Hsieh}, and
  {Ziegler}\/}}]{hoffert-etal-1981}
{Hoffert}, M.~I., A.~J. {Callegari}, C.~T. {Hsieh}, and W.~{Ziegler}, 1981:
\newblock {Liquid water on Mars: an energy balance climate model for
  CO$_2$/H$_2$O atmospheres}.
\newblock {\it Icarus\/}, {\bf 47}, 112--129.

\bibitem[{{Hoffman} and {Schrag}(2002){\it {Hoffman} and
  {Schrag}\/}}]{hoffman-schrag-2002}
{Hoffman}, P.~F., and D.~{Schrag}, 2002:
\newblock {Review article: The snowball Earth hypothesis: testing the limits of
  global change}.
\newblock {\it Terra Nova\/}, {\bf 14}, 129--155.

\bibitem[{{Holton}(2004){\it {Holton}\/}}]{holton-2004}
{Holton}, J.~R., 2004:
\newblock {\it An Introduction to Dynamic Meteorology, 4th Ed.\/}.
\newblock Academic Press, San Diego.

\bibitem[{{Huang} and {Robinson}(1998){\it {Huang} and
  {Robinson}\/}}]{huang-robinson-1998}
{Huang}, H.-P., and W.~A. {Robinson}, 1998:
\newblock {Two-Dimensional Turbulence and Persistent Zonal Jets in a Global
  Barotropic Model.}
\newblock {\it Journal of Atmospheric Sciences\/}, {\bf 55}, 611--632.

\bibitem[{{Hubbard} et~al.(1991){\it {Hubbard}, {Nellis}, {Mitchell}, {Holmes},
  {McCandless}, and {Limaye}\/}}]{hubbard-etal-1991}
{Hubbard}, W.~B., W.~J. {Nellis}, A.~C. {Mitchell}, N.~C. {Holmes}, P.~C.
  {McCandless}, and S.~S. {Limaye}, 1991:
\newblock {Interior structure of Neptune - Comparison with Uranus}.
\newblock {\it Science\/}, {\bf 253}, 648--651.

\bibitem[{{Hunt}(1979){\it {Hunt}\/}}]{hunt-1979}
{Hunt}, B.~G., 1979:
\newblock {The Influence of the Earth's Rotation Rate on the General
  Circulation of the Atmosphere.}
\newblock {\it Journal of Atmospheric Sciences\/}, {\bf 36}, 1392--1408.

\bibitem[{{Ingersoll}(1969){\it {Ingersoll}\/}}]{ingersoll-1969}
{Ingersoll}, A.~P., 1969:
\newblock {The Runaway Greenhouse: A History of Water on Venus.}
\newblock {\it Journal of Atmospheric Sciences\/}, {\bf 26}, 1191--1198.

\bibitem[{{Ingersoll}(1990){\it {Ingersoll}\/}}]{ingersoll-1990}
{Ingersoll}, A.~P., 1990:
\newblock {Atmospheric dynamics of the outer planets}.
\newblock {\it Science\/}, {\bf 248}, 308--315.

\bibitem[{{Ingersoll} and {Cuzzi}(1969){\it {Ingersoll} and
  {Cuzzi}\/}}]{ingersoll-cuzzi-1969}
{Ingersoll}, A.~P., and J.~N. {Cuzzi}, 1969:
\newblock {Dynamics of Jupiter's cloud bands.}
\newblock {\it Journal of Atmospheric Sciences\/}, {\bf 26}, 981--985.

\bibitem[{{Ingersoll} and {Porco}(1978){\it {Ingersoll} and
  {Porco}\/}}]{ingersoll-porco-1978}
{Ingersoll}, A.~P., and C.~C. {Porco}, 1978:
\newblock {Solar heating and internal heat flow on Jupiter}.
\newblock {\it Icarus\/}, {\bf 35}, 27--43.

\bibitem[{{Ingersoll} et~al.(1981){\it {Ingersoll}, {Beebe}, {Mitchell},
  {Garneau}, {Yagi}, and {Muller}\/}}]{ingersoll-etal-1981}
{Ingersoll}, A.~P., R.~F. {Beebe}, J.~L. {Mitchell}, G.~W. {Garneau}, G.~M.
  {Yagi}, and J.-P. {Muller}, 1981:
\newblock {Interaction of eddies and mean zonal flow on Jupiter as inferred
  from Voyager 1 and 2 images}.
\newblock {\it \jgr\/}, {\bf 86}, 8733--8743.

\bibitem[{{Iro} et~al.(2005){\it {Iro}, {B{\'e}zard}, and
  {Guillot}\/}}]{iro-etal-2005}
{Iro}, N., B.~{B{\'e}zard}, and T.~{Guillot}, 2005:
\newblock {A time-dependent radiative model of HD 209458b}.
\newblock {\it \aap\/}, {\bf 436}, 719--727.

\bibitem[{{Irwin} et~al.(2009){\it {Irwin}, {Charbonneau}, {Nutzman}, and
  {Falco}\/}}]{irwin-etal-2009}
{Irwin}, J., D.~{Charbonneau}, P.~{Nutzman}, and E.~{Falco}, 2009:
\newblock {The MEarth project: searching for transiting habitable super-Earths
  around nearby M dwarfs}.
\newblock  {\it IAU Symposium\/}, vol. 253 of {\it IAU Symposium\/}, pp.
  37--43.

\bibitem[{{Ishioka} et~al.(1999){\it {Ishioka}, {Yamada}, {Hayashi}, and
  {Yoden}\/}}]{ishioka-etal-1999}
{Ishioka}, K., M.~{Yamada}, Y.-Y. {Hayashi}, and S.~{Yoden}, 1999:
\newblock {Pattern formation from two-dimensional decaying turbulence on a
  rotating sphere}.
\newblock {\it Nagare Multimedia, The Japan Society of Fluid Mechanics\/}, {\bf
  [available online at http://www.nagare.or.jp/mm/99/ishioka/]}.

\bibitem[{{James}(1994){\it {James}\/}}]{james-1994}
{James}, I.~N., 1994:
\newblock {\it Introduction to Circulating Atmospheres\/}.
\newblock Cambridge Atmospheric and Space Science Series, Cambridge University
  Press, UK.

\bibitem[{{Joshi}(2003){\it {Joshi}\/}}]{joshi-2003}
{Joshi}, M., 2003:
\newblock {Climate Model Studies of Synchronously Rotating Planets}.
\newblock {\it Astrobiology\/}, {\bf 3}, 415--427.

\bibitem[{{Joshi} et~al.(1997){\it {Joshi}, {Haberle}, and
  {Reynolds}\/}}]{joshi-etal-1997}
{Joshi}, M.~M., R.~M. {Haberle}, and R.~T. {Reynolds}, 1997:
\newblock {Simulations of the Atmospheres of Synchronously Rotating Terrestrial
  Planets Orbiting M Dwarfs: Conditions for Atmospheric Collapse and the
  Implications for Habitability}.
\newblock {\it Icarus\/}, {\bf 129}, 450--465.

\bibitem[{{Kaspi} et~al.(2009){\it {Kaspi}, {Flierl}, and
  {Showman}\/}}]{kaspi-etal-2009}
{Kaspi}, Y., G.~R. {Flierl}, and A.~P. {Showman}, 2009:
\newblock {The deep wind structure of the giant planets: results from an
  anelastic general circulation model}.
\newblock {\it Icarus\/}, {\bf 202}, 525--542.

\bibitem[{{Kasting}(1988){\it {Kasting}\/}}]{kasting-1988}
{Kasting}, J.~F., 1988:
\newblock {Runaway and moist greenhouse atmospheres and the evolution of earth
  and Venus}.
\newblock {\it Icarus\/}, {\bf 74}, 472--494.

\bibitem[{{Kasting} and {Catling}(2003){\it {Kasting} and
  {Catling}\/}}]{kasting-catling-2003}
{Kasting}, J.~F., and D.~{Catling}, 2003:
\newblock {Evolution of a Habitable Planet}.
\newblock {\it \araa\/}, {\bf 41}, 429--463.

\bibitem[{{Kasting} et~al.(1993){\it {Kasting}, {Whitmire}, and
  {Reynolds}\/}}]{kasting-etal-1993}
{Kasting}, J.~F., D.~P. {Whitmire}, and R.~T. {Reynolds}, 1993:
\newblock {Habitable Zones around Main Sequence Stars}.
\newblock {\it Icarus\/}, {\bf 101}, 108--128.

\bibitem[{{Kawamura} et~al.(2007){\it {Kawamura},
  et~al.\/}}]{kawamura-etal-2007}
{Kawamura}, K., et~al., 2007:
\newblock {Northern hemisphere forcing of climatic cycles in Antarctica over
  the past 360,000 years}.
\newblock {\it \nat\/}, {\bf 448}, 912--916.

\bibitem[{{Kim} and {Lee}(2001){\it {Kim} and {Lee}\/}}]{kim-lee-2001b}
{Kim}, H.-K., and S.~{Lee}, 2001:
\newblock {Hadley Cell Dynamics in a Primitive Equation Model. Part II:
  Nonaxisymmetric Flow.}
\newblock {\it Journal of Atmospheric Sciences\/}, {\bf 58}, 2859--2871.

\bibitem[{{Kirk} and {Stevenson}(1987){\it {Kirk} and
  {Stevenson}\/}}]{kirk-stevenson-1987}
{Kirk}, R.~L., and D.~J. {Stevenson}, 1987:
\newblock {Hydromagnetic constraints on deep zonal flow in the giant planets}.
\newblock {\it \apj\/}, {\bf 316}, 836--846.

\bibitem[{{Knutson} et~al.(2007){\it {Knutson}, {Charbonneau}, {Allen},
  {Fortney}, {Agol}, {Cowan}, {Showman}, {Cooper}, and
  {Megeath}\/}}]{knutson-etal-2007b}
{Knutson}, H.~A., D.~{Charbonneau}, L.~E. {Allen}, J.~J. {Fortney}, E.~{Agol},
  N.~B. {Cowan}, A.~P. {Showman}, C.~S. {Cooper}, and S.~T. {Megeath}, 2007:
\newblock {A map of the day-night contrast of the extrasolar planet HD
  189733b}.
\newblock {\it \nat\/}, {\bf 447}, 183--186.

\bibitem[{{Knutson} et~al.(2008){\it {Knutson}, {Charbonneau}, {Allen},
  {Burrows}, and {Megeath}\/}}]{knutson-etal-2008a}
{Knutson}, H.~A., D.~{Charbonneau}, L.~E. {Allen}, A.~{Burrows}, and S.~T.
  {Megeath}, 2008:
\newblock {The 3.6-8.0 {$\mu$}m Broadband Emission Spectrum of HD 209458b:
  Evidence for an Atmospheric Temperature Inversion}.
\newblock {\it \apj\/}, {\bf 673}, 526--531.

\bibitem[{{Knutson} et~al.(2009{\natexlab{a}}){\it {Knutson}, {Charbonneau},
  {Burrows}, {O'Donovan}, and {Mandushev}\/}}]{knutson-etal-2009b}
{Knutson}, H.~A., D.~{Charbonneau}, A.~{Burrows}, F.~T. {O'Donovan}, and
  G.~{Mandushev}, 2009{\natexlab{a}}:
\newblock {Detection of A Temperature Inversion in the Broadband Infrared
  Emission Spectrum of TrES-4}.
\newblock {\it \apj\/}, {\bf 691}, 866--874.

\bibitem[{{Knutson} et~al.(2009{\natexlab{b}}){\it {Knutson}, {Charbonneau},
  {Cowan}, {Fortney}, {Showman}, {Agol}, {Henry}, {Everett}, and
  {Allen}\/}}]{knutson-etal-2009a}
{Knutson}, H.~A., D.~{Charbonneau}, N.~B. {Cowan}, J.~J. {Fortney}, A.~P.
  {Showman}, E.~{Agol}, G.~W. {Henry}, M.~E. {Everett}, and L.~E. {Allen},
  2009{\natexlab{b}}:
\newblock {Multiwavelength Constraints on the Day-Night Circulation Patterns of
  HD 189733b}.
\newblock {\it \apj\/}, {\bf 690}, 822--836.

\bibitem[{{Korty} and {Schneider}(2008){\it {Korty} and
  {Schneider}\/}}]{korty-schneider-2008}
{Korty}, R.~L., and T.~{Schneider}, 2008:
\newblock {Extent of Hadley circulations in dry atmospheres}.
\newblock {\it \grl\/}, {\bf 35}, L23,803.

\bibitem[{{Langton} and {Laughlin}(2007){\it {Langton} and
  {Laughlin}\/}}]{langton-laughlin-2007}
{Langton}, J., and G.~{Laughlin}, 2007:
\newblock {Observational Consequences of Hydrodynamic Flows on Hot Jupiters}.
\newblock {\it \apjl\/}, {\bf 657}, L113--L116.

\bibitem[{{Langton} and {Laughlin}(2008{\natexlab{a}}){\it {Langton} and
  {Laughlin}\/}}]{langton-laughlin-2008}
{Langton}, J., and G.~{Laughlin}, 2008{\natexlab{a}}:
\newblock {Hydrodynamic Simulations of Unevenly Irradiated Jovian Planets}.
\newblock {\it \apj\/}, {\bf 674}, 1106--1116.

\bibitem[{{Langton} and {Laughlin}(2008{\natexlab{b}}){\it {Langton} and
  {Laughlin}\/}}]{langton-laughlin-2008b}
{Langton}, J., and G.~{Laughlin}, 2008{\natexlab{b}}:
\newblock {Persistent circumpolar vortices on the extrasolar giant planet HD
  37605 b}.
\newblock {\it \aap\/}, {\bf 483}, L25--L28.

\bibitem[{{Laughlin} et~al.(2009){\it {Laughlin}, {Deming}, {Langton}, {Kasen},
  {Vogt}, {Butler}, {Rivera}, and {Meschiari}\/}}]{laughlin-etal-2009}
{Laughlin}, G., D.~{Deming}, J.~{Langton}, D.~{Kasen}, S.~{Vogt}, P.~{Butler},
  E.~{Rivera}, and S.~{Meschiari}, 2009:
\newblock {Rapid heating of the atmosphere of an extrasolar planet}.
\newblock {\it \nat\/}, {\bf 457}, 562--564.

\bibitem[{{Lee} et~al.(2007){\it {Lee}, {Lewis}, and {Read}\/}}]{lee-etal-2007}
{Lee}, C., S.~R. {Lewis}, and P.~L. {Read}, 2007:
\newblock {Superrotation in a Venus general circulation model}.
\newblock {\it Journal of Geophysical Research (Planets)\/}, {\bf 112}({E11}).

\bibitem[{{Lewis} et~al.(2009){\it {Lewis}, {Showman}, {Fortney}, and
  {Marley}\/}}]{lewis-etal-2009}
{Lewis}, N., A.~P. {Showman}, J.~J. {Fortney}, and M.~S. {Marley}, 2009:
\newblock {Three-Dimensional Atmospheric Dynamics of Eccentric Extrasolar
  Planets}.
\newblock  {\it American Astronomical Society Meeting Abstracts\/}, vol. 213 of
  {\it American Astronomical Society Meeting Abstracts\/}, p. 346.01.

\bibitem[{{Lian} and {Showman}(2008){\it {Lian} and
  {Showman}\/}}]{lian-showman-2008}
{Lian}, Y., and A.~P. {Showman}, 2008:
\newblock {Deep jets on gas-giant planets}.
\newblock {\it Icarus\/}, {\bf 194}, 597--615.

\bibitem[{{Lian} and {Showman}(2009){\it {Lian} and
  {Showman}\/}}]{lian-showman-2009}
{Lian}, Y., and A.~P. {Showman}, 2009:
\newblock {Generation of equatorial jets by large-scale latent heating on the
  giant planets}.
\newblock {\it Icarus, in press\/}.

\bibitem[{{Lindzen} and {Farrell}(1980{\natexlab{a}}){\it {Lindzen} and
  {Farrell}\/}}]{lindzen-farrell-1980}
{Lindzen}, R.~S., and B.~{Farrell}, 1980{\natexlab{a}}:
\newblock {The role of the polar regions in climate, and a new parameterization
  of global heat transport}.
\newblock {\it Montly Weather Review\/}, {\bf 108}, 2064--2079.

\bibitem[{{Lindzen} and {Farrell}(1980{\natexlab{b}}){\it {Lindzen} and
  {Farrell}\/}}]{lindzen-farrell-1980b}
{Lindzen}, R.~S., and B.~{Farrell}, 1980{\natexlab{b}}:
\newblock {A simple approximate result for the maximum growth rate of
  baroclinic instabilities}.
\newblock {\it Journal of the Atmospheric Sciences\/}, {\bf 37}, 1648--1654.

\bibitem[{{Lindzen} and {Hou}(1988){\it {Lindzen} and
  {Hou}\/}}]{lindzen-hou-1988}
{Lindzen}, R.~S., and A.~V. {Hou}, 1988:
\newblock {Hadley Circulations for Zonally Averaged Heating Centered off the
  Equator.}
\newblock {\it Journal of Atmospheric Sciences\/}, {\bf 45}, 2416--2427.

\bibitem[{{Little} et~al.(1999){\it {Little}, {Anger}, {Ingersoll}, {Vasavada},
  {Senske}, {Breneman}, {Borucki}, and {The Galileo SSI
  Team}\/}}]{little-etal-1999}
{Little}, B., C.~D. {Anger}, A.~P. {Ingersoll}, A.~R. {Vasavada}, D.~A.
  {Senske}, H.~H. {Breneman}, W.~J. {Borucki}, and {The Galileo SSI Team},
  1999:
\newblock {Galileo Images of Lightning on Jupiter}.
\newblock {\it Icarus\/}, {\bf 142}, 306--323.

\bibitem[{{Liu} et~al.(2008){\it {Liu}, {Goldreich}, and
  {Stevenson}\/}}]{liu-etal-2008}
{Liu}, J., P.~M. {Goldreich}, and D.~J. {Stevenson}, 2008:
\newblock {Constraints on deep-seated zonal winds inside Jupiter and Saturn}.
\newblock {\it Icarus\/}, {\bf 196}, 653--664.

\bibitem[{{Lorenz}(1967){\it {Lorenz}\/}}]{lorenz-1967}
{Lorenz}, E.~N., 1967:
\newblock {\it The Nature and Theory of the General Circulation of the
  Atmosphere\/}.
\newblock World Meteorological Org., Geneva.

\bibitem[{{Lu} et~al.(2007){\it {Lu}, {Vecchi}, and
  {Reichler}\/}}]{lu-etal-2007}
{Lu}, J., G.~A. {Vecchi}, and T.~{Reichler}, 2007:
\newblock {Expansion of the Hadley cell under global warming}.
\newblock {\it \grl\/}, {\bf 34 (L06805)}.

\bibitem[{{Machalek} et~al.(2008){\it {Machalek}, {McCullough}, {Burke},
  {Valenti}, {Burrows}, and {Hora}\/}}]{machalek-etal-2008}
{Machalek}, P., P.~R. {McCullough}, C.~J. {Burke}, J.~A. {Valenti},
  A.~{Burrows}, and J.~L. {Hora}, 2008:
\newblock {Thermal Emission of Exoplanet XO-1b}.
\newblock {\it \apj\/}, {\bf 684}, 1427--1432.

\bibitem[{{Marcus} et~al.(2000){\it {Marcus}, {Kundu}, and
  {Lee}\/}}]{marcus-etal-2000}
{Marcus}, P.~S., T.~{Kundu}, and C.~{Lee}, 2000:
\newblock {Vortex dynamics and zonal flows}.
\newblock {\it Physics of Plasmas\/}, {\bf 7}, 1630--1640.

\bibitem[{{Marcy} and {Butler}(1996){\it {Marcy} and
  {Butler}\/}}]{marcy-butler-1996}
{Marcy}, G.~W., and R.~P. {Butler}, 1996:
\newblock {A Planetary Companion to 70 Virginis}.
\newblock {\it \apjl\/}, {\bf 464}, L147+.

\bibitem[{{Mayor} and {Queloz}(1995){\it {Mayor} and
  {Queloz}\/}}]{mayor-queloz-1995}
{Mayor}, M., and D.~{Queloz}, 1995:
\newblock {A Jupiter-Mass Companion to a Solar-Type Star}.
\newblock {\it \nat\/}, {\bf 378}, 355--+.

\bibitem[{{Menou} and {Rauscher}(2009){\it {Menou} and
  {Rauscher}\/}}]{menou-rauscher-2009}
{Menou}, K., and E.~{Rauscher}, 2009:
\newblock {Atmospheric Circulation of Hot Jupiters: A Shallow Three-Dimensional
  Model}.
\newblock {\it \apj\/}, {\bf 700}, 887--897.

\bibitem[{{Menou} et~al.(2003){\it {Menou}, {Cho}, {Seager}, and
  {Hansen}\/}}]{menou-etal-2003}
{Menou}, K., J.~Y.-K. {Cho}, S.~{Seager}, and B.~M.~S. {Hansen}, 2003:
\newblock {``Weather'' Variability of Close-in Extrasolar Giant Planets}.
\newblock {\it \apjl\/}, {\bf 587}, L113--L116.

\bibitem[{{Morales-Calder{\'o}n} et~al.(2006){\it {Morales-Calder{\'o}n},
  et~al.\/}}]{morales-calderon-etal-2006}
{Morales-Calder{\'o}n}, M., et~al., 2006:
\newblock {A Sensitive Search for Variability in Late L Dwarfs: The Quest for
  Weather}.
\newblock {\it \apj\/}, {\bf 653}, 1454--1463.

\bibitem[{{Navarra} and {Boccaletti}(2002){\it {Navarra} and
  {Boccaletti}\/}}]{navarra-boccaletti-2002}
{Navarra}, A., and G.~{Boccaletti}, 2002:
\newblock {Numerical general circulation experiments of sensitivity to Earth
  rotation rate}.
\newblock {\it Climate Dynamics\/}, {\bf 19}, 467--483.

\bibitem[{{Noll} et~al.(1997){\it {Noll}, {Geballe}, and
  {Marley}\/}}]{noll-etal-1997}
{Noll}, K.~S., T.~R. {Geballe}, and M.~S. {Marley}, 1997:
\newblock {Detection of Abundant Carbon Monoxide in the Brown Dwarf Gliese
  229B}.
\newblock {\it \apjl\/}, {\bf 489}, L87+.

\bibitem[{{North}(1975){\it {North}\/}}]{north-1975}
{North}, G.~R., 1975:
\newblock {Analytical solution to a simple climate model with diffusive heat
  transport}.
\newblock {\it Journal of the Atmospheric Sciences\/}, {\bf 32}, 1301--1307.

\bibitem[{{North} et~al.(1981){\it {North}, {Cahalan}, and
  {Coakley}\/}}]{north-etal-1981}
{North}, G.~R., R.~F. {Cahalan}, and J.~A.~J. {Coakley}, 1981:
\newblock {Energy balance climate models}.
\newblock {\it Rev. Geophysics Space Physics\/}, {\bf 19}, 91--121.

\bibitem[{{Nozawa} and {Yoden}(1997){\it {Nozawa} and
  {Yoden}\/}}]{nozawa-yoden-1997a}
{Nozawa}, T., and S.~{Yoden}, 1997:
\newblock {Formation of zonal band structure in forced two-dimensional
  turbulence on a rotating sphere}.
\newblock {\it Physics of Fluids\/}, {\bf 9}, 2081--2093.

\bibitem[{{Okuno} and {Masuda}(2003){\it {Okuno} and
  {Masuda}\/}}]{okuno-masuda-2003}
{Okuno}, A., and A.~{Masuda}, 2003:
\newblock {Effect of horizontal divergence on the geostrophic turbulence on a
  beta-plane: Suppression of the Rhines effect}.
\newblock {\it Physics of Fluids\/}, {\bf 15}, 56--65.

\bibitem[{{Pedlosky}(1987){\it {Pedlosky}\/}}]{pedlosky-1987}
{Pedlosky}, J., 1987:
\newblock {\it Geophysical Fluid Dynamics, 2nd Ed.\/}.
\newblock Springer-Verlag, New York.

\bibitem[{{Peixoto} and {Oort}(1992){\it {Peixoto} and
  {Oort}\/}}]{peixoto-oort-1992}
{Peixoto}, J.~P., and A.~H. {Oort}, 1992:
\newblock {\it Physics of Climate\/}.
\newblock American Institute of Physics, New York.

\bibitem[{{Podolak} et~al.(1991){\it {Podolak}, {Hubbard}, and
  {Stevenson}\/}}]{podolak-etal-1991}
{Podolak}, M., W.~B. {Hubbard}, and D.~J. {Stevenson}, 1991:
\newblock {Models of Uranus' interior and magnetic field}.
\newblock  {\it Uranus\/}, J.~T. {Bergstralh}, E.~D. {Miner}, and M.~S.
  {Matthews}, Eds., Univ. Arizona Press, Tucson, AZ, pp. 29--61.

\bibitem[{{Prinn} and {Barshay}(1977){\it {Prinn} and
  {Barshay}\/}}]{prinn-barshay-1977}
{Prinn}, R.~G., and S.~S. {Barshay}, 1977:
\newblock {Carbon monoxide on Jupiter and implications for atmospheric
  convection}.
\newblock {\it Science\/}, {\bf 198}, 1031--1034.

\bibitem[{{Rauscher} et~al.(2007){\it {Rauscher}, {Menou}, {Cho}, {Seager}, and
  {Hansen}\/}}]{rauscher-etal-2007a}
{Rauscher}, E., K.~{Menou}, J.~Y.-K. {Cho}, S.~{Seager}, and B.~M.~S. {Hansen},
  2007:
\newblock {Hot Jupiter Variability in Eclipse Depth}.
\newblock {\it \apjl\/}, {\bf 662}, L115--L118.

\bibitem[{{Rauscher} et~al.(2008){\it {Rauscher}, {Menou}, {Cho}, {Seager}, and
  {Hansen}\/}}]{rauscher-etal-2008}
{Rauscher}, E., K.~{Menou}, J.~Y.-K. {Cho}, S.~{Seager}, and B.~M.~S. {Hansen},
  2008:
\newblock {On Signatures of Atmospheric Features in Thermal Phase Curves of Hot
  Jupiters}.
\newblock {\it \apj\/}, {\bf 681}, 1646--1652.

\bibitem[{{Reinaud} et~al.(2003){\it {Reinaud}, {Dritschel}, and
  {Koudella}\/}}]{reinaud-etal-2003}
{Reinaud}, J.~N., D.~G. {Dritschel}, and C.~R. {Koudella}, 2003:
\newblock {The shape of vortices in quasi-geostrophic turbulence}.
\newblock {\it Journal of Fluid Mechanics\/}, {\bf 474}, 175--192.

\bibitem[{{Rhines}(1975){\it {Rhines}\/}}]{rhines-1975}
{Rhines}, P.~B., 1975:
\newblock {Waves and turbulence on a beta-plane}.
\newblock {\it Journal of Fluid Mechanics\/}, {\bf 69}, 417--443.

\bibitem[{{Richardson} et~al.(2007){\it {Richardson}, {Deming}, {Horning},
  {Seager}, and {Harrington}\/}}]{richardson-etal-2007}
{Richardson}, L.~J., D.~{Deming}, K.~{Horning}, S.~{Seager}, and
  J.~{Harrington}, 2007:
\newblock {A spectrum of an extrasolar planet}.
\newblock {\it \nat\/}, {\bf 445}, 892--895.

\bibitem[{{Salby}(1996){\it {Salby}\/}}]{salby-1996}
{Salby}, M.~L., 1996:
\newblock {\it {Fundamentals of Atmospheric Physics}\/}.
\newblock San Diego: Academic Press, 1996.

\bibitem[{{Sanchez-Lavega} et~al.(2000){\it {Sanchez-Lavega}, {Rojas}, and
  {Sada}\/}}]{sanchez-lavega-etal-2000}
{Sanchez-Lavega}, A., J.~F. {Rojas}, and P.~V. {Sada}, 2000:
\newblock {Saturn's Zonal Winds at Cloud Level}.
\newblock {\it Icarus\/}, {\bf 147}, 405--420.

\bibitem[{{S{\'a}nchez-Lavega} et~al.(2008){\it {S{\'a}nchez-Lavega},
  et~al.\/}}]{sanchez-lavega-etal-2008}
{S{\'a}nchez-Lavega}, A., et~al., 2008:
\newblock {Depth of a strong jovian jet from a planetary-scale disturbance
  driven by storms}.
\newblock {\it \nat\/}, {\bf 451}, 437--440.

\bibitem[{{Saravanan}(1993){\it {Saravanan}\/}}]{saravanan-1993}
{Saravanan}, R., 1993:
\newblock {Equatorial Superrotation and Maintenance of the General Circulation
  in Two-Level Models.}
\newblock {\it Journal of Atmospheric Sciences\/}, {\bf 50}, 1211--1227.

\bibitem[{{Saumon} et~al.(1996){\it {Saumon}, {Hubbard}, {Burrows}, {Guillot},
  {Lunine}, and {Chabrier}\/}}]{saumon-etal-1996}
{Saumon}, D., W.~B. {Hubbard}, A.~{Burrows}, T.~{Guillot}, J.~I. {Lunine}, and
  G.~{Chabrier}, 1996:
\newblock {A Theory of Extrasolar Giant Planets}.
\newblock {\it \apj\/}, {\bf 460}, 993--1018.

\bibitem[{{Saumon} et~al.(2000){\it {Saumon}, {Geballe}, {Leggett}, {Marley},
  {Freedman}, {Lodders}, {Fegley}, and {Sengupta}\/}}]{saumon-etal-2000}
{Saumon}, D., T.~R. {Geballe}, S.~K. {Leggett}, M.~S. {Marley}, R.~S.
  {Freedman}, K.~{Lodders}, B.~{Fegley}, Jr., and S.~K. {Sengupta}, 2000:
\newblock {Molecular Abundances in the Atmosphere of the T Dwarf GL 229B}.
\newblock {\it \apj\/}, {\bf 541}, 374--389.

\bibitem[{{Saumon} et~al.(2006){\it {Saumon}, {Marley}, {Cushing}, {Leggett},
  {Roellig}, {Lodders}, and {Freedman}\/}}]{saumon-etal-2006}
{Saumon}, D., M.~S. {Marley}, M.~C. {Cushing}, S.~K. {Leggett}, T.~L.
  {Roellig}, K.~{Lodders}, and R.~S. {Freedman}, 2006:
\newblock {Ammonia as a Tracer of Chemical Equilibrium in the T7.5 Dwarf Gliese
  570D}.
\newblock {\it \apj\/}, {\bf 647}, 552--557.

\bibitem[{{Saumon} et~al.(2007){\it {Saumon}, et~al.\/}}]{saumon-etal-2007}
{Saumon}, D., et~al., 2007:
\newblock {Physical Parameters of Two Very Cool T Dwarfs}.
\newblock {\it \apj\/}, {\bf 656}, 1136--1149.

\bibitem[{{Sayanagi} et~al.(2008){\it {Sayanagi}, {Showman}, and
  {Dowling}\/}}]{sayanagi-etal-2008}
{Sayanagi}, K.~M., A.~P. {Showman}, and T.~E. {Dowling}, 2008:
\newblock {The Emergence of Multiple Robust Zonal Jets from Freely Evolving,
  Three-Dimensional Stratified Geostrophic Turbulence with Applications to
  Jupiter}.
\newblock {\it Journal of Atmospheric Sciences\/}, {\bf 65}, 3947--3962.

\bibitem[{{Schneider}(2006){\it {Schneider}\/}}]{schneider-2006}
{Schneider}, T., 2006:
\newblock {The General Circulation of the Atmosphere}.
\newblock {\it Annual Review of Earth and Planetary Sciences\/}, {\bf 34},
  655--688.

\bibitem[{{Schneider} and {Liu}(2009){\it {Schneider} and
  {Liu}\/}}]{schneider-liu-2009}
{Schneider}, T., and J.~{Liu}, 2009:
\newblock {Formation of Jets and Equatorial Superrotation on Jupiter}.
\newblock {\it J. Atmos. Sci.\/}, {\bf 66}, 579--601.

\bibitem[{{Schneider} and {Walker}(2008){\it {Schneider} and
  {Walker}\/}}]{schneider-walker-2008}
{Schneider}, T., and C.~C. {Walker}, 2008:
\newblock {Scaling laws and regime transitions of macroturbulence in dry
  atmospheres}.
\newblock {\it J. Atmos. Sci.\/}, {\bf 65}, 2153--2173.

\bibitem[{{Scott} and {Polvani}(2007){\it {Scott} and
  {Polvani}\/}}]{scott-polvani-2007}
{Scott}, R.~K., and L.~{Polvani}, 2007:
\newblock {Forced-dissipative shallow water turbulence on the sphere and the
  atmospheric circulation of the giant planets}.
\newblock {\it J. Atmos. Sci\/}, {\bf 64}, 3158--3176.

\bibitem[{{Scott} and {Polvani}(2008){\it {Scott} and
  {Polvani}\/}}]{scott-polvani-2008}
{Scott}, R.~K., and L.~M. {Polvani}, 2008:
\newblock {Equatorial superrotation in shallow atmospheres}.
\newblock {\it \grl\/}, {\bf 35}, L24,202.

\bibitem[{{Seager} et~al.(2008){\it {Seager}, {Deming}, and
  {Valenti}\/}}]{seager-etal-2008}
{Seager}, S., D.~{Deming}, and J.~A. {Valenti}, 2008:
\newblock {Transiting Exoplanets with JWST}.
\newblock {\it ArXiv e-prints\/}.

\bibitem[{{Showman}(2007){\it {Showman}\/}}]{showman-2007}
{Showman}, A.~P., 2007:
\newblock {Numerical simulations of forced shallow-water turbulence: effects of
  moist convection on the large-scale circulation of Jupiter and Saturn}.
\newblock {\it J. Atmos. Sci.\/}, {\bf 64}, 3132--3157.

\bibitem[{{Showman} and {Guillot}(2002){\it {Showman} and
  {Guillot}\/}}]{showman-guillot-2002}
{Showman}, A.~P., and T.~{Guillot}, 2002:
\newblock {Atmospheric circulation and tides of ``51 Pegasus b-like'' planets}.
\newblock {\it \aap\/}, {\bf 385}, 166--180.

\bibitem[{{Showman} et~al.(2006){\it {Showman}, {Gierasch}, and
  {Lian}\/}}]{showman-etal-2006}
{Showman}, A.~P., P.~J. {Gierasch}, and Y.~{Lian}, 2006:
\newblock {Deep zonal winds can result from shallow driving in a giant-planet
  atmosphere}.
\newblock {\it Icarus\/}, {\bf 182}, 513--526.

\bibitem[{{Showman} et~al.(2008{\natexlab{a}}){\it {Showman}, {Cooper},
  {Fortney}, and {Marley}\/}}]{showman-etal-2008a}
{Showman}, A.~P., C.~S. {Cooper}, J.~J. {Fortney}, and M.~S. {Marley},
  2008{\natexlab{a}}:
\newblock {Atmospheric Circulation of Hot Jupiters: Three-dimensional
  Circulation Models of HD 209458b and HD 189733b with Simplified Forcing}.
\newblock {\it \apj\/}, {\bf 682}, 559--576.

\bibitem[{{Showman} et~al.(2008{\natexlab{b}}){\it {Showman}, {Menou}, and
  {Cho}\/}}]{showman-etal-2008b}
{Showman}, A.~P., K.~{Menou}, and J.~Y.-K. {Cho}, 2008{\natexlab{b}}:
\newblock {Atmospheric Circulation of Hot Jupiters: A Review of Current
  Understanding}.
\newblock  {\it Extreme Solar Systems\/}, D.~{Fischer}, F.~A. {Rasio}, S.~E.
  {Thorsett}, and A.~{Wolszczan}, Eds., vol. 398 of {\it Astronomical Society
  of the Pacific Conference Series\/}, pp. 419--441.

\bibitem[{{Showman} et~al.(2009){\it {Showman}, {Fortney}, {Lian}, {Marley},
  {Freedman}, {Knutson}, and {Charbonneau}\/}}]{showman-etal-2009}
{Showman}, A.~P., J.~J. {Fortney}, Y.~{Lian}, M.~S. {Marley}, R.~S. {Freedman},
  H.~A. {Knutson}, and D.~{Charbonneau}, 2009:
\newblock {Atmospheric Circulation of Hot Jupiters: Coupled Radiative-Dynamical
  General Circulation Model Simulations of HD 189733b and HD 209458b}.
\newblock {\it \apj\/}, {\bf 699}, 564--584.

\bibitem[{{Smith}(2004){\it {Smith}\/}}]{smith-2004}
{Smith}, K.~S., 2004:
\newblock {A local model for planetary atmospheres forced by small-scale
  convection}.
\newblock {\it J. Atmos. Sciences\/}, {\bf 61}, 1420--1433.

\bibitem[{{Spiegel} et~al.(2008){\it {Spiegel}, {Menou}, and
  {Scharf}\/}}]{spiegel-etal-2008}
{Spiegel}, D.~S., K.~{Menou}, and C.~A. {Scharf}, 2008:
\newblock {Habitable Climates}.
\newblock {\it \apj\/}, {\bf 681}, 1609--1623.

\bibitem[{{Spiegel} et~al.(2009){\it {Spiegel}, {Menou}, and
  {Scharf}\/}}]{spiegel-etal-2009}
{Spiegel}, D.~S., K.~{Menou}, and C.~A. {Scharf}, 2009:
\newblock {Habitable Climates: The Influence of Obliquity}.
\newblock {\it \apj\/}, {\bf 691}, 596--610.

\bibitem[{{Stevens}(2005){\it {Stevens}\/}}]{stevens-2005}
{Stevens}, B., 2005:
\newblock {Atmospheric Moist Convection}.
\newblock {\it Annual Review of Earth and Planetary Sciences\/}, {\bf 33},
  605--643.

\bibitem[{{Stevenson}(1979){\it {Stevenson}\/}}]{stevenson-1979}
{Stevenson}, D.~J., 1979:
\newblock {Turbulent thermal convection in the presence of rotation and a
  magnetic field - A heuristic theory}.
\newblock {\it Geophysical and Astrophysical Fluid Dynamics\/}, {\bf 12},
  139--169.

\bibitem[{{Stevenson}(1991){\it {Stevenson}\/}}]{stevenson-1991}
{Stevenson}, D.~J., 1991:
\newblock {The search for brown dwarfs}.
\newblock {\it \araa\/}, {\bf 29}, 163--193.

\bibitem[{{Stone}(1972){\it {Stone}\/}}]{stone-1972}
{Stone}, P.~H., 1972:
\newblock {A simplified radiative-dynamical model for the static stability of
  rotating atmospheres}.
\newblock {\it Journal of the Atmospheric Sciences\/}, {\bf 29}, 406--418.

\bibitem[{{Stone}(1978){\it {Stone}\/}}]{stone-1978}
{Stone}, P.~H., 1978:
\newblock {Baroclinic Adjustment.}
\newblock {\it Journal of Atmospheric Sciences\/}, {\bf 35}, 561--571.

\bibitem[{{Suarez} and {Duffy}(1992){\it {Suarez} and
  {Duffy}\/}}]{suarez-duffy-1992}
{Suarez}, M.~J., and D.~G. {Duffy}, 1992:
\newblock {Terrestrial Superrotation: A Bifurcation of the General
  Circulation.}
\newblock {\it Journal of Atmospheric Sciences\/}, {\bf 49}, 1541--1556.

\bibitem[{{Sukoriansky} et~al.(2007){\it {Sukoriansky}, {Dikovskaya}, and
  {Galperin}\/}}]{sukoriansky-etal-2007}
{Sukoriansky}, S., N.~{Dikovskaya}, and B.~{Galperin}, 2007:
\newblock {On the "arrest" of inverse energy cascade and the Rhines scale}.
\newblock {\it J. Atmos. Sci.\/}, {\bf 64}, 3312--3327.

\bibitem[{{Swain} et~al.(2008){\it {Swain}, {Vasisht}, and
  {Tinetti}\/}}]{swain-etal-2008}
{Swain}, M.~R., G.~{Vasisht}, and G.~{Tinetti}, 2008:
\newblock {The presence of methane in the atmosphere of an extrasolar planet}.
\newblock {\it \nat\/}, {\bf 452}, 329--331.

\bibitem[{{Swain} et~al.(2009){\it {Swain}, {Vasisht}, {Tinetti}, {Bouwman},
  {Chen}, {Yung}, {Deming}, and {Deroo}\/}}]{swain-etal-2009}
{Swain}, M.~R., G.~{Vasisht}, G.~{Tinetti}, J.~{Bouwman}, P.~{Chen}, Y.~{Yung},
  D.~{Deming}, and P.~{Deroo}, 2009:
\newblock {Molecular Signatures in the Near-Infrared Dayside Spectrum of HD
  189733b}.
\newblock {\it \apjl\/}, {\bf 690}, L114--L117.

\bibitem[{{Tabeling}(2002){\it {Tabeling}\/}}]{tabeling-2002}
{Tabeling}, P., 2002:
\newblock {Two-dimensional turbulence: a physicist approach}.
\newblock {\it \physrep\/}, {\bf 362}, 1--62.

\bibitem[{{Thompson} and {Young}(2007){\it {Thompson} and
  {Young}\/}}]{thompson-young-2007}
{Thompson}, A.~F., and W.~R. {Young}, 2007:
\newblock {Two-layer baroclinic eddy heat fluxes: zonal flows and energy
  balance}.
\newblock {\it J. Atmos. Sci.\/}, {\bf 64}, 3214--3231.

\bibitem[{{Tinetti} et~al.(2007){\it {Tinetti}, et~al.\/}}]{Tinetti-etal-2007b}
{Tinetti}, G., et~al., 2007:
\newblock {Water vapour in the atmosphere of a transiting extrasolar planet}.
\newblock {\it \nat\/}, {\bf 448}, 169--171.

\bibitem[{{Vallis}(2006){\it {Vallis}\/}}]{vallis-2006}
{Vallis}, G.~K., 2006:
\newblock {\it Atmospheric and Oceanic Fluid Dynamics: Fundamentals and
  Large-Scale Circulation\/}.
\newblock Cambridge Univ. Press, Cambridge, UK.

\bibitem[{{Vasavada} and {Showman}(2005){\it {Vasavada} and
  {Showman}\/}}]{vasavada-showman-2005}
{Vasavada}, A.~R., and A.~P. {Showman}, 2005:
\newblock {Jovian atmospheric dynamics: an update after Galileo and Cassini}.
\newblock {\it Reports of Progress in Physics\/}, {\bf 68}, 1935--1996.

\bibitem[{{Walker} and {Schneider}(2005){\it {Walker} and
  {Schneider}\/}}]{walker-schneider-2005}
{Walker}, C.~C., and T.~{Schneider}, 2005:
\newblock {Response of idealized Hadley circulations to seasonally varying
  heating}.
\newblock {\it \grl\/}, {\bf 32}, L06,813.

\bibitem[{{Walker} and {Schneider}(2006){\it {Walker} and
  {Schneider}\/}}]{walker-schneider-2006}
{Walker}, C.~C., and T.~{Schneider}, 2006:
\newblock {Eddy Influences on Hadley Circulations: Simulations with an
  Idealized GCM}.
\newblock {\it Journal of Atmospheric Sciences\/}, {\bf 63}, 3333--3350.

\bibitem[{{Ward}(1974){\it {Ward}\/}}]{ward-1974}
{Ward}, W.~R., 1974:
\newblock {Climatic variations on Mars. I. Astronomical theory of insolation.}
\newblock {\it \jgr\/}, {\bf 79}, 3375--3386.

\bibitem[{{Williams} and {Kasting}(1997){\it {Williams} and
  {Kasting}\/}}]{williams-kasting-1997}
{Williams}, D.~M., and J.~F. {Kasting}, 1997:
\newblock {Habitable Planets with High Obliquities}.
\newblock {\it Icarus\/}, {\bf 129}, 254--267.

\bibitem[{{Williams} and {Pollard}(2002){\it {Williams} and
  {Pollard}\/}}]{williams-pollard-2002}
{Williams}, D.~M., and D.~{Pollard}, 2002:
\newblock {Earth-like worlds on eccentric orbits: excursions beyond the
  habitable zone}.
\newblock {\it International Journal of Astrobiology\/}, {\bf 1}, 61--69.

\bibitem[{{Williams} and {Pollard}(2003){\it {Williams} and
  {Pollard}\/}}]{williams-pollard-2003}
{Williams}, D.~M., and D.~{Pollard}, 2003:
\newblock {Extraordinary climates of Earth-like planets: three-dimensional
  climate simulations at extreme obliquity}.
\newblock {\it International Journal of Astrobiology\/}, {\bf 2}, 1--19.

\bibitem[{{Williams}(1978){\it {Williams}\/}}]{williams-1978}
{Williams}, G.~P., 1978:
\newblock {Planetary circulations. I - Barotropic representation of Jovian and
  terrestrial turbulence}.
\newblock {\it Journal of Atmospheric Sciences\/}, {\bf 35}, 1399--1426.

\bibitem[{{Williams}(1979){\it {Williams}\/}}]{williams-1979}
{Williams}, G.~P., 1979:
\newblock {Planetary circulations. II - The Jovian quasi-geostrophic regime}.
\newblock {\it Journal of Atmospheric Sciences\/}, {\bf 36}, 932--968.

\bibitem[{{Williams}(1988{\natexlab{a}}){\it {Williams}\/}}]{williams-1988a}
{Williams}, G.~P., 1988{\natexlab{a}}:
\newblock {The dynamical range of global circulations -- I}.
\newblock {\it Climate Dynamics\/}, {\bf 2}, 205--260.

\bibitem[{{Williams}(1988{\natexlab{b}}){\it {Williams}\/}}]{williams-1988b}
{Williams}, G.~P., 1988{\natexlab{b}}:
\newblock {The dynamical range of global circulations -- II}.
\newblock {\it Climate Dynamics\/}, {\bf 3}, 45--84.

\bibitem[{{Williams}(2003){\it {Williams}\/}}]{williams-2003a}
{Williams}, G.~P., 2003:
\newblock {Jovian Dynamics. Part III: Multiple, Migrating, and Equatorial
  Jets.}
\newblock {\it Journal of Atmospheric Sciences\/}, {\bf 60}, 1270--1296.

\bibitem[{{Williams} and {Holloway}(1982){\it {Williams} and
  {Holloway}\/}}]{williams-holloway-1982}
{Williams}, G.~P., and J.~L. {Holloway}, 1982:
\newblock {The range and unity of planetary circulations}.
\newblock {\it \nat\/}, {\bf 297}, 295--299.

\bibitem[{{Yamamoto} and {Takahashi}(2006){\it {Yamamoto} and
  {Takahashi}\/}}]{yamamoto-takahashi-2006}
{Yamamoto}, M., and M.~{Takahashi}, 2006:
\newblock {Superrotation Maintained by Meridional Circulation and Waves in a
  Venus-Like AGCM}.
\newblock {\it Journal of Atmospheric Sciences\/}, {\bf 63}, 3296--3314.

\bibitem[{{Yoden} et~al.(1999){\it {Yoden}, {Ishioka}, {Hayashi}, and
  {Yamada}\/}}]{yoden-etal-1999}
{Yoden}, S., K.~{Ishioka}, Y.-Y. {Hayashi}, and M.~{Yamada}, 1999:
\newblock {A further experiment on two-dimensional decaying turbulence on a
  rotating sphere}.
\newblock {\it Nuovo Cimento C Geophysics Space Physics C\/}, {\bf 22},
  803--812.

\bibitem[{{Zurita-Gotor} and {Lindzen}(2006){\it {Zurita-Gotor} and
  {Lindzen}\/}}]{zurita-gotor-lindzen-2006}
{Zurita-Gotor}, P., and R.~S. {Lindzen}, 2006:
\newblock {Theories of baroclinic adjustment and eddy equilibration}.
\newblock  {\it The Global Circulation of the Atmosphere\/}, T.~{Schneider} and
  A.~H. {Sobel}, Eds., Princeton University Press, pp. 22--46.

\bibitem[{{Zurita-Gotor} and {Vallis}(2009){\it {Zurita-Gotor} and
  {Vallis}\/}}]{zurita-gotor-vallis-2009}
{Zurita-Gotor}, P., and G.~K. {Vallis}, 2009:
\newblock {Equilibration of baroclinic turbulence in primitive equations and
  quasigeostrophic models}.
\newblock {\it J. Atmos. Sci.\/}, {\bf 66}, 837--863.

\end{thebibliography}

\end{document}